\documentclass[twoside,12pt]{article}
\usepackage{graphicx}

\usepackage{epsfig}
\usepackage{pst-plot}
\usepackage{amssymb}
\usepackage{dcolumn}  
\usepackage{bm}       
\usepackage{amsmath}
\usepackage{sidecap}  %

\def\Journal#1#2#3#4{{#1} {#2} (#4) #3 }

\def\NPA{{\em Nucl. Phys.} A}

\def\PLB{{\em Phys. Lett.} B}

\def\PRL{\em Phys. Rev. Lett.}

\def\PREP{\em Phys. Rep.}

\def\PRD{{\em Phys. Rev.} D}
\def\PRC{{\em Phys. Rev.} C}

\newcommand{\be}{\begin{equation}}
\newcommand{\ee}{\end{equation}}
\newcommand{\bea}{\begin{eqnarray}}
\newcommand{\eea}{\end{eqnarray}}

\begin{document}

\title{Shell-model calculations and realistic effective interactions}

\author{L. Coraggio$^{1}$, A. Covello$^{1,2}$, A. Gargano$^{1}$, 
N. Itaco$^{1,2}$ and T. T. S. Kuo$^{3}$\\
  $^1$
  Istituto Nazionale di Fisica Nucleare,\\
   Complesso  Universitario di  Monte S. Angelo,
   Via Cintia, I-80126 Napoli, Italy \\
  $^2$
  Dipartimento di Scienze Fisiche,
  Universit\`a di Napoli Federico II,\\
  Complesso Universitario di Monte S. Angelo, 
  Via Cintia, I-80126 Napoli, Italy\\
  $^3$
  Department of Physics, SUNY, Stony Brook,
   New York 11794, USA\\
}

\maketitle

\begin{abstract}
A review is presented of the development and current status of nuclear 
shell-model calculations in which the two-body effective interaction 
between the valence nucleons is 
derived from the free nucleon-nucleon potential.  
The significant progress made in this field within the last decade is 
emphasized, in particular as regards the so-called $V_{\rm low-k}$ approach 
to the renormalization of the bare nucleon-nucleon interaction. 
In the last part 
of the review we first give a survey of realistic shell-model calculations 
from early to present days. 
Then, we  report recent results for neutron-rich nuclei near doubly magic 
$^{132}$Sn and for the whole even-mass $N=82$ isotonic chain.
These illustrate how shell-model effective interactions derived from modern 
nucleon-nucleon potentials are able to provide an accurate description of 
nuclear structure properties.
\end{abstract}

\tableofcontents
\section{Introduction} \label{sec:intr}
The shell model is the basic framework for nuclear structure calculations 
in terms of nucleons. This model, which entered into nuclear physics more 
than fifty years
ago \cite{Mayer49,Haxel49}, is based on the assumption that, as a first 
approximation, each nucleon inside the nucleus moves independently from the 
others in a spherically symmetric potential including a strong spin-orbit 
term. 
Within this approximation the nucleus is considered as an inert core, made up 
by shells filled up with neutrons and protons paired to angular momentum $J=0$,
 plus a certain number of external nucleons, the ``valence" nucleons.  As is 
well known, this extreme single-particle shell model, supplemented by 
empirical coupling rules, proved very soon to be able to account for 
various nuclear properties \cite{Mayer55}, like the angular momentum and
 parity of the ground-states of odd-mass nuclei. It was clear from the 
beginning \cite{Mayer50}, however, that for a description of nuclei with 
two or more valence nucleons the ``residual" two-body interaction between 
the valence nucleons had to be taken explicitly into account, the term
 residual meaning that part of the interaction which is not absorbed into the 
central potential. This removes the degeneracy of the states belonging to the 
same configuration and produces a mixing of different configurations. A
 fascinating account of the early stages of the nuclear shell model is given 
in the comprehensive review by Talmi \cite{Talmi03}.

In any shell-model calculation one has to start by defining a ``model space", 
namely by
specifying a set of active single-particle (SP) orbits. It is in this 
truncated Hilbert
space that the Hamiltonian matrix has to be set up and diagonalized.  A 
basic input, as mentioned above,
is the residual interaction between valence nucleons. This is in reality
 a ``model-space effective interaction",
which differs from the interaction between free nucleons in various respects. 
In fact, besides being residual
in the sense mentioned above, it must account for the configurations excluded 
from the model space.

It goes without saying that a fundamental goal of nuclear physics is 
to understand the properties of nuclei starting from the forces between 
nucleons. 
Nowadays, the $A$ nucleons in a nucleus are understood as non-relativistic
particles interacting via a Hamiltonian consisting of two-body, three-body, 
and  higher-body potentials, with the nucleon-nucleon ($NN$) term being the 
dominant one. 

In this context, it is worth emphasizing that there are two  main lines of 
attack to attain this ambitious goal. The first one is comprised of the 
so-called {\em ab initio} calculations where nuclear properties, such as 
binding and 
excitation energies, are calculated directly from the first principles of 
quantum mechanics using appropriate computational scheme.  To this category 
belong the Green's function Montecarlo Method 
(GFMC)~\cite{Pieper01a,Pieper01b},
the no-core shell-model (NCSM))~\cite{Navratil96,Navratil00}, and 
the coupled-cluster methods (CCM))~\cite{Suzuki94,Dean04}. The GFMC 
calculations, on which we shall briefly comment in Sec.~\ref{sec:NN3}, are 
at present limited to nuclei with A $\leq$ 12. This limit may be overcome 
by using limited Hilbert spaces and introducing effective interactions, which 
is done in the NCSM and CCM. Actually, coupled-cluster calculations employing 
modern $NN$ potentials have been recently performed for $^{16}$O and its 
immediate neighbours~\cite{Kowalski04,Wloch05,Gour06}.

The  main feature of the NCSM is the use of  a large, but finite number of  
harmonic oscillator basis states to diagonalize an effective Hamiltonian 
containing realistic two- and three- nucleon interactions~\cite{Nogga06}. 
In this way, no 
closed core is assumed and  all nucleons in the nucleus are treated as 
active particles.  Very recently, this approach has ben applied to the study 
of $A= 10-13$ nuclei~\cite{Navratil07}.
Clearly, all {\em ab initio} calculations need huge amount of computational 
resources and therefore, as mentioned above,  are currently limited to light
nuclei.

The second line of attack is just the theme of this review paper, namely the
use of the traditional shell-model with two-nucleon effective interactions 
derived from  the bare $NN$ potential. In this case, only the valence nucleons
 are treated as active particles. However, as we shall discuss in detail
 in Sec..~\ref{sec:SMEI}, core polarization effects are taken  into account  
perturbatively in the derivation of the effective interaction.
Of course, this approach allows to perform calculations for  
medium- and heavy-mass nuclei which are far beyond the reach of ab 
initio calculations. Many-body forces beyond the $NN$ potential may also 
play a role in the shell-model effective Hamiltonian. However, this is 
still an open problem and it is the main scope of this review to assess t
he progress made in the derivation of a two-body effective interaction 
from the free $NN$ potential.

While efforts in this direction started some forty years ago 
\cite{Dawson62,Kuo66}, for 
a long time there was widespread  skepticism  about 
the practical value of what had become known as ``realistic shell model" 
calculations. 
This was mainly related to the highly
complicated nature of the nucleon-nucleon force, in particular to the 
presence of a very strong repulsion at short distances, which 
made especially difficult solving the nuclear many-body problem. As a 
consequence, 
a major problem, and correspondingly a main source of uncertainty, in shell-model 
calculations has long been the two-body effective interaction between the valence nucleons. 
An early survey of the various approaches to this problem can be found in 
the Carg\`ese lectures by Elliott \cite{Elliott69a}, where a classification of the various 
categories of nuclear structure calculations was made by the number of free parameters 
in the two-body interaction being used in the calculation. 

Since the early 1950s through the mid 1990s hundreds of shell-model calculations have been carried out, 
many of them being very successful in describing a variety of nuclear structure phenomena. In the vast 
majority of these calculations either empirical effective interactions containing adjustable parameters 
have been used or the two body-matrix elements themselves have been treated as free parameters. The latter 
approach, which was pioneered
by Talmi \cite{Talmi63}, is generally limited to relatively small spaces owing to the large number of parameters
to be least-squares fitted to  experimental data. 

Early calculations of this kind  were performed for the $p$-shell nuclei by 
Cohen and Kurath \cite{Cohen65}, who determined fifteen matrix elements  of
the two-body interaction and two single-particle energies from 35 experimental
energies of nuclei from $A=8$ up to $A=16$. The results of these calculations
turned out to be  in very satisfactory  agreement with experiment.

A more recent and very successful application of this approach 
is that by Brown and Wildenthal \cite{Wildenthal77,Wildenthal84,Brown88}, where
the  $A=17-39$ nuclei were studied in the complete $sd$-shell space. In the 
final version of this study, 
which spanned about 10 years, 66 parameters were determined by a least squares fit to 447 binding and 
excitation energies. A measure of the quality of the results is given by the rms deviation, whose value 
turned out to be 185 keV. In this connection, it is worth mentioning that in 
the recent  work of  
Brown and Richter \cite{BrownBA06} the determination of a new effective interaction for 
the $sd$-shell has been pursued by the inclusion of an updated set of 
experimental data.
As regards the empirical effective interactions, they may be schematically divided into two categories. The first is based on the use of simple potentials such as a Gaussian or Yukawa central force with various exchange operators consistent with those present in the interaction between free nucleons. These contain several parameters, such as the 
strengths and ranges of the singlet and triplet interactions, which are usually determined by a fit to the spectroscopic properties under study. 
In the second category one may put the so-called ``schematic interactions". These contain few free parameters, typically one or two, at the price of being an oversimplified representation of the real potential. Into this category fall the well known pairing \cite{Belayev59,Kisslinger60,Kerman61}, pairing plus quadrupole \cite{Kisslinger63}, and surface delta \cite{Green65,Arvieu66,Plastino66} interactions. The successful spin and isospin
dependent Migdal interaction  \cite{Migdal67} also belongs to this category. While these simple interactions are able to reproduce some specific nuclear properties 
[see, e.g., \cite{Andreozzi90,Andreozzi92}], they are clearly  inadequate for detailed quantitative studies. 

We have given above a brief sketch of the various approaches to the determination of the shell-model effective interaction that have dominated the field for more than four
decades. This we have done to place in its proper perspective the great progress made over the last decade by the more fundamental approach employing realistic effective interactions derived from modern $NN$ potentials. We refer to 
Ref.~\cite{Talmi03} for a detailed review of shell-model calculations based on the above empirical approaches.

As we shall discuss in detail in the following sections, from the late 1970s on there 
has been substantial progress toward a microscopic approach to nuclear structure calculations starting from the free $NN$ potential $V_{NN}$. This has concerned both the two basic ingredients which come into play in this approach, namely the $NN$ potential and the many-body methods for deriving the model space effective interaction, 
$V_{\rm eff}$. As regards the first point, $NN$ potentials have been constructed, which reproduce quite accurately the $NN$ scattering data. As regards
the derivation of $V_{\rm eff}$, the first problem one is confronted with is that all
realistic $NN$ potentials have a strong repulsive core which prevents their direct use in nuclear structure calculations. As is well known, this difficulty can be overcome by resorting to the Brueckner $G$-matrix 
method (see Sec.~\ref{sec:GM}), which was originally designed for nuclear matter 
calculations. 

While in earlier calculations one had to make some approximations in calculating the $G$ matrix for finite 
nuclei, the developments of improved techniques, as for instance  the Tsai-Kuo method \cite{Tsai72,Krenciglowa74} (see Sec.~\ref{sec:GM}), allowed to 
calculate it in a practically exact way. 
Another major improvement consisted in the inclusion in the perturbative expansion for $V_{\rm eff}$ 
of folded diagrams, whose important role had been recognized by many authors, 
within the framework of the so-called $\hat Q$-box formulation (\cite{Kuo90}, 
see Sec.~\ref{sec:SMEI}).

These improvements brought about a revival of interest in shell-model 
calculations with realistic effective interactions. This started in the early 
1990s and continued to increase during the following years. Given the 
skepticism mentioned above, the main aim  of this new generation of 
realistic calculations was to give an answer to the key question of whether
they could provide an accurate description of nuclear structure properties. 
By the end of the 1990s the answer to this question turned out to be in the
affirmative. 
In fact, a substantial body of results for nuclei in various mass regions 
(see, for instance \cite{Covello01}) proved the ability of shell-model 
calculations employing two-body matrix elements derived from modern $NN$ 
potentials to provide a description of nuclear structure properties at least 
as accurate as that provided by traditional, empirical interactions. It should 
be noted that in this approach the single-nucleon energies are generally taken 
from experiment (see, e.g., Sec.~\ref{sec:res3}), so that the calculation contains 
essentially no free parameters. Based on these results, in the last few years 
the use of realistic effective interactions has been rapidly gaining ground 
in nuclear structure theory. 

As will be discussed in detail in Sec.~\ref{sec:SMC}, the $G$ matrix  has been 
routinely used in practically 
all realistic shell-model calculations through 2000. However, the $G$ matrix is model-space dependent 
as well as energy dependent; these dependences make its actual calculation 
rather involved (see Sec.~\ref{sec:GM}). 
In this connection, it may be recalled that an early criticism of the $G$-matrix method 
dates back to the 1960s \cite{Elliott68}. which led to the development of a
method for deriving directly from the phase shifts a set of matrix elements of 
$V_{NN}$ in oscillator wave functions \cite{Elliott68,Elliott69a,Elliott69b}. This resulted in the 
well-known Sussex interaction which has been used in several nuclear structure calculations.

Recently, a new approach to overcome the difficulty posed by the strong short-range repulsion contained 
in the free $NN$ potential  has been proposed \cite{Bogner01,Bogner02,Bogner03}.
The basic idea underlying this approach  is inspired by the recent applications of effective field
 theory and renormalization group to low-energy nuclear systems. Starting from 
a realistic $NN$ potential, a low-momentum $NN$ potential, $V_{\rm low-k}$, is constructed 
that preserves the physics of the original potential $V_{NN}$ up to a certain cutoff momentum 
$\Lambda$. In particular, the scattering phase shifts and deuteron binding energy calculated from $V_{NN}$
 are reproduced by $V_{\rm low-k}$. This is achieved by integrating out, in the sense of the 
renormalization group,  the high-momentum components of $V_{NN}$.
The resulting $V_{\rm low-k}$ is a smooth potential that can be used directly in nuclear structure 
calculations without first calculating the $G$ matrix.
The practical value of the $V_{\rm low-k}$ approach has been assessed by several calculations, which have shown that it provides an advantageous alternative 
to the $G$-matrix one (see Sec~\ref{sec:res1}). 

The purpose of the present paper is to give a review of the basic formalism and the current status 
of realistic shell-model calculations and a self-contained survey of the major developments in the
 history of the field as regards both the $NN$ potential and the many-body approach to the derivation 
of the effective interaction. During the last four decades there have been several reviews focused on 
either of these two subjects and we shall have cause to refer to most of them in the following sections. 
Our review is similar in spirit to the one by Hjorth-jensen {\em et al.} 
\cite{Hjorth95}, in the sense that it aims at giving 
an overall view of the various aspects of realistic shell-model calculations, including recent selected 
results. The novelty of the present paper is that it  covers the developments of the last decade 
which have brought these calculations into the mainstream of nuclear structure 
[see, for instance, \cite{Talmi03}]. 
As mentioned above, from the mid 1990s  on there has been a growing success in explaining 
experimental data by means of two-body effective interactions derived from the free $NN$ potential,
which  has evidenced the practical value of realistic shell-model calculations.
It is worth emphasizing that a major step in this direction has been the introduction of
the low-momentum potential $V_{\rm low-k}$, which greatly simplifies the microscopic derivation
of the shell-model effective interaction.
On these grounds, we may consider that a first important phase  in the  
microscopic approach to shell model, started more than 40 years ago, 
has been completed. It is just this consideration   
at the origin of the present review.

Four more recent reviews 
\cite{Brown01,Dean04,Caurier05,Otsuka01} reporting on progress in shell-model studies 
are in some ways complementary to ours, in that they discuss aspects which we have 
considered to be beyond the scope of the present review. These regard, for instance, 
a phenomenologically oriented survey of shell-model
applications \cite{Brown01} or large-scale shell-model calculations 
\cite{Dean04,Caurier05,Otsuka01}.

We start in Sec.~\ref{sec:NN} with a review of the $NN$ interaction, trying 
to give an idea  of the long-standing, painstaking work that lies behind the 
development of the 
modern high-precision potentials. In Sec.~\ref{sec:SMEI} we discuss the 
derivation of the 
shell-model effective interaction within the framework of degenerate 
perturbation 
theory. The crucial role of folded diagrams is emphasized. Sec.~\ref{sec:G&V} 
is 
devoted to the handling of the short-range repulsion contained in the free 
$NN$ potential. 
We first discuss in Sec.~\ref{sec:GM} the traditional Brueckner $G$-matrix 
method and then 
introduce in Sec.~\ref{sec:vlowk} the new approach based on the construction 
of a low-momentum 
$NN$ potential. In Sec.~\ref{sec:SMC} we first give a survey of realistic 
shell-model calculations performed over the last four decades 
(Secs.~\ref{sec:SMCEP} and .~\ref{sec:SMCMC}) and then present some results 
of recent calculations. More precisely, in Sec.~\ref{sec:res1} a comparison 
is made between the  $G$-matrix and $V_{\rm low-k}$ approaches while in 
Sec.~\ref{sec:res2} results obtained with 
different $NN$ potentials are presented. In Sec.~\ref{sec:res3} we 
report selected results of calculations for nuclei neighboring doubly magic 
$^{132}$Sn  and compare them with experiment. Finally, in Sec.~\ref{sec:res4} 
we discuss the role  of the many-body contributions to the effective 
interaction by investigating the results of a study of the even $N=82$ isotones
The last section, Sec.~\ref{sec:sum}, contains a brief summary and concluding 
remarks. 

\section{Nucleon-nucleon interaction} \label{sec:NN}

\subsection{Historical overview} \label{sec:NN1}

The nucleon-nucleon interaction has been extensively studied since the discovery of the neutron and in the course of time there have been a number of Conferences
\cite{Green67,Vinh83,Vinh90} and review papers \cite{Machleidt89,Machleidt94,Machleidt01a} marking the advances in the understanding of its nature. Here, we shall start by giving a brief historical account and a survey of the main aspects relevant to nuclear structure, the former serving the purpose to look back and recall how hard it has been making progress in this field. 

As is well known, the theory of nuclear forces started with the meson exchange idea introduced by Yukawa \cite{Yukawa35}. Following the discovery of the pion, in the 1950s many efforts were made to describe the nucleon-nucleon ($NN$) interaction in terms of pion-exchange models. However, while by the end of the 1950s the one-pion exchange (OPE) had been experimentally established as the long-range part of  $V_{NN}$, the calculations of the two-pion exchange were plagued by serious ambiguities. This led to several pion-theoretical potentials differing quite widely in the two-pion exchange effects. This unpleasant situation is well reflected in various review papers of the period of the 1950s, for instance the article by  Phillips \cite{Phillips59}; a comprehensive list of references can be found 
in Ref.~\cite{Machleidt89}. 

While the theoretical efforts mentioned above were not very successful, a substantial progress in the experimental study of the properties of the $NN$ interaction was made during the course of the 1950s. In particular, from the examination of the 
$pp$ scattering data at 340 MeV in the laboratory system \cite{Jastrow51} 
inferred the existence of a strong short-range repulsion, which he represented by a hard sphere for convenience in calculation.
As we shall discuss in detail later, this feature, which prevents the direct use of 
$V_{NN}$ in nuclear structure calculations, has been at the origin of the Brueckner $G$-matrix method (Sec.~\ref{sec:GM})  and of the recent 
$V_{\rm low-k}$ approach (Sec.~\ref{sec:vlowk}). 

At this point it must be recalled that as early as 1941 an investigation of the possible types of nonrelativistic $NN$ interaction at most linear in the relative momentum 
\mbox{\boldmath$p$} of the two nucleons and limited by invariance conditions was carried out by Eisenbud and Wigner \cite{Eisenbud41}. It turned out that the 
general form of $V_{NN}$ consists of central, spin-spin, tensor and spin-orbit terms. Some twenty years later, the most general $V_{NN}$ when all powers of \mbox{\boldmath$p$} are allowed was given by Okubo \cite{Okubo58}, which added a 
quadratic spin-orbit term. When sufficiently reliable phase-shift analyses of $NN$ scattering data became available (see for instance Ref.~\cite{Stapp57}), 
these studies were a key guide for the construction of phenomenological $NN$ potentials. In the early stages of this approach, the inclusion of all the four types of interaction resulting from the study of Eisenbud and Wigner (1941), with the assumption of charge independence, led to the Gammel-Thaler potential \cite{Gammel57}, which may be considered the first quantitative $NN$ potential. In this potential, following the suggestion of Jastrow (1951), a strong short-range repulsion represented by a hard core (infinite repulsion) at about 0.4 fm was used. As we shall see later, it took a decade before soft-core potentials were considered.  

In the early 1960s two vastly improved phenomenological potentials appeared, both going beyond the Eisenbud-Wigner form with addition of a quadratic spin-orbit term. These were developed by the Yale group \cite{Lassila62} and by 
Hamada and Johnston (HJ) \cite{Hamada62}. Both potentials have infinite 
repulsive cores and approach the one-pion-exchange-potential at large distances.
Historically, the HJ potential occupies a special place in the field of microscopic nuclear structure. In fact, it was used in the mid 1960s in the work 
of Kuo and Brown  \cite{Kuo66}, which was the first successful attempt to derive the shell-model effective interaction from the free $NN$ potential. We therefore find it appropriate to summarize here its main features.
This may also allow a comparison with the today's high-quality phenomenological potentials, as for instance Argonne $V_{18}$ (see Sec.~\ref{sec:NN2}).
The HJ potential has the form
\begin{equation}
V = V_{\rm C}(r) + V_{\rm T}(r)S_{12} + V_{LS}(r)\mbox{\boldmath ${L \cdot S}$}
+ V_{LL}(r)L_{12},
\end{equation} 
where C, T, $LS$ and $LL$ denote respectively central , tensor, spin-orbit and quadratic spin-orbit terms. The operator $S_{12}$ is the ordinary tensor operator and the quadratic spin-orbit operator is defined by
\begin{equation}
L_{12}=[\delta_{LJ} + (\mbox{\boldmath$\sigma$}_1 \cdot \mbox{\boldmath$\sigma$}_2)]
\mbox{\boldmath$L$}^2 - (\mbox{\boldmath$L \cdot S$})^2.
\end{equation}
The $V_i$ ($i$=C, T, $LS$ and $LL$) are spin-parity dependent, and hard cores, with a common radius of 0.485 fm, are present in all states. With about 30 parameters the HJ potential model reproduced in a quantitative way the $pp$ and $np$ data below 315 MeV.

As mentioned above, the era of soft-core potentials started in the late 1960s with the work of Reid \cite{Reid68} and Bressel {\em et al.} \cite{Bressel69}. 
The original Reid soft-core potential Reid68 has been updated some 25 years later \cite{Stoks94} producing a high-quality potential denoted as Reid93 (see Sec.~\ref{sec:NN2}).

Let us now come back to the meson-theory based potentials. The discovery of heavy mesons in the early 1960s revived the field. This resulted in the development of various one-boson-exchange (OBE) potentials and in a renewed confidence in the theoretical approach to the study of the $NN$ interaction.  
The optimistic view of the field brought about by the advances made during the 1960s is  reflected in the Summary \cite{Green67} of the 1967 International Conference on the Nucleon-Nucleon Interaction held at the University of Florida in Gainesville. A concise and clear account of the early OBE potentials (OBEP), including a list of relevant references, can be 
found in the review of the meson theory of nuclear forces by Machleidt 
\cite{Machleidt89}.

During the 1960s sustained efforts were made to try to understand the properties of complex nuclei in terms of the fundamental $NN$ interaction. This brought in focus the problem of how to handle the serious difficulty resulting from the strong short-range repulsion contained in the free $NN$ potential. We shall discuss this point in detail in Sec.~\ref{sec:G&V}. Here, it should be mentioned that the idea of overcoming the above difficulty by constructing a smooth, yet realistic, $NN$ potential that could be used directly in nuclear structure calculations was actively explored in the mid 1960s. This led to the development of a non-local, separable potential fitting two-nucleon scattering data with reasonable 
accuracy \cite{Tabakin64,Tabakin68}. This potential, known as Tabakin potential, was used by the MIT group in several calculations of the structure of finite nuclei within the framework of the Hartree-Fock method \cite{Kerman66,Bassichis67,Kerman67,Kerman69}. An early account of the results of nuclear structure calculations using realistic $NN$ interactions was given at the above mentioned Gainesville Conference by Moszkowski \cite{Moszkowski67}.

As regards the experimental study of the $NN$ scattering, this was also actively pursued in the 1960s (see \cite{Green67}), leading to the much improved phase-shift analysis of McGregor {\em et al.} \cite{McGregor69}, which included 
2066 $pp$ and $np$ data up to 450 MeV. This set the stage for the theoretical efforts of the 1970s, which were addressed to the construction of a quantitative $NN$ potential  (namely, able to reproduce with good accuracy all the known $NN$ scattering data) within the framework of the meson theory. In this context, a main goal was to go beyond the OBE model by taking into account multi-meson exchange, in particular the 2$\pi$-exchange contribution. These efforts were essentially based on two different approaches: dispersion relations and field theory. 

The work along these two lines, which went on for more than one decade, resulted eventually in the Paris potential 
\cite{Cottingham73,Vinh-Mau73,Lacombe75,Lacombe80,Vinh83} and in the so called ``Bonn full model" \cite{Machleidt87}, the latter including also contributions beyond 2$\pi$. In the sector of the OBE model a significant progress was made through the work of the Nijmegen group \cite{Nagels78}. This was based on Regge-pole theory and led to a quite sophisticated OBEP  which is known as the Nijmegen78 potential. The Nijmegen, Paris, and Bonn potentials fitted the world $NN$ data below 300 MeV available in 1992  with a $\chi^2$/datum = 5.12, 3.71, and 1.90, respectively \cite{Machleidt94}. 

To have a firsthand idea of the status of the theory of the $NN$ interaction around 1990 we refer to Ref.~\cite{Vinh90} while a detailed discussion of the 
above three potentials can be found in \cite{Machleidt94}. Here we would like to emphasize that they mark the beginning of a new era in the field of nuclear forces and may be considered as the first generation of $NN$ realistic potentials. In particular, as will be discussed in Sec.~\ref{sec:SMCEP}, the Paris and Bonn potentials have played an important role in the revival of interest in nuclear structure calculations starting from the bare $NN$ interaction. We shall therefore give here a brief outline of the main characteristics of these two potentials as well as of the energy-independent OBE parametrization of the Bonn full model, which has been generally employed in nuclear structure applications.  

In addition to the $2\pi$-exchange contribution, the Paris potential contains the OPE and $\omega$-meson exchange. This gives the long-range and medium-range part of the $NN$ interaction, while the short-range part is of purely phenomenological nature. In its final version \cite{Lacombe80} the Paris potential is parametrized in an analytical form consisting of a regularized discrete superposition of Yukawa-type terms. This introduces a large number of free parameters, about 60 \cite{Machleidt94}, that are determined by fitting the $NN$ scattering data.

As already mentioned, the Bonn full model is a field-theoretical meson-exchange model for the $NN$ interaction. In addition to the 2$\pi$-exchange contribution, this model contains single $\pi$, $\omega$, and $\delta$ exchanges and $\pi\rho$ contributions. It has been shown \cite{Machleidt87} that the latter are essential for a quantitative description of the phase shifts in the lower partial waves while additional $3\pi$ and $4\pi$ contributions are not very important. The Bonn full model has in all 12 parameters which are the coupling constants and cutoff masses of the meson-nucleon vertices involved. 
This model is an energy-dependent potential, which makes it inconvenient for application in nuclear structure calculations. Therefore, an energy independent one-boson parametrization of this potential has been developed within the framework of the relativistic three-dimensional Blanckenbecler-Sugar (BbS) reduction of the Bethe-Salpeter equation \cite{Machleidt87,Machleidt89}. This OBEP includes exchanges of two pseudoscalar ($\pi$ and $\eta$), two scalar ($\sigma$ and $\delta$), and two vector ($\rho$ and $\omega$) mesons. 
As in the Bonn full model, there are only twelve parameters which have to be determined through a fit of the $NN$ scattering data.

At this point, it must be pointed out that there are three variants of the above relativistic OBE potential, denoted by Bonn A, Bonn B and Bonn C. The parameters of these potentials and the predictions of Bonn B for the two-nucleon system are given in \cite{Machleidt89}. The latter are very similar to the ones by the Bonn full model.
The main difference between the three potentials is the strength of the tensor force as reflected in the predicted $D$-state probability of the deuteron $P_D$. With $P_D$=4.4\% 
Bonn A has the weakest tensor force. Bonn B and Bonn C predict 5\% and 5.6\%, respectively. Note that for the Paris potential $P_D$=5.8\%. We shall have cause to come back to this important point later.

We should now mention that there also exist three other variants of the OBE parametrization of the Bonn full model. These are formulated within the framework of the Thompson equation \cite{Machleidt89,Brockmann90} and uses the pseudovector coupling for $\pi$ and $\eta$, while the potential defined within the BbS equation uses the pseudoscalar coupling. It may be mentioned that the results obtained with the Thompson choice differ little from those obtained with the BbS reduction. A detailed discussion on this point is given in \cite{Brockmann90}. 

As we shall see later, the potential with the weaker tensor force, namely Bonn A, has turned out to give the best results in nuclear structure calculations. Unless otherwise stated, in the following we shall denote by Bonn A, B, and C  the three variants of the energy-independent approximation to the Bonn full model defined within the BbS equation.  
However, to avoid any confusion when consulting the literature on this subject, the reader may take a look at Tables A.1 and A.2 in \cite{Machleidt89}.  	
  
\subsection{High-precision $NN$ potentials} \label{sec:NN2}

From the early 1990s on there has been much progress in the field of nuclear forces. In the first place, the $NN$ phase shift analysis was greatly improved by the Nijmegen group
\cite{Bergervoet90,Klomp91,Klomp92,Stoks93}. They performed a multienergy partial-wave analysis of all $NN$ scattering data below 350 MeV laboratory energy after rejection of a rather large number of data (about 900 and 300 for the $np$ and $pp$ data, respectively) on the basis of statistical criteria. In this way, the final database consisted of 1787 $pp$ and 2514 $np$ data. The $pp$, $np$ and combined $pp + np$ analysis all yielded a $\chi^2$/datum $\approx 1$, significantly lower than any previous multienergy partial-wave analysis.
This analysis has paved the way to a new generation of high-quality $NN$ potentials which, similar to the analysis, fit the $NN$ data with the almost perfect
$\chi^2$/datum $\approx 1$. These are the potentials constructed in the mid 1990s by the Nijmegen group, NijmI, NijmII and Reid93 \cite{Stoks94}, the Argonne $V_{18}$ potential \cite{Wiringa95}, and the CD-Bonn potential \cite{Machleidt96,Machleidt01b}. 

The two potentials  NijmI and NijmII are based on the original Nijm78 potential 
\cite{Nagels78} discussed in the previous section. They are termed Reid-like potentials since, as is the case for the Reid68 potential \cite{Reid68}, each partial wave is parametrized independently. At very short distances these potentials are regularized by exponential form factors. The Reid93 potential is an updated version of the Reid68 potential, where the singularities have been removed by including a dipole form factor. While the NijmII and the Reid93 are totally local potentials, the NijmI contains momentum-dependent terms which in configuration space give rise to nonlocalities in the central force component.
Except for the OPE tail, these potentials are purely phenomenological with a total of 41, 47 and 50 parameters for NijmI, NijmII and Reid93, respectively. They all fit the $NN$ scattering data with an excellent $\chi^2$/datum=1.03 \cite{Stoks94}. As regards the $D$-state probability of the deuteron, this is practically the same for the three potentials,
namely $P_D$ in \%= 5.66 for NijmI, 5.64 for NijmII, and 5.70 for Reid93. 
It is worth mentioning that in the work by Stoks {\em et al.} \cite{Stoks94} 
an improved version of the Nijm78 potential, dubbed Nijm93, was also presented, which with 15 parameters produced a $\chi^2$/datum of 1.87.

The CD-Bonn potential \cite{Machleidt01b} is a charge-dependent OBE potential. It includes the $\pi$, $\rho$, and $\omega$ mesons plus two effective scalar-isoscalar $\sigma$ bosons, the parameters of which are partial-wave dependent. As is the case for the early OBE Bonn potentials, CD-Bonn is a nonlocal potential. It predicts a deuteron $D$-state probability substantially lower than that yielded by the potentials of the Nijmegen family,
namely $P_D$=4.85\%. This may be traced to the nonlocalities contained in the tensor force \cite{Machleidt01b}. While the CD-Bonn potential reproduces important predictions by the Bonn full model, the additional fit freedom obtained by adjusting the parameters of the $\sigma_1$ and $\sigma_2$ bosons in each partial wave produces a $\chi^2$/datum of 1.02 for the 4301 data of the Nijmegen database, the total number of free parameters being 43. In this connection, it may be mentioned that the Nijmegen database has been updated \cite{Machleidt01b} by adding the $pp$ and $np$ data between January 1993 and December 1999. This 1999 database contains 2932 $pp$ data and 3058 $np$ data, namely 5990 data in total.  The $\chi^2$/datum for the CD-Bonn potential in regard to the latter database remains 1.02. 

The Argonne $V_{18}$ model \cite{Wiringa95}, so named for its operator content, is a purely phenomenological (except for the correct OPE tail) nonrelativistic $NN$ potential with a local operator structure. It is an updated version of the Argonne $V_{14}$ potential  \cite{Wiringa84}, which was constructed in the early 1980s, with the addition of three charge-dependent and one charge-asymmetric operators. In operator form the $V_{18}$ potential is written as a sum of 18 terms,
\begin{equation} V_{ij} = \sum_{p=1,18}V_p(r_{ij})O^p_{ij}.
\end{equation}
To give an idea of the degree of sophistication reached by  modern phenomenological potentials, it may be instructive to write here explicitly the operator structure of the $V_{18}$ potential 
\cite{Wiringa95}. 
The first 14 charge independent operators are given by:

\begin{eqnarray}
O_{ij}^{p=1,14}= 1, (\mbox {\boldmath $\tau$}_i \cdot \mbox{\boldmath $\tau$}_j), (\mbox{\boldmath $\sigma$}_i \cdot \mbox{\boldmath  $\sigma$}_j), 
(\mbox{\boldmath $\sigma$}_i \cdot \mbox{\boldmath $\sigma$}_j)(\mbox{\boldmath $\tau$}_i 
\cdot \mbox{\boldmath $\tau$}_j),  S_{ij}, S_{ij}(\mbox {\boldmath $\tau$}_i \cdot \mbox{\boldmath $\tau$}_j), \mbox{\boldmath$L \cdot S$}, \mbox{\boldmath$L \cdot S$} (\mbox{\boldmath $\tau$}_i \cdot \mbox{\boldmath $\tau$}_j),  \nonumber\\ 
L^2, L^2(\mbox{\boldmath $\tau$}_i \cdot \mbox{\boldmath $\tau$}_j),
L^2(\mbox{\boldmath $\sigma$}_i \cdot \mbox{\boldmath  $\sigma$}_j), 
L^2(\mbox{\boldmath $\sigma$}_i \cdot \mbox{\boldmath  $\sigma$}_j)(\mbox{\boldmath $\tau$}_i \cdot \mbox{\boldmath  $\tau$}_j), 
(\mbox{\boldmath $L \cdot S$})^2, (\mbox{\boldmath$L \cdot S$})^2(\mbox{\boldmath $\tau$}_i \cdot \mbox{\boldmath $\tau$}_j).
\end{eqnarray}

The four additional operators breaking charge independence are given by
\begin{equation}
O_{ij}^{p=15,18}=T_{ij}, (\mbox{\boldmath $\sigma$}_i \cdot \mbox{\boldmath  $\sigma$}_j)T_{ij},
S_{ij}T_{ij}, (\tau_{zi} + \tau_{zj}),
\end{equation}
where $T_{ij} = 3 \tau_{zi}\tau_{zj} - \mbox{\boldmath $\tau$}_i \cdot \mbox {\boldmath $\tau$}_j $ is the isotensor operator analogous to the $S_{ij}$ operator.
As is the case for the NijmI and NijmII potentials, at very short distances the $V_{18}$ potential is regularized by exponential form factors. With 40 adjustable parameters this potential gives a $\chi^2$/datum of 1.09 for the 4301 data of the Nijmegen database. As regards the deuteron $D$-state probability, this is $P_D$= 5.76\%, very close to that predicted by the potentials of the Nijmegen family. 

All the high-precision $NN$ potentials described above have a large number of free parameters, say about 45, which is the price one has to pay to achieve a very accurate fit of the world $NN$ data. This makes it clear that, to date, high-quality potentials with an excellent $\chi^2$/datum $\approx$ 1 can only be obtained within the framework of a substantially phenomenological approach. 
Since these potentials fit almost equally well the $NN$ data up to the inelastic threshold, their on-shell properties are essentially identical, namely they are phase-shift equivalent. In addition, they all predict almost identical deuteron observables (quadrupole moment and $D/S$-state ratio) \cite{Machleidt01a}.
While they have also in common the inclusion of the OPE contribution, their off-shell behavior may be quite different. In fact, the short-range (high-momentum) components of these potentials are indeed quite different, as we shall discuss later in Sec.~\ref{sec:vlowk}. This raises a central question of how much nuclear structure results may depend on the $NN$ potential one starts with. We shall consider this important point in Sec.~\ref{sec:res2}.

The brief review of the $NN$ interaction given above has been mainly aimed at highlighting the  progress made in this field over a period of about 50 years. As already pointed out in the Introduction, and as we shall discuss in detail 
in Secs.~\ref{sec:SMCEP} and ~\ref{sec:SMCMC}, this has been instrumental in paving the way to a more fundamental approach to nuclear structure calculations than the traditional, empirical one. It is clear, however, that from a first-principle point of view a substantial theoretical progress in the field of the $NN$ interaction is still in demand. It seems fair to say that this is not likely to be achieved along the lines of the traditional meson theory. Indeed, in the past few years efforts in this direction have been made within the framework of the chiral effective theory.
The literature on this subject, which is still actively pursued, is by now very extensive and there are several comprehensive reviews \cite{vanKolck99,Beane01,Bedaque02},  to which we refer the reader. Therefore, in the next section we shall only give a brief survey focusing attention on chiral potentials which have been recently employed in nuclear structure calculations. 

\subsection{Chiral potentials} \label{sec:NN3}

The approach to the $NN$ interaction  based upon chiral effective field theory was started by  Weinberg \cite{Weinberg90,Weinberg91} some fifteen years ago, 
and since then it has been developed by several authors. The basic idea \cite{Weinberg90} is to derive the $NN$ potential starting from the most general Lagrangian for  low-energy pions and nucleons consistent with the symmetries of quantum chromodynamics (QCD), in particular the spontaneously broken chiral symmetry. All other particle types are ``integrated out", their effects being contained in the coefficients of the series of terms in the pion-nucleon Lagrangian. The chiral Lagrangian 
provides a perturbative framework for the derivation of the nucleon-nucleon potential.  In fact, it was shown by Weinberg \cite{Weinberg91} that a systematic expansion of the nuclear potential exists in powers of the small parameter $Q/\Lambda_\chi$, where 
$Q$ denotes a generic low-momentum and $\Lambda_\chi \approx$  1 GeV is the chiral symmetry breaking scale. This perturbative low-energy theory is called chiral perturbation theory 
($\chi PT$). The contribution of any diagram to the perturbation expansion is characterized by the power $\nu$ of the momentum $Q$, and the expansion is organized by counting powers of $Q$.
This procedure \cite{Weinberg91} is referred to as power counting.

Soon after the pioneering work by Weinberg, where only the lowest order $NN$ potential was obtained, Ord{\'o}{\~n}ez {\em et al.}\cite{Ordonez92} extended the effective chiral potential to order ($Q/\Lambda_\chi$)$^3$ [next-to-next-to-leading order (NNLO), $\nu$=3)] showing that this accounted, at least qualitatively, for the most relevant features of the nuclear potential. Later on, this approach was further pursued by Ord\'{o}\~{n}ez, Ray and van Kolck \cite{Ordonez94,Ordonez96}, who derived at NNLO a $NN$  potential both in momentum and coordinate space. With 26 free parameters this potential model gave a satisfactory description of the Nijmegen phase shifts up to about 100 MeV \cite{Ordonez96}. 
These initial achievements prompted extensive efforts to understand the $NN$ force within the framework of chiral effective field theory. 

A clean test of chiral symmetry in the two-nucleon system was provided by the work of Kaiser, Brockmann and Weise \cite{Kaiser97} and Kaiser, Gerstend{\"o}rfer and Weise \cite{Kaiser98}.
Restricting themselves to the peripheral nucleon-nucleon interaction, these authors  obtained at NNLO, without adjustable parameters, an accurate description of the empirical phase shifts in the partial waves with $L \geq 3$ up to 350 MeV and up to about (50-80) MeV for the D-waves. 

Based on a modified Weinberg power counting, Epelbaum, Gl{\"o}ckle and Meissner constructed a chiral $NN$ potential at NNLO consisting of one- and two-pion exchange diagrams and contact interactions (which represent the short-range force) \cite{Epelbaum98,Epelbaum00}. The nine parameters related to the contact interactions were determined by a fit to the $np$ $S$- and $P$-waves and the mixing parameter 
$\epsilon_1$ for $E_{\rm Lab}$ $<$ 100 MeV. This potential gives a $\chi^2$/datum for the $NN$ data of the 1999 database below 290 MeV laboratory energy of more than 20 \cite{Entem03}.

In their program to develop a $NN$ potential based upon chiral effective theory, Entem and Machleidt set themselves the task to achieve an accuracy for the reproduction of the $NN$ data comparable to that of the high-precision potentials constructed in the 1990s, which have been discussed in Sec.~\ref{sec:NN1}.
The first outcome of this program was a NNLO potential, called Idaho potential \cite{Entem02a}.
This model includes one- and two-pion exchange contributions up to chiral order three and contact terms up to order four. For the latter, partial wave dependent cutoff parameters are used, which introduces more parameters bringing the total number up to 46. This potential gives a $\chi^2$/datum for the reproduction of the 1999 $np$ database up to $E_{\rm Lab}$=210 MeV of 0.98 \cite{Entem02c}.

The next step taken by Entem and Machleidt was the investigation of the chiral $2\pi$-exchange contributions to the $NN$ interaction at fourth order, which was based on the work by Kaiser \cite{Kaiser01,Kaiser02}, who gave analytical results for these contributions in a form suitable for implementation in a next-to-next-to-next-to-leading (N$^3$LO, fourth order) calculation. This eventually resulted in the first chiral $NN$ potential at N$^3$LO \cite{Entem03}. This model includes 24 contact terms (24 parameters) which contribute to the partial waves with $L \leq 2$. With 29 parameters in all, it gives a 
$\chi^2$/datum for the reproduction of the 1999 $np$ and $pp$ data below 290 MeV of 1.10  and 1.50, respectively. The deuteron $D$-state probability is $P_D$= 4.51\%.

Very recently a $NN$ potential at N$^3$LO has been constructed by Epelbaum, Gl{\"o}ckle and Meissner (2005) which differs in various ways from that of Entem and Machleidt, as discussed in detail in \cite{Epelbaum05}. It consists of one-, two- and three-pion exchanges and a set of 24 contact interactions. The total number of free parameters is 26. These have been determined by a combined fit to some $nn$ and $pp$ phase shifts from the Nijmegen analysis together with the $nn$ scattering length. The description of the phase shifts and deuteron properties at N$^3$LO turns out to be improved compared to that previously obtained by the same authors at NLO and NNLO \cite{Epelbaum04}. As regards the deuteron
$D$-state probability, this N$^3$LO potential gives $P_D$=2.73--3.63\%, a value which is significantly smaller than  that predicted by any other modern $NN$ potential.  

In regard to potentials at N$^3$LO, it is worth mentioning that it has been shown 
\cite{Kaiser99,Kaiser00} that the effects of three-pion exchange, which starts to contribute at this order, are very small and  therefore of no practical relevance. Accordingly, they have been neglected in both the above studies.

The foregoing discussion has all been focused on the two-nucleon force. 
The role of three-nucleon interactions in 
light nuclei has been,  and is currently, actively investigated within the 
framework of {\it ab initio} approaches, such as the GFMC and the NCSM.  
Let us only remark here that in recent years the Green's function Monte Carlo
method has proved to be a valuable tool for calculations of properties of 
light nuclei using realistic two-nucleon and three-nucleon potentials 
\cite{Pieper01a,Pieper02}. 
In particular, the combination of the Argonne $V_{18}$ potential and Illinois-2 three-nucleon potential has yielded good results for energies of nuclei up to $^{12}$C \cite{Pieper05}. For a review of the GFMC method and applications up to A=8 we refer the interested reader to the paper 
by Pieper and Wiringa \cite{Pieper01b}.

In this context, it should be pointed out that an important advantage of the 
chiral perturbation theory is that at NNLO and higher orders it generates 
three-nucleon forces. This has prompted applications of the complete chiral 
interaction at NNLO to the three- and four-nucleon systems \cite{Epelbaum02}. 
These applications are currently being extended to light nuclei with $A > 4$ 
\cite{Nogga04}.

However, as regards the derivation of a realistic shell-model effective interaction the $3N$
forces have not been taken into account up to now. As mentioned in the Introduction, in
this review we shall give a brief discussion of the three-body effects, as inferred
from the study of many valence-nucleon systems

\section{Shell-model effective interaction}\label{sec:SMEI}

\subsection{Generalities} \label{sec:SMEI1}

As mentioned in the Introduction, a basic input to  
nuclear shell-model calculations is the model-space effective interaction.
It is worth recalling that this interaction differs 
from the interaction between two free nucleons in several respects.
In the first place, a large part of the $NN$ interaction is absorbed
into the mean field  which is due to the average interaction between the 
nucleons.
In the second place, the $NN$ interaction in the nuclear medium is affected 
by the presence of the other nucleons; one has certainly to take into
account the Pauli exclusion principle, which forbids two
interacting nucleons to scatter into states occupied by other
nucleons. Finally, the effective interaction has to account for effects of 
the configurations excluded from the model space. 
Ideally, the eigenvalues of the shell-model Hamiltonian in the model space 
should be a subset of the eigenvalues of the full nuclear Hamiltonian in the 
entire Hilbert space.

In a microscopic approach this shell-model Hamiltonian may be constructed 
starting  from a realistic $NN$ potential by means of many-body perturbation
techniques. This approach has long been a central topic of nuclear theory.
The following subsections are devoted to a detailed discussion of it.

First, let us introduce the general formalism which is needed in 
the effective interaction theory.
We would like  to solve the Schr\"odinger equation for the $A$-nucleon system:
\begin{equation}
H | \Psi_{\nu} \rangle  = E_{\nu} | \Psi_{\nu} \rangle \label{eq1},
\end{equation}

\noindent
where
\begin{equation}
 H=H_0+H_1  \label{defh},
\end{equation}
and 
\begin{equation}
H_0= \sum_{i=1}^A (t_i+U_i)  \label{defh0},
\end{equation}
\begin{equation}
H_1=\sum_{i<j=1}^{A} V^{NN}_{ij}-\sum_{i=1}^AU_i \label{defh1}~. 
\end{equation}
An  auxiliary one-body potential $U_i$ has been introduced in order to
break up the nuclear Hamiltonian as the sum of a one-body term
$H_0$, which describes the independent motion of the nucleons, and 
the interaction $H_1$.

In the shell model,  the nucleus is represented  as
an inert core plus $n$ valence nucleons moving  in 
a limited number of SP  orbits above  the closed core  and 
interact through a model-space effective interaction. 
The valence or model space is defined in terms of the eigenvectors     
of $H_0$
\begin{equation}
|\Phi_i \rangle = [ a^{\dagger}_1 a^{\dagger}_2 ~...~a^{\dagger}_n ]_i
| c \rangle ,~~~ i=1,...,d,
\end{equation}

\noindent
where $|c \rangle$ represents the inert core and  the subscripts 1, 2, ..., 
$n$ denote the SP valence states. The index $i$ stands for all the other 
quantum numbers needed to specify the state.

The aim of the effective interaction theory is to reduce the eigenvalue 
problem of Eq. (\ref{eq1}) to a model-space eigenvalue problem
\begin{equation}
PH_{\rm eff}P | \Psi_{\alpha} \rangle  = E_{\alpha} P | \Psi_{\alpha}
\rangle \label{eq3},
\end{equation}

\noindent
where the operator $P$,
\begin{equation}
P= \sum_{i=1}^d | \Phi_i \rangle \langle \Phi_i |,
\end{equation}
projects from the complete Hilbert space onto the model space.
The operator $Q=1-P$ is its complement.  
The projection operators $P$ and $Q$
satisfy the properties

$$ P^2=P, ~~ Q^2=Q, ~~ PQ=QP=0 ~.$$

In the following, the concept of the effective interaction is introduced by 
a very general and simple method \cite{Bloch58,Feshbach62}.
Let us define the operators 

$$PHP = H_{PP}, ~~ PHQ = H_{PQ},$$
$$ QHP = H_{QP},~~  QHQ = H_{QQ}~.$$

\noindent Then the Schr\"odinger equation
(\ref{eq1}) can be written as
\begin{eqnarray}
H_{PP}P|\Psi_{\nu} \rangle + H_{PQ} Q |\Psi_{\nu} \rangle & = &
E_{\nu} P |\Psi_{\nu} \rangle  \label{eq4} \\
H_{QP}P|\Psi_{\nu} \rangle + H_{QQ} Q |\Psi_{\nu} \rangle & = &
E_{\nu} Q |\Psi_{\nu} \rangle \label{eq5} ~.
\end{eqnarray}
From the latter  equation we obtain
\begin{equation}
Q | \Psi_{\nu} \rangle = \frac{1}{E_{\nu}-H_{QQ}} H_{QP} P
|\Psi_{\nu} \rangle ,
\end{equation}

\noindent and substituting the r.h.s. of this  equation into
Eq. (\ref{eq4}) we have 
\begin{equation}
\left( H_{PP}+H_{PQ}\frac{1}{E_{\nu}-H_{QQ}} H_{QP} \right)P |\Psi_{\nu}
  \rangle = E_{\nu} P |\Psi_{\nu} \rangle \label{eq7}~.
\end{equation}
If the l.h.s. operator, which  acts only within the model space, is
denoted as 
\begin{equation}
H_{\rm eff} (E_{\nu}) =  H_{PP}+H_{PQ}\frac{1}{E_{\nu}-H_{QQ}}
H_{QP} \label{eq8},
\end{equation}

\noindent
Eq. (\ref{eq7}) reads
\begin{equation}
H_{\rm eff} (E_{\nu}) P |\Psi_{\nu} \rangle = E_{\nu} P 
|\Psi_{\nu} \rangle \label{eq9},
\end{equation}
which is of the form of Eq. (\ref{eq3}). Moreover, since the
operators $P$ and $Q$ commute with $H_0$, we can write Eq. (\ref{eq8}) as
\begin{equation}
H_{\rm eff} (E_{\nu}) = PH_0P+V_{\rm eff}(E_{\nu}),
\end{equation}

\noindent
with 
\begin{equation}
V_{\rm eff} (E_{\nu}) =  PH_1P+PH_1Q\frac{1}{E_{\nu}-H_{QQ}}
QH_1P \label{eqfesh}~.
\end{equation}
This  equation defines the effective interaction as
derived by Feshbach in nuclear reaction studies \cite{Feshbach62}.

Now, on expanding  $(E_{\nu}-H_{QQ})^{-1}$ we can write
\begin{equation}
V_{\rm eff} (E_{\nu}) = PH_1P + PH_1 \frac{Q}{E_{\nu}-QH_0Q}
H_1P 
+ PH_1 \frac{Q}{E_{\nu}-QH_0Q} H_1 \frac{Q}{E_{\nu}-QH_0Q} H_1 P
+ ...\label{eqblochhor1} ,
\end{equation}

\noindent
which is equivalent to the Bloch-Horowitz form of the effective
interaction \cite{Bloch58}:
\begin{equation}
V_{\rm eff} (E_{\nu}) =  PH_1P+PH_1 \frac{Q}{E_{\nu}-QH_0Q} 
V_{\rm eff} (E_{\nu}) ~.\label{eqblochhor2}
\end{equation}

Eqs. (\ref{eqfesh}) and (\ref{eqblochhor2}) are the desired result.
In fact, they represent effective interactions which, used in a
truncated model space, give a subset of the true
eigenvalues. Bloch and Horowitz \cite{Bloch58} have studied 
the analytic properties
of the eigenvalue problem in terms of the effective interaction of
Eq. (\ref{eqblochhor2}).
It should be noted, however,  that the above effective interactions
depend on the eigenvalue $E_{\nu}$. This energy dependence is a
serious drawback, since one has different Hamiltonians  for different
eigenvalues. 

Some forty years ago, the theoretical basis for an energy-independent 
effective Hamiltonian was set down by Brandow in the frame of a 
time-independent perturbative method \cite{Brandow67}.
Starting from the degenerate version of the Brillouin-Wigner
perturbation theory, the energy terms were expanded out of the energy
denominators.
Then a rearrangement of the series was performed leading
to a completely linked-cluster expansion. 
The energy dependence was eliminated by introducing a special type of diagrams,
the so-called folded diagrams.

A linked-cluster expansion for the shell-model effective interaction
was also derived in Refs.~\cite{Morita63,Oberlechner70,Johnson71,Kuo71}
within the framework of the time-dependent perturbation theory.
In the following subsection we shall describe in some detail the
time-dependent perturbative approach by Kuo, Lee and Ratcliff \cite{Kuo71}. 
We have tried to give a brief, self-contained  presentation of this
subject which as matter of fact is rather complex and multi-faceted. 
To this end, we have discussed the various elements entering this approach
whithout going into the details of the proofs. Furthermore, we have found 
it useful to first introduce in Secs.~\ref{sec:SMEI21}  and 
\ref{sec:SMEI22} the concept of 
folded diagrams  and the decomposition theorem, respectively, which are 
two basic tools for the derivation of  the effective interaction, as 
is shown in Sec.~\ref{sec:SMEI23}. 
A complete review of this approach can be found in \cite{Kuo90}, to which 
we refer the reader for details.

\subsection{Degenerate time-dependent perturbation theory: 
folded-diagram approach}\label{sec:SMEI2}

\subsubsection{Folded diagrams} \label{sec:SMEI21}

In this section,  we focus on the case of two-valence
nucleons and therefore the nucleus is a doubly closed core plus
two valence nucleons. We denote as active states those SP levels 
above the core which are made accessible to the two valence nucleons.
The higher-energy SP levels and the filled ones in the core are called
passive states.
In such a frame the  basis vector $|\Phi_i \rangle$ is

\begin{equation}
|\Phi_i \rangle = [ a^{\dagger}_1 a^{\dagger}_2 ]_i | c \rangle ~.
\end{equation}

In the complex time limit  the time-development operator 
in the interaction representation  is given by

\begin{equation} 
U(0,-\infty) =  \lim_{\epsilon \rightarrow 0^+} \lim_{t \rightarrow
  -\infty (1-i\epsilon)} U(0,t) 
=  \lim_{\epsilon \rightarrow 0^+} \lim_{t \rightarrow -\infty 
(1-i\epsilon)} e^{iHt}e^{-iH_0t}, \nonumber
\end{equation} 

\noindent
which  can be expanded as 
\begin{equation}
U(0,-\infty)= \lim_{\epsilon \rightarrow 0^+} \lim_{t' \rightarrow
  -\infty (1-i\epsilon)}  \sum_{n=0}^{+\infty} (-i)^n \int_{t'}^{0}
dt_1 
\int_{t'}^{t_1} dt_2 ... 
\int_{t'}^{t_{n-1}} dt_n~H_1(t_1) H_1(t_2) ... H_1(t_n),
\end{equation}
where 
\[ H_1(t)= e^{iH_0t} H_1 e^{-iH_0t} ~.\]

Let us now act  on $|\Phi_i \rangle$ with $U(0,-\infty)$: 
\begin{equation}
U(0,-\infty) |\Phi_i \rangle = U(0,-\infty) [ a^{\dagger}_1
a^{\dagger}_2 ]_i | c \rangle ,\label{ufi}
\end{equation}

\noindent
The action of the time-development operator on the unperturbed wave 
function  $|\Phi_i \rangle$ may be represented by  an infinite
collection of diagrams. The type of diagrams we consider here is referred to as 
time-ordered Goldstone diagrams (for a description of Goldstone
diagrams see for instance Ref.~\cite{Brandow67})

\begin{figure}[hbtp]
\begin{center}

\includegraphics[scale=0.8,angle=0]{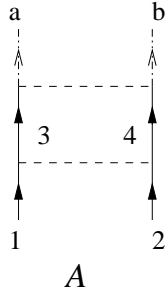}
\end{center}
\caption{A second-order time-ordered Goldstone diagram.\label{prima}}
\end{figure}

As an example, we show  in Fig. \ref{prima}  a second-order time-ordered
Goldstone diagram , which is  one of the diagrams appearing in (\ref{ufi}).
The dashed vertex lines denote $V_{NN}$-interactions (for the sake of
simplicity we take $H_1=V_{NN}$), $a$ and $b$ represent two passive
particle states while  1, 2, 3, and 4 are valence states.
From now on in the present section, the passive particle states will be
represented by letters and  dashed-dotted lines. 

\begin{figure}[hbtp]
\begin{center}
\includegraphics[scale=0.65,angle=0]{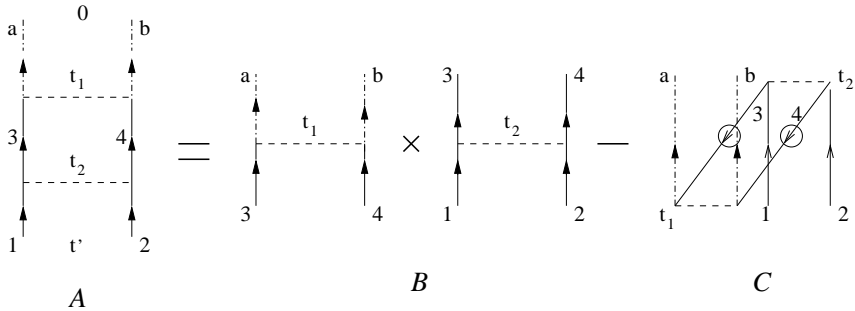}
\end{center}
\caption{A diagrammatic identity which defines the folded diagram. 
\label{seconda}}
\end{figure}

The diagram $A$ of Fig. \ref{prima} gives a contribution
\begin{equation}
A= a^{\dagger}_a a^{\dagger}_b  | c \rangle \times \frac{1}{4}
V_{ab34} V_{3412} I(A),
\end{equation}

\noindent
where $V_{\alpha \beta \gamma \delta}$ are non-antisymmetrized
matrix elements of $V_{NN}$.
$I(A)$ is the time integral 
\begin{equation}
I(A) = \lim_{\epsilon \rightarrow 0^+} \lim_{t' \rightarrow
  -\infty (1-i\epsilon)} (-i)^2
\int_{t'}^{0} dt_1 \int_{t'}^{t_1} dt_2 
e^{-i(\epsilon_3 + \epsilon_4
- \epsilon_a -\epsilon_b)t_1} ~e^{-i(\epsilon_1 + \epsilon_2
- \epsilon_3 -\epsilon_4)t_2} ,
\end{equation}

\noindent
where the $\epsilon_{\alpha}$'s are SP energies.
A folded diagram arises upon factorization of diagram $A$, as shown 
in Fig. \ref{seconda}. From this figure, we see that
diagram $B$ represents a factorization
of diagram $A$ into the product of two independent diagrams.
The time sequence for diagram $A$ is $0 \geq t_1 \geq t_2 \geq t'$,
while in diagram $B$ it is $0 \geq t_1 \geq t'$ and $0 \geq
t_2 \geq t'$, with no constraint on the relative ordering of
$t_1$ and $t_2$.

\noindent Therefore, diagrams $A$ and $B$ are not equal, unless
subtracting from $B$ the time-incorrect contribution represented by
the folded diagram $C$.
It is worth noting that lines 3 and 4 in $C$ are not hole lines,
but folded active particle lines. 
From now on the folded lines will be denoted by drawing a little
circle.
Explicitly, diagram $C$ is given by
\begin{equation}
C= a^{\dagger}_a a^{\dagger}_b  | c \rangle \times \frac{1}{4}
V_{ab34} V_{3412} I(C),
\end{equation}
where
\begin{equation}
I(C) = \lim_{\epsilon \rightarrow 0^+} \lim_{t' \rightarrow
  -\infty (1-i\epsilon)} (-i)^2
\int_{t'}^{0} dt_1 \int_{t_1}^{0} dt_2 
e^{-i(\epsilon_3 + \epsilon_4
- \epsilon_a -\epsilon_b)t_1} ~e^{-i(\epsilon_1 + \epsilon_2
- \epsilon_3 -\epsilon_4)t_2} ~.
\end{equation}
Note that the rules to evaluate the folded diagrams are
identical to those for standard Goldstone diagrams, except
counting as hole lines the folded active lines in the energy denominator.

\begin{figure}[hbtp]
\begin{center}
\includegraphics[scale=0.6,angle=0]{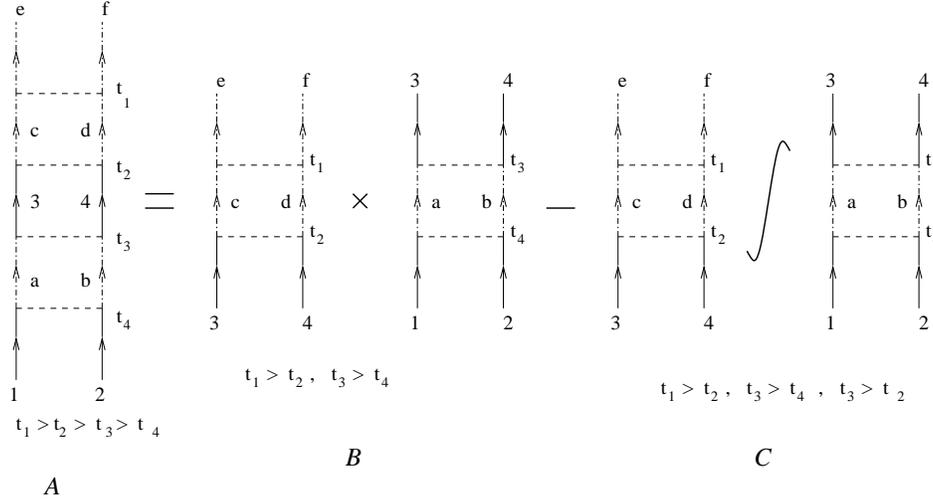}
\end{center}
\caption{A diagrammatic identity which illustrates the generalized 
folded diagram.\label{terza}}
\end{figure}

\begin{figure}[hbtp]
\begin{center}

\includegraphics[scale=0.68,angle=0]{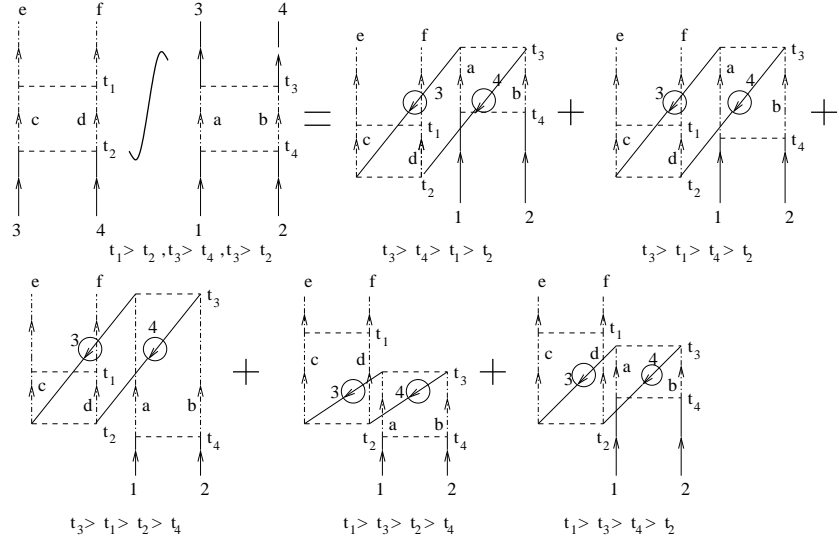}
\end{center}
\caption{Generalized folded diagram expressed as sum of ordinary folded
diagrams.
\label{quarta}}
\end{figure}

We now introduce the concept of generalized folded
diagram and derive a convenient method to compute it in a
degenerate model space.
Let us consider, for example, the diagrams shown in Fig. \ref{terza}.
All the three diagrams $A$, $B$, and $C$ have identical integrands and
constant factors, the only difference being in the integration limits.
The integration limits of diagram $A$ correspond to the time ordering
$0 \geq t_1 \geq t_2 \geq t_3 \geq t_4 \geq t'$.
As pointed out before, the factorization of $A$ into two independent
diagrams (diagram $B$) violates the above time ordering.
To correct for the time ordering in $B$, one has to
subtract  the folded diagram $C$, whose time constraints are $0
\geq t_1 \geq t_2 \geq t'$, $0 \geq t_3 \geq t_4 \geq t'$,  and $t_3
\geq t_2$.
Five different time sequences satisfy the above three constraints,
thus $C$ consists of five ordinary folded diagrams (see
Fig. \ref{quarta}) and is called generalized folded diagram.
From now on the integral sign will denote the generalized folding
operation.

An advantageous method to evaluate generalized
folded diagrams in a degenerate model space is as follows.
Let us consider diagram $C$ of Fig. \ref{terza}.
As pointed out before, $A$, $B$, and $C$ have identical integrands and
constant factors, so that we may write the time integral $I(C)$ as 
(\cite{Kuo90}, pp. 16-18) 
\begin{eqnarray}
I(C)=I(B)-I(A) & = & \frac{1}{\epsilon_1 + \epsilon_2 - \epsilon_3 -
  \epsilon_4}  
\left[ \frac{1}{(\epsilon_3 + \epsilon_4 - \epsilon_e - 
\epsilon_f)(\epsilon_3 + \epsilon_4 - \epsilon_c - \epsilon_d)}
 \right. \\ \nonumber 
&& -  \left. \frac{1}{(\epsilon_1 + \epsilon_2 - \epsilon_e - \epsilon_f)
(\epsilon_1 + \epsilon_2 - \epsilon_c - \epsilon_d)} \right] 
\frac{1}{\epsilon_1 + \epsilon_2 - \epsilon_a - \epsilon_b}~.
\label{denominatore}
\end{eqnarray}

\noindent
In a degenerate model space, the first factor is infinite, while the
second is zero.
However, $I(C)$ can be determined by a limiting procedure.
If we write $\epsilon_1+\epsilon_2 = \epsilon_3 + \epsilon_4 +
\Delta$, where $\Delta \rightarrow 0$, we can put
Eq.~(\ref{denominatore}) into the form

\begin{eqnarray}
I(C) & = & \lim_{\Delta \rightarrow 0} \frac{1}{\Delta} \left[
  \frac{1}{(\epsilon_3 + \epsilon_4 - \epsilon_e -
  \epsilon_f)(\epsilon_3 + \epsilon_4 - \epsilon_c - \epsilon_d)} -
 \frac{1}{(\epsilon_3 + \epsilon_4 - \epsilon_e - \epsilon_f
  +\Delta) (\epsilon_3 + \epsilon_4 - \epsilon_c - \epsilon_d +
  \Delta)} \right] \\ \nonumber
&& \times    \frac{1}{(\epsilon_3 + \epsilon_4 - \epsilon_a - 
\epsilon_b + \Delta)} \\ \nonumber
 & = & \left[ \frac{1}{ (\epsilon_3 + \epsilon_4 - \epsilon_e -
    \epsilon_f)^2 (\epsilon_3 + \epsilon_4 - \epsilon_c - \epsilon_d)} 
 + \frac{1}{ (\epsilon_3 + \epsilon_4 - \epsilon_e -
    \epsilon_f)(\epsilon_3 + \epsilon_4 - \epsilon_c - \epsilon_d)^2}
  \right] \\ \nonumber
&& \times \frac{1}{(\epsilon_3 + \epsilon_4 - \epsilon_a - \epsilon_b)}~.
\label{intermedio}
\end{eqnarray}
\noindent
Thus, the energy denominator of the generalized folded diagram $C$ can
be expressed as the derivative of the energy denominator of the
l.h. part of the diagram with respect to the energy variable
$\omega$, calculated at $\epsilon_1+\epsilon_2$:
\begin{equation}
I(C)= - \frac{1}{\epsilon_3 + \epsilon_4 -\epsilon_a - \epsilon_b} 
 ~~\frac{d}{d \omega}\left( \frac{1}{\omega - \epsilon_c -
  \epsilon_d}~\frac{1}{\omega - \epsilon_e - \epsilon_f}
  \right)_{\omega = \epsilon_1+\epsilon_2}~.
\label{derivata}
\end{equation}

\noindent
The last equation will prove to be very useful to evaluate the folded diagrams.

\subsubsection{The decomposition theorem} \label{sec:SMEI22}

Let us consider again the wave function $U(0,-\infty) |\Phi_i \rangle$.
We can rewrite it as
\begin{equation}
U(0,-\infty) |\Phi_i \rangle = U_L (0,-\infty) | \Phi_i \rangle 
\times U (0,-\infty) | c \rangle , 
\label{facufi1}
\end{equation}

\noindent
where the subscript $L$ indicates that all the $H_1$ vertices in $U_L 
(0,-\infty)|\Phi_i\rangle$ are valence linked, i.e. are linked
directly or indirectly to at least one of the valence lines.
We now factorize each of the two terms on the r.h.s. of
Eq. (\ref{facufi1}) in order to write $U(0,-\infty) |\Phi_i \rangle$
in a form useful for the  derivation of  the model-space secular
equation.

\begin{figure}
\begin{center}

\includegraphics[scale=0.50,angle=0]{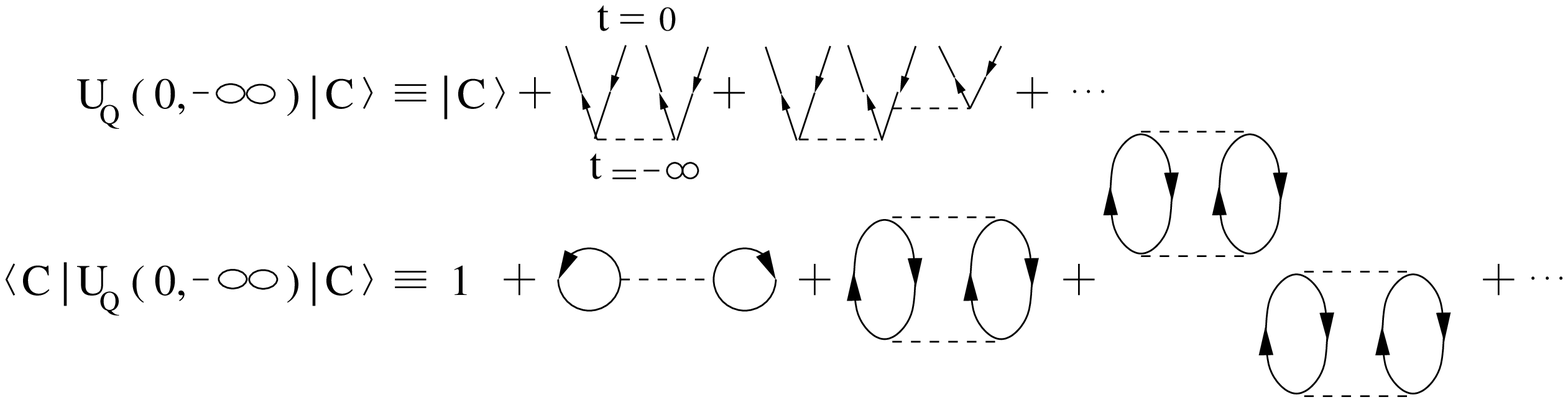}
\end{center}
\caption{Diagrammatic representation of $U_Q (0,-\infty) | c \rangle$ 
and $\langle c | U (0,-\infty) | c \rangle$.\label{fluc}}
\end{figure}

We first consider $U (0,-\infty) | c \rangle$, which can rewritten as 
\begin{equation}
U (0,-\infty) | c \rangle = U_Q (0,-\infty) | c \rangle \times \langle
c | U (0,-\infty) | c \rangle ,\label{facufi3}
\end{equation}
\noindent
where $U_Q (0,-\infty) | c \rangle$ denotes 
the collection of
diagrams in which every vertex is connected to the
time $t=0$ boundary. In fact, $U_Q (0,-\infty) | c \rangle$ is
proportional to the true ground-state wave function of the closed-shell
system, while
$\langle c | U (0,-\infty) | c \rangle$ represents all the
vacuum fluctuation diagrams. These two terms are illustrated in 
the first and second line of  Fig. \ref{fluc}, respectively.

\begin{figure}[hbtp]
\begin{center}

\includegraphics[scale=0.75,angle=0]{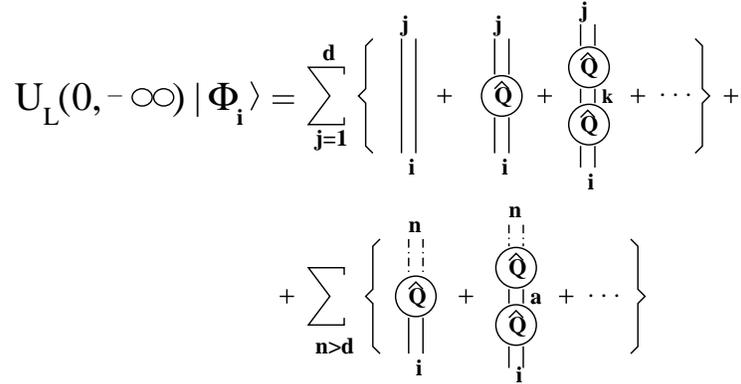}
\end{center}
\caption{Diagrammatic representation of $U_L (0,-\infty) | \Phi_i \rangle$.
\label{facufi}}
\end{figure}

A similar factorization of $U_L (0,-\infty) | 
\Phi_i \rangle$ can be performed, which can be expressed in terms of
the so-called $\hat{Q}$-boxes.
The $\hat{Q}$-box, which should not be confused
with the projection operator $Q$ introduced in Sec.~\ref{sec:SMEI1},
is defined as the sum of all diagrams that have
at least one $H_1$-vertex, are valence linked and irreducible 
(i.e., with at least one
passive line between two successive vertices).
Clearly, $U_L (0,-\infty) | \Phi_i \rangle$ must terminate either in
an active or passive state at $t=0$, thus we can write 
\begin{equation}
U_L (0,-\infty) | \Phi_i \rangle = | \chi^P_i \rangle + | \chi^Q_i
\rangle ,\label{newequation}
\end{equation}

\noindent
as shown in Fig. \ref{facufi}.
Note that in this figure the intermediate indices k, a, 
... represent summations over all $P$-space states.

\begin{figure}[hbtp]
\begin{center}
\includegraphics[scale=0.55,angle=-90]{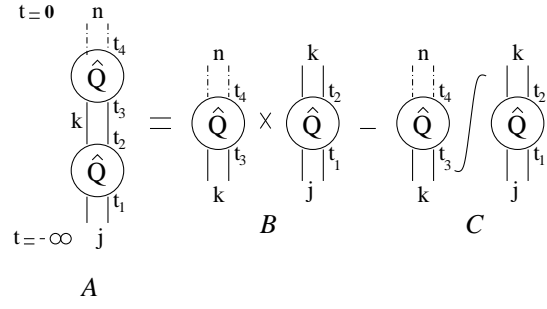}
\end{center}
\caption{Factorization of a 2-$\hat{Q}$-box sequence.
\label{facQ1}}
\end{figure}

It is also possible to factorize out of $| \chi^Q_i \rangle$ a term
belonging to $| \chi^P_i \rangle$ by means of the folded-diagram 
factorization.
In Fig. \ref{facQ1} we show, as an example, how a 2-$\hat{Q}$-box
sequence can be factorized.

It is worth noting  the similarity between Fig. \ref{facQ1} and Fig. 
\ref{terza}. 
As a matter of fact, when factorizing diagram $A$ time incorrect
contributions arise in diagram $B$, that are compensated by
subtracting from  it the generalized folded diagram $C$.

\begin{figure}[hbtp]
\begin{center}
\includegraphics[scale=0.50,angle=-90]{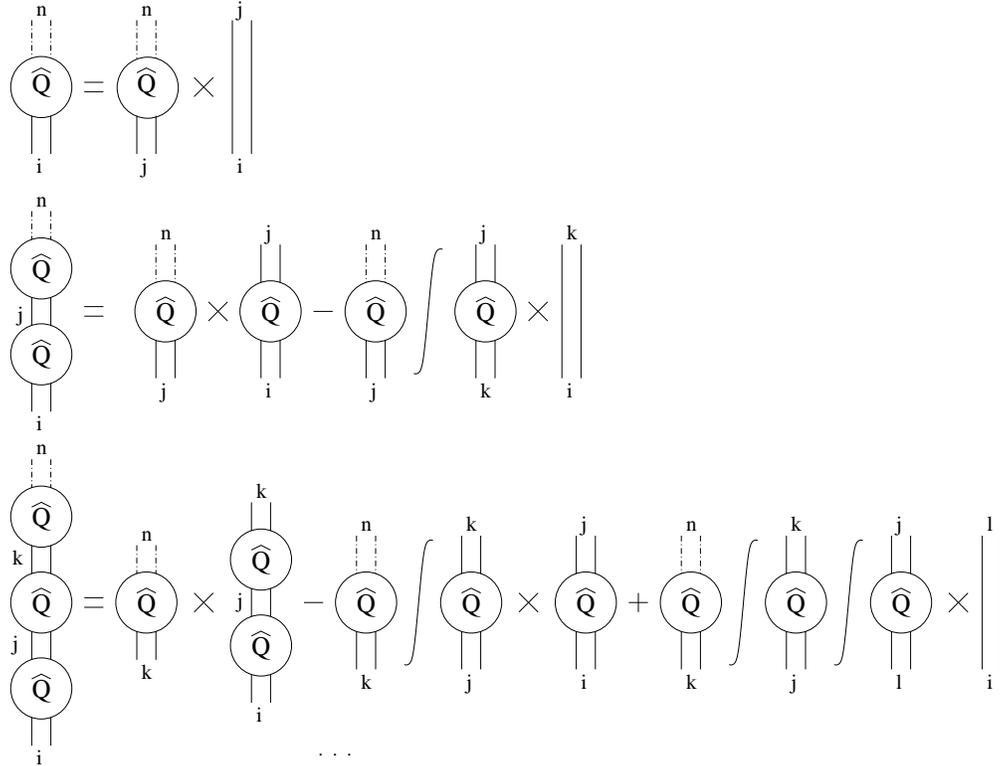}
\end{center}
\caption{Factorization of $| \chi^Q_i \rangle$ using the generalized 
folding procedure.
\label{facQ2}}
\end{figure}

\begin{figure}[hbtp]
\begin{center}
\includegraphics[scale=0.55,angle=0]{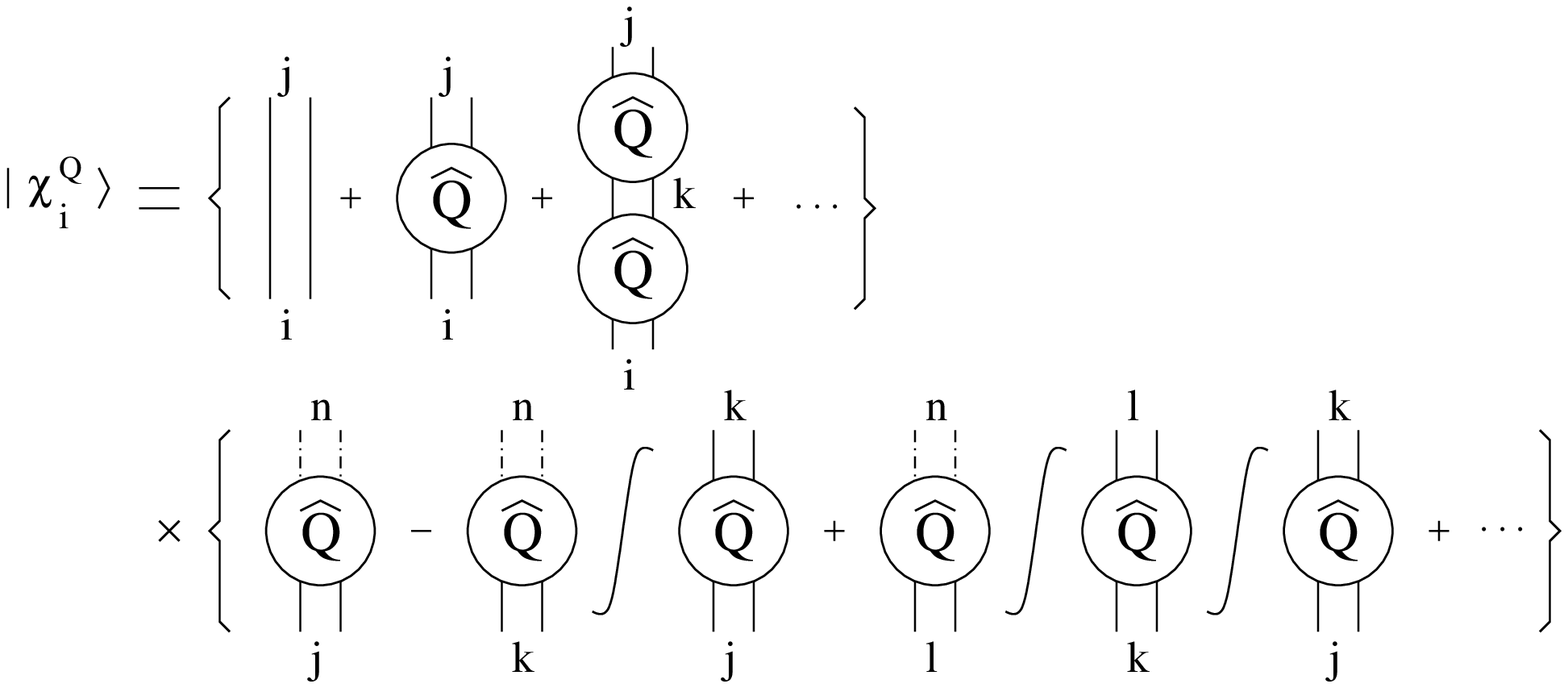}
\end{center}
\caption{$| \chi^Q_i \rangle$ wave function.
\label{facQ3}}
\end{figure}

Using the generalized folding procedure, we are able to factorize out
of each term in $| \chi^Q_i \rangle$ a diagram belonging to $|
\chi^P_i \rangle$ (see Fig. \ref{facQ2}).
Applying this factorization to all the terms in $| \chi^Q_i \rangle$, 
collecting columnwise the diagrams on the r.h.s. in Fig. \ref{facQ2}
and adding them up, we may represent $| \chi^Q_i \rangle$ as shown in
Fig. \ref{facQ3}.

\begin{figure}[hbtp]
\begin{center}
\includegraphics[scale=0.55,angle=-90]{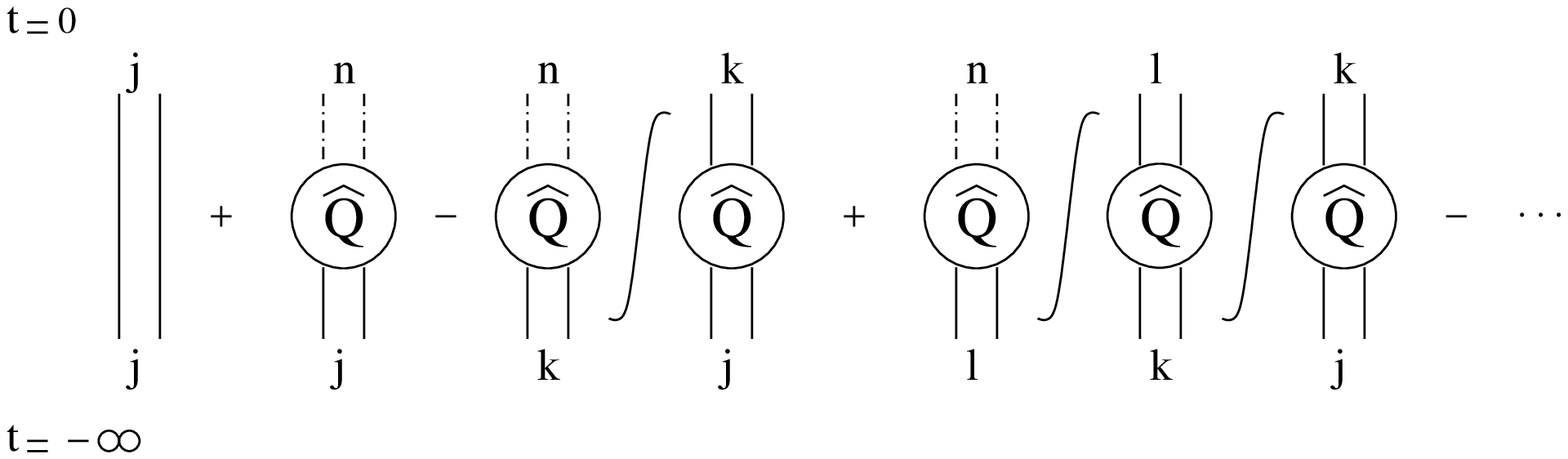}
\end{center}
\caption{Diagrammatic representation of $U_{QL} (0,-\infty) | \Phi_j 
\rangle$. 
\label{figuql}}
\end{figure}

The collection of diagrams in the upper parenthesis of
Fig. \ref{facQ3} is simply $\langle \Phi_j | U_L(0,-\infty) | \Phi_i
\rangle$, which, according to Eq. (\ref{newequation}), is
related to $| \chi^P_i \rangle$ by
\begin{equation}
|\chi^P_i \rangle = \sum_{j=1}^d | \Phi_j \rangle \langle \Phi_j | U_L
(0,-\infty) | \Phi_i \rangle ~.\label{eqchi}
\end{equation}
Therefore, taking into account Figs. \ref{facufi} and \ref{facQ3}, and 
Eq. (\ref{eqchi})  we can express $U_L(0,-\infty) | \Phi_i
\rangle$ as
\begin{equation}
U_L (0,-\infty) | \Phi_i \rangle = \sum_{j=1}^d U_{QL} (0,-\infty) |
\Phi_j \rangle \langle \Phi_j | U_L (0,-\infty) | \Phi_i \rangle , 
\label{equl}
\end{equation}

\noindent
where we have represented  diagrammatically $U_{QL} (0,-\infty) | \Phi_j 
\rangle$ in Fig. \ref{figuql}.

Eqs. (\ref{facufi1}), (\ref{facufi3}), and (\ref{equl}) define the
decomposition theorem, which we rewrite as follows:
\begin{equation}
U (0,-\infty) | \Phi_i \rangle = \sum_{j=1}^d U_Q (0,-\infty) | \Phi_j
\rangle \langle \Phi_j | U (0,-\infty) | \Phi_i \rangle ,\label{dectheo}
\end{equation}

\noindent
where 
\begin{equation}
\langle \Phi_j | U (0,-\infty) | \Phi_i \rangle = \langle \Phi_j | 
U_L (0,-\infty) | \Phi_i \rangle \times \langle c | U (0,-\infty) | c 
\rangle ,\label{dectheo1}
\end{equation}

\noindent
and
\begin{equation}
U_Q (0,-\infty) | \Phi_j \rangle = U_{QL} (0,-\infty) | \Phi_j \rangle
\times U_Q (0,-\infty) | c \rangle ~.\label{dectheo2}
\end{equation}

The decomposition theorem, as given by Eq. (\ref{dectheo}),
states that the action of $U (0,-\infty)$ on $| \Phi_i \rangle$ can be 
represented as the sum of the wave functions $U_Q (0,-\infty) | \Phi_j
\rangle$ weighted with the matrix elements $\langle \Phi_j | U
(0,-\infty) | \Phi_i \rangle$.
Eq. (\ref{dectheo}) will play a crucial role in the next Sec.~\ref{sec:SMEI23},
 where we shall give an expression for the shell-model effective Hamiltonian.

\subsubsection{The model-space secular equation} \label{sec:SMEI23}

For the sake of clarity, let us recall the model-space secular equation 
(\ref{eq3})
\[
PH_{\rm eff}P | \Psi_{\alpha} \rangle  = E_{\alpha} P | \Psi_{\alpha}
\rangle ,
\]

\noindent
where $\alpha=1,..,d$, $| \Psi_{\alpha} \rangle$ and $E_{\alpha}$
are the true eigenvectors and eigenvalues of the full Hamiltonian $H$.

From now on, we shall use, for the convenience of the proof, the
Schr\"odinger representation.
However, it is worth to point out that the results obtained hold
equally well in the interaction picture.

First of all, we establish a one-to-one correspondence between some
model-space parent states $| \rho_{\alpha} \rangle$ and $d$ true
eigenfunctions $| \Psi_{\alpha} \rangle$. 
Let us start with a trial parent state
\begin{equation}
| \rho_{1} \rangle = \frac{1}{\sqrt{d}} \sum_{i=1}^{d} | \Phi_{i}
  \rangle,
\end{equation} 

\noindent
and act with the time development operator $U (0,-\infty)$ on it.
More precisely, we construct the wave function
\begin{equation}
\frac{U (0,-\infty) | \rho_{1} \rangle }{\langle \rho_{1} | U
  (0,-\infty) | \rho_{1} \rangle } = \lim_{\epsilon \rightarrow 0^+}
  \lim_{t' \rightarrow -\infty (1-i\epsilon)} \frac{e^{i H t'} |
  \rho_{1} \rangle }{\langle \rho_{1} |e^{i H t'}  | \rho_{1}
  \rangle }~.
\end{equation}

\noindent
By inserting a complete set of eigenstates of $H$ 
between the time evolution operator and $|\rho_1\rangle$, we obtain
\begin{eqnarray}
\frac{U (0,-\infty) | \rho_{1} \rangle }{\langle \rho_{1} | U
  (0,-\infty) | \rho_{1} \rangle }  
= \lim_{\epsilon \rightarrow 0^+}
  \lim_{t' \rightarrow -\infty } \frac{\sum_{\lambda}
  e^{i E_{\lambda} t'} e^{E_{\lambda} \epsilon t'} | \Psi_{\lambda} \rangle
\langle \Psi_{\lambda} |
  \rho_{1} \rangle }{\sum_{\beta}
  e^{i E_{\beta} t'} e^{E_{\beta} \epsilon t'} \langle \rho_{1} | 
\Psi_{\beta} \rangle \langle \Psi_{\beta} |
  \rho_{1} \rangle}
=\frac{| \Psi_1 \rangle}{\langle \rho_1 | \Psi_1 \rangle} \equiv | 
\tilde{\Psi}_1 \rangle ~.
 \end{eqnarray}

\noindent
Here, $|\Psi_1\rangle $ is the lowest eigenstate of $H$ for which $\langle
\Psi_1 | \rho_1 \rangle \neq 0$, this stems from the fact that the
real exponential damping factor in the above equation suppresses all
the other non-vanishing terms.

This procedure can be easily continued, thus obtaining a set of wave
functions
\begin{equation}
| \tilde{\Psi}_{\alpha} \rangle = \frac { U(0, -\infty) |
  \rho_{\alpha} \rangle}{ \langle \rho_{\alpha} | U(0, -\infty) |
  \rho_{\alpha} \rangle},
\label{parent}
\end{equation}

\noindent
where
\begin{equation}
\langle \rho_{\alpha} | \rho_{\alpha} \rangle = 1,
\end{equation}
\begin{equation}
\langle \rho_{\alpha} | \Psi_{\alpha} \rangle \neq 0,
\end{equation}
\begin{equation}
\langle \rho_{\alpha} | \Psi_{1} \rangle = \langle \rho_{\alpha} 
| \Psi_{2} \rangle = ... = \langle \rho_{\alpha} | \Psi_{\alpha -1}
\rangle = 0~.
\label{parentlast}
\end{equation}
The above correspondence (Eqs. \ref{parent}-\ref{parentlast})
holds if the parent states $| \rho_{\alpha} \rangle$ are linearly
independent. Under this assumption, we can write
\begin{equation}
| \rho_{\alpha} \rangle = \sum_{i=1}^{d} C_i^{\alpha} | \Phi_i \rangle,
\label{orthoa}
\end{equation}
with 
\begin{equation}
\sum_{i=1}^{d} C_i^{\alpha} C_i^{\beta} = \delta_{\alpha \beta}~.
\end{equation}

By construction, $| \tilde{\Psi}_{\alpha} \rangle$ is an eigenstate of
$H$, so, using Eq. (\ref{parent}), we can write
\begin{equation}
 H \frac { U(0, -\infty) |
  \rho_{\alpha} \rangle}{ \langle \rho_{\alpha} | U(0, -\infty) |
  \rho_{\alpha} \rangle} = E_{\alpha} \frac { U(0, -\infty) |
  \rho_{\alpha} \rangle}{ \langle \rho_{\alpha} | U(0, -\infty) |
  \rho_{\alpha} \rangle}~.
\end{equation}

\noindent
Now, making use of  Eq. (\ref{orthoa}) and applying the decomposition
theorem as expressed by Eq. (\ref{dectheo}), the above equation becomes
\begin{equation}
H \frac{\sum_{i j} U_Q(0, -\infty) | \Phi_j \rangle \langle \Phi_{j} | 
U(0, -\infty) |
  \Phi_{i} \rangle C_i^{\alpha}}{\sum_{k m} \langle \Phi_{k} | 
U(0, -\infty) |
  \Phi_{m} \rangle C_k^{\alpha} C_m^{\alpha}} 
= E_{\alpha} \left. \frac{\sum_{i j} U_Q(0, -\infty) | \Phi_j \rangle \langle
  \Phi_{j} | U(0, -\infty) |
  \Phi_{i} \rangle C_i^{\alpha}}{\sum_{k m} \langle \Phi_{k} | 
U(0, -\infty) |
  \Phi_{m} \rangle C_k^{\alpha} C_m^{\alpha}} \right..
\label{secul1}
\end{equation}

\noindent
In order to simplify the above expression, we define the coefficients 
$b_j^{\alpha}$
\begin{equation}
b_j^{\alpha} = \frac{\sum_{i} \langle \Phi_{j} | 
U(0, -\infty) |
  \Phi_{i} \rangle C_i^{\alpha}}{\sum_{k m} \langle \Phi_{k} | 
U(0, -\infty) |
  \Phi_{m} \rangle C_k^{\alpha} C_m^{\alpha}}  
= \frac{\sum_{i} \langle \Phi_{j} | 
U_L(0, -\infty) |
  \Phi_{i} \rangle C_i^{\alpha}}{\sum_{k m} \langle \Phi_{k} | 
U_L(0, -\infty) |
  \Phi_{m} \rangle C_k^{\alpha} C_m^{\alpha}},
\end{equation}

\noindent 
where the r.h.s. of the above equation has been obtained by use of 
Eq. (\ref{dectheo1}), which cancels out the vacuum fluctuations
diagrams $\langle c | U(0, -\infty) | c \rangle$ of Fig. \ref{fluc}.
Multiplying Eq. (\ref{secul1}) by $\langle \Phi_k |$, it becomes
\begin{equation}
\sum_{j = 1}^{d} \langle \Phi_k | H U_Q(0, -\infty) |
  \Phi_{j} \rangle b_j^{\alpha} = E_{\alpha} b_k^{\alpha},
\label{secul2}
\end{equation}

\noindent 
where use has been made of the relation $\langle \Phi_k | U_Q(0,
-\infty) | \Phi_{j} \rangle = \delta_{kj}$ (see Ref.~\cite{Kuo71}).
The above equation is the model-space secular equation we needed,
where $H_{\rm eff}$ is given by $H U_Q(0, -\infty)$ and 
$b_j^{\alpha}$ represents the projection of $| \tilde{\Psi}_{\alpha}
\rangle$ onto the model-space wave function $| \Phi_j \rangle$.
  
\begin{figure}[hbtp]
\begin{center}
\includegraphics[scale=0.42,angle=-90]{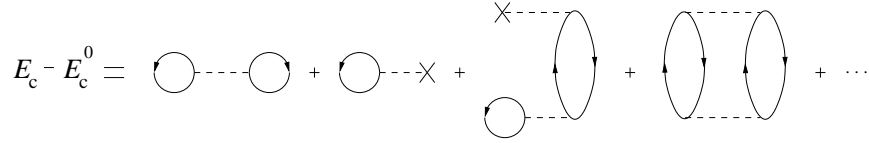}
\end{center}
\caption{Goldstone linked-diagram expansion of $E_c-E^0_c$. 
The cross insertions represent the $H_0$ operator.}
\label{goldstone}
\end{figure}

In Eq. (\ref{secul2}) we can write the Hamiltonian $H$ as $H=H_0+H_1$.
First, let us consider the contribution from $H_0$.
Since $|\Phi_i\rangle$ is an eigenstate of $H_0$ and $\langle \Phi_i |
U_Q(0,-\infty) | \Phi_j \rangle = \delta_{ij}$, we obtain 
\begin{equation}
\langle \Phi_i | H_0 U_Q(0,-\infty) | \Phi_j \rangle = \langle 
\Phi_i | H_0 | \Phi_j \rangle = \delta_{ij} ( E^0_c + E^0_v ),
\end{equation}

\noindent
where $E^0_c$ is  the unperturbed core energy and $E^0_v$ is the 
unperturbed energy of the two valence nucleons with respect to $E^0_c$.

As for the matrix element $\langle
\Phi_i| H_1 U_Q(0,-\infty) | \Phi_j \rangle$, we see from inspection of 
Eq. (\ref{dectheo2})  that it  contains a collection of
diagrams in which $H_1$ is not linked to any valence line at $t=0$.
These diagrams are obtained acting with  $H_1$ on $U_Q (0,-\infty) | c \rangle$
 and their contribution  to the l.h.s. of
Eq. (\ref{secul2}) is $\delta_{ij} \langle c | H_1 U_Q(0,-\infty) 
| c \rangle = \delta_{ij}(E_c-E^0_c)$, $E_c$ being the true 
ground-state energy of the closed-shell system.
The diagram expansion of $(E_c-E^0_c)$ is given in  Fig. \ref{goldstone}
as illustrated in~\cite{Goldstone57}.

The other terms of $\langle \Phi_i| H_1 U_Q(0,-\infty) | \Phi_j
\rangle$ are all linked to the external active lines. 
By denoting, for simplicity, the collection of these 
terms as
\begin{equation}
\langle \Phi_i | [ H_1 U_Q (0,-\infty) ]_L | \Phi_j \rangle ,
\label{h1uq}
\end{equation}

\noindent
the secular equation (\ref{secul2}) can be rewritten in the
following form:
\begin{equation}
 E^0_v + 
\sum_{j = 1}^{d} \langle \Phi_i | [ H_1 U_Q(0, -\infty)]_L |
 \Phi_{j} \rangle b_j^{\alpha} 
= (E_{\alpha}-E_c) b_i^{\alpha}~.
\label{secul}
\end{equation}

\begin{figure}[hbtp]
\begin{center}
\includegraphics[scale=0.47,angle=-90]{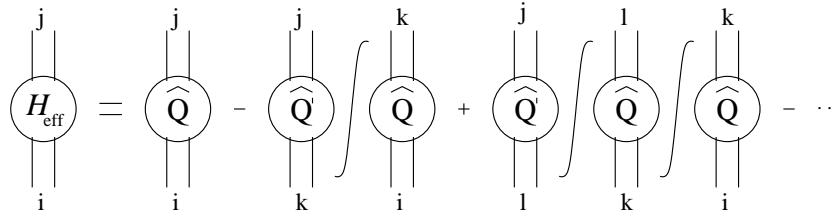}
\end{center}
\caption{Diagrammatic representation of $H_{\rm eff}$ matrix elements.
\label{heff}}
\end{figure}

We now define

\begin{equation}
H^1_{\rm eff} = [ H_1 U_Q(0, -\infty)]_L ,
\end{equation}

\noindent
and show in Fig. \ref{heff}   a diagrammatic representation of its matrix
elements, which has been obtained 
starting from the definition of $U_{QL}
(0,-\infty) | \Phi_j \rangle$ given in Fig. \ref{figuql}.
It should be noted that we have two kinds of $\hat{Q}$-box,
$\hat{Q}$ and $\hat{Q'}$. 
$\hat{Q}$ and $\hat{Q'}$ are a collection of irreducible, valence-linked
diagrams with at least one and two $H_1$-vertices,
respectively. 
The fact that the lowest order term in $\hat{Q'}$ is of second order
in $H_1$ is just because of the presence of $H_1$ in the matrix elements of
Eq. (\ref{h1uq}).  

Formally, $H^1_{\rm eff}$ can be written in operator form as 
\begin{equation}
H^1_{\rm eff} = \hat{Q} - \hat{Q}' \int \hat{Q} + \hat{Q}' \int \hat{Q} \int
\hat{Q} - \hat{Q}' \int \hat{Q} \int \hat{Q} \int \hat{Q} + ~...~~,
\label{heffop}
\end{equation}

\noindent
where the integral sign represents a generalized folding operation.
It is worth noting that, by definition, the $\hat{Q}$-box contains
diagrams at any order in $H_1$.
Actually, when performing realistic shell-model calculations
it is customary to include diagrams up to a finite order in
$H_1$. 
A complete list of all the $\hat{Q}$-box diagrams up to third order
can be found in Ref.~\cite{Hjorth95}.

The Schr\"{o}dinger equation (\ref{eq1}) is finally reduced to the 
model-space eigenvalue problem of Eq.~(\ref{secul}), whose 
eigenvalues are the energies of the $A$-nucleus
relative to the core ground-state energy $E_{c}$. As mentioned above, the
latter  can be calculated by way of
the Goldstone expansion \cite{Goldstone57}, expressed as the sum of
diagrams shown  in Fig.~\ref{goldstone}. It is also worth noting 
that the operator $H^1_{\rm eff}$, as defined by Eq.~(\ref{heffop}), 
contains both one- and two-body contributions since in our derivation of the 
of Eq.~(\ref{secul}) we have considered nuclei with two-valence nucleons.
All the one-body contributions, the so-called $\hat{S}$-box
\cite{Shurpin83}, once summed to the eigenvalues of $H_{0}$,  $E^0_v$, 
give  the SP term of the effective shell-model Hamiltonian.
The  eigenvalues  of this term represent the energies of the nucleus with 
one-valence nucleon 
relative to the core. This identification justifies the commonly used
subtraction procedure
\cite{Shurpin83} where only the two-body terms of $H^1_{\rm eff}$ (the
effective two-body interaction $V_{\rm eff}$) are retained while the
single-particle energies are taken from experiment.

In this context, it is worth to mentioning  that in some recent papers
\cite{Coraggio05c,Coraggio07} the SP energies  and the two-body
interaction employed in realistic shell-model calculations are
derived consistently in the framework of the linked-cluster expansion.
In particular, in Ref.~\cite{Coraggio05c}, where the light 
$p$-shell nuclei have been studied using the  CD-Bonn potential renormalized
through the $V_{\rm low-k}$ procedure (see Sec.~\ref{sec:vlowk}),
a Hartree-Fock basis is derived which is 
used to calculate the binding energy of $^{4}$He and the 
effective shell-model Hamiltonian composed of one- and two-body terms.

In concluding this brief discussion of Eq.~(\ref{secul}), it is worth to
point out  that for systems with more than 
two valence nucleons  $H^{1}_{\rm eff}$  contains 1-, 2-, 3-,$\cdots$, 
$n$-body components, 
even if the original Hamiltonian of Eq. (\ref{defh}) contains only a 
two-body force~\cite{Ellis77}. The role of these effective many-body
forces as well as that of a genuine 3-body potential in the shell model 
is still an open  problem (see \cite{Ellis05} and 
references therein) and is outside the scope of this review.
However, we shall  come back to this point in Sec.~\ref{sec:res4}.

In Sec.~\ref{sec:SMEI21} we have shown how, in a degenerate model space, a
generalized folding diagram  can be evaluated.
In \cite{Kuo80}, it has been shown that a term
like $- \hat{Q}' \int \hat{Q}$ may be written as 
\begin{equation}
- \hat{Q}' \int \hat{Q} = \frac{d \hat{Q}' (\omega)}{d \omega} \hat{Q}
  (\omega),
\end{equation}

\noindent
where $\omega$ is equal to the energy of the incoming particles.

The above result may be extended to obtain a convenient prescription
to calculate $H^1_{\rm eff}$ as given by Eq. (\ref{heffop}). 
We can write 
\begin{equation}
H^1_{\rm eff} = \sum_{i=0}^{\infty} F_i~,
\label{sumfi}
\end{equation}

\noindent
where
\begin{eqnarray}
F_0 &=& \hat{Q},  \nonumber \\
F_1 &=& \hat{Q}_1 \hat{Q} ,\nonumber\\
F_2 &=& \hat{Q}_2 \hat{Q} \hat{Q} + \hat{Q}_1 \hat{Q}_1 \hat{Q} ,
\label{fm} \\
&~~~& ... \nonumber
\end{eqnarray}

\noindent
and
\begin{equation}
\hat{Q}_m = \frac {1}{m!} \frac {d^m \hat{Q} (\omega)}{d \omega^m} \biggl| 
_{\omega=\omega_0} ~, 
\label{qm}
\end{equation}

\noindent
$\omega_0$ being the energy of the incoming particles at $t =
- \infty$.

Note that in Eq. (\ref{fm}) we have made use of the
fact that, by definition, 
\begin{equation}
 \frac{d \hat{Q}' (\omega)}{d \omega} = \frac{d \hat{Q} (\omega)}{d
 \omega}~.
\end{equation}

The number of terms in $F_i$ grows dramatically with $i$. Two
iteration methods to partially sum up the folded diagram series have
been introduced in Refs.~\cite{Krenciglowa74,Lee80,Suzuki80}.
These methods are known as the Krenciglowa-Kuo (KK) and the Lee-Suzuki
(LS) procedure, respectively. In \cite{Suzuki80},
it has been shown that, when converging, the
KK partial summation converges to those states with the largest model
space overlap, while the LS one converges to the lowest  states in
energy.

The LS iteration procedure   was proposed within the framework of an 
approach to the 
construction of the effective interaction known as Lee-Suzuki 
method, which is based on the similarity transformation theory. 
In the following subsection we shall briefly present this 
method to illustate the LS iterative technique used  
to sum up the folded-diagram series (\ref{sumfi}).

\subsection{The Lee-Suzuki method }\label{sec:SMEI3}

Let us start with  the Schr\"odinger equation for the $A$-nucleon system
as given in Eq.~(\ref{eq1}) and consider  the similarity transformation
\begin{equation}
\mathcal{H}=X^{-1} H X ,
\end{equation}
\noindent
where $X$ is a transformation operator defined in the whole Hilbert
space.

If we require that 
\begin{equation}
Q \mathcal{H} P=0 , \label{deceq1}
\end{equation}
\noindent
then it can be easily proved that the $P$-space effective Hamiltonian
satisfying  Eq. (\ref{eq3}) is just $P \mathcal{H} P$.
Equation (\ref{deceq1}) is the so-called decoupling equation, whose solution 
leads to the determination of $H_{\rm eff}$.

There is, of course, more than one choice for  the transformation
operator $X$. We take
\begin{equation}
X=e^{\Omega}, \label{omegaop}
\end{equation}

\noindent
where the wave operator $\Omega$ satifies the conditions:
\begin{equation}
\Omega= Q \Omega P ,
\label{omegapro1}
\end{equation}
\begin{equation}
P \Omega P= Q \Omega Q = P \Omega Q =0 ~. 
\label{omegapro2}
\end{equation}

Taking into account Eq. (\ref{omegapro1}), we can write $X=1 + \Omega$
and consequently
\begin{equation}
H_{\rm eff} = P \mathcal{H} P = PHP +PH_1 Q \Omega ,
\end{equation}
\noindent
with $H_{1}$ defined in Eq. (\ref{defh1}), while 
the decoupling equation (\ref{deceq1}) becomes
\begin{equation}
Q H_1 P + Q H Q \Omega - \Omega P H P - \Omega P H_1 Q \Omega = 0~. 
\label{deceq2} 
\end{equation}

We now introduce the $P$-space effective interaction $R$ by
subtracting the unperturbed energy $P H_0 P$ from $H_{\rm eff}$:
\begin{equation}
R=P H_1 P + P H_1 Q \Omega ~. \label{effint}
\end{equation}

In a degenerate model space $PH_0P=\omega_0P$, we can consider a linearized
iterative equation for the solution of the decoupling equation
(\ref{deceq2})
\begin{equation}
[ E_0 - (Q H Q - \Omega_{n-1} P H_1 Q) ] \Omega_n = Q H_1 P -
\Omega_{n-1} P H_1 P ~. \label{iter1}
\end{equation}

We now write the $\hat{Q}$-box 
introduced in Sec.~\ref{sec:SMEI22} in operatorial form as 
given in~\cite{Suzuki80} 
\begin{equation}
\hat{Q} = P H_1 P + P H_1 Q \frac{1}{E_0 - Q H Q} Q H_1 P , \label{qbox}
\end{equation}
\noindent 
and define $R_n$ as the $n$-th order iterative effective interaction 
\begin{equation}
R_n= P H_{1} P + P H_1 Q \Omega_n ~.
\end{equation}

Then, if we start with $\Omega_0=0$ in
Eq. (\ref{iter1}), $R_n$ can be written in terms of the $\hat{Q}$-box
and its derivatives:
\begin{eqnarray}
R_1 &=& \hat{Q},  \nonumber \\
R_2 &=& [1-\hat{Q}_1]^{-1} \hat{Q} ,\nonumber\\
R_3 &=& [1-\hat{Q}_1-\hat{Q}_2 R_2]^{-1} \hat{Q} ,\nonumber\\
&~~~& ... \nonumber \\
R_n &=& \left[ 1 - \hat{Q}_1 - \sum_{m=2}^{n-1} \hat{Q}_m
  \prod_{k=n-m+1}^{n-1} R_k \right]^{-1} \hat{Q} 
\end{eqnarray}

\noindent
where $\hat{Q}_n$ is defined in Eq. (\ref{qm}).

The solution $R_n$ so obtained corresponds to a certain resummation of
the folded diagrams to infinite order.
In fact, if we consider, for example, $R_2$ and expand it in power
series of $\hat{Q}_1$, we obtain
\begin{equation}
R_2 = [1-\hat{Q}_1]^{-1} \hat{Q}=1 + \hat{Q}_1\hat{Q}+ \hat{Q}_1
\hat{Q}_1 \hat{Q} + ...
\end{equation}
It is clear from the above expression that $R_2$ contains terms
corresponding to an infinite number of folds.

It can be shown that the Lee-Suzuki method yields converged results after
a small number of iterations \cite{Shurpin83}, making this
procedure very advantageous to sum up the folded-diagram series.

\section{Handling the short-range repulsion of the $NN$ potential} 
\label{sec:G&V}

As already pointed out, a most important goal
of nuclear shell-model theory is to derive the effective interaction
between valence nucleons directly from the free $NN$
potential. In Sec.~\ref{sec:SMEI},
we have shown how this effective interaction may be
calculated microscopically within the framework of a many-body 
theory. As is well known, 
however, $V_{NN}$ is  not suitable for this kind of approach. In fact, 
owing to the contribution from the repulsive core, the
matrix elements of  $V_{NN}$ are generally very large and an order-by-order 
perturbative calculation of the effective interaction in terms of $V_{NN}$ is
clearly not meaningful. 
A resummation method has to be employed in order to take care
of  the  strong short-range repulsion contained in  $V_{NN}$. 
\begin{figure}[hbtp]
\begin{center}
\includegraphics[scale=0.45,angle=-90]{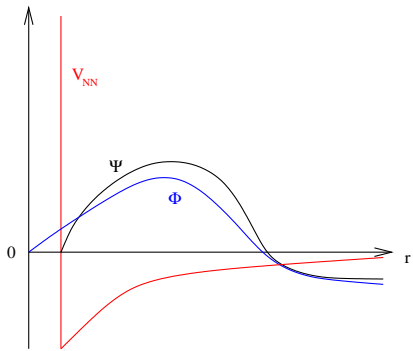}
\end{center}
\caption{(Color online) Radial dependence of the relative wave function $\Psi$ 
of two nucleons interacting via a hard-core potential $V_{NN}$. 
$\Phi$ refers to the uncorrelated wave function.}
\label{fig:defg}
\end{figure}

This point is more evident if we consider the extreme model of a
hard-core potential, as is the case of the 
potential models developed in the early 1960s. 
This situation is illustrated in Fig. \ref{fig:defg}. 
For these potentials the perturbation expansion of the effective
interaction in terms of $V_{NN}$ is  meaningless, since
each term of the series involving
matrix elements of the $NN$ potential between unperturbed two-body
states is infinite. 
This is because the unperturbed wave function, in contrast to the
true wave function, gives a non-zero probability of finding a particle
located inside the hard-core distance. 

The traditional way out of this problem is the
so-called Brueckner reaction matrix $G$, which is based on the idea of
treating exactly the interaction between a given pair of nucleons
\cite{Brueckner54}.
The $G$ matrix, defined as a sum of all ladder-type 
interactions (see Sec.~\ref{sec:GM2}),
is used to replace the $NN$ interaction vertices once a
rearrangement of the effective interaction perturbative series has
been performed.

Recently, a new method to
renormalize the $NN$ interaction has been proposed \cite{Bogner01,Bogner02}.
A low-momentum model space defined up to a cutoff momentum $\Lambda$ is
introduced and an effective potential $V_{\rm low-k}$ is derived from
$V_{NN}$.
This $V_{\rm low-k}$ satisfies a decoupling condition between the low-
and high-momentum spaces. 
Moreover, it is a smooth potential which preserves exactly the
on-shell properties of the original potential and it is thus suitable
to advantageously replace $V_{NN}$ in realistic many-body
calculations.

Secs.~\ref{sec:GM} and \ref{sec:vlowk}  are devoted to the description of the
reaction $G$ matrix and $V_{\rm low-k}$ potential, respectively.

Before doing this, however, it may be worth recalling that a method to avoid 
the $G$-matrix treatment to eliminate effects of the repulsive core in the
$NN$ potential was proposed 
in the late 1960s \cite{Elliott67,Elliott68}. As already mentioned in the 
Introduction, this consists in using the experimental $NN$ phase shifts to 
deduce matrix elements of the $NN$ potential in a basis of relative harmonic 
oscillator states. These matrix elements, which have become known as the 
Sussex matrix elements (SME), have been used in several nuclear structure 
calculations, but the agreement with experiment has been generally only 
semi-quantitative. A comparison between the results obtained by 
Sinatkas {\em et al.} \cite{Sinatkas92} using the SME and those obtained 
with a realistic 
effective interaction derived from the Bonn A potential is  made  for the 
N=50 isotones in \cite{Coraggio00}. 

\subsection{The Brueckner $G$-matrix approach} \label{sec:GM}

\subsubsection{Historical introduction} \label{sec:GM1} 

The concept of $G$ matrix originates from the theory of multiple
scattering of Watson \cite{Watson53,Francis53}. 
In this approach, the elastic scattering of a fast particle by a 
nucleus was described by way  of a transformed
potential obtained in terms of the Lippmann-Schwinger matrix
\cite{Lippmann50} for the two-body scattering.
The procedure of Watson for constructing such an ``equivalent two-body 
potential'' was generalized to the study of nuclear many-body systems
by Brueckner and co-workers  \cite{Brueckner54, Brueckner55}. 
They introduced a reaction matrix for the scattering of two nucleons
while they are moving in the nuclear medium. 
This matrix, which is known as the Brueckner reaction matrix, includes
all two-particle correlations via summing all ladder-type
interactions,  and made it  possible to perform Hartree-Fock
self-consistent calculations for nuclear matter.

Only a few years later Goldstone  \cite{Goldstone57} proved a new 
perturbation
formula for the ground-state energy of nuclear matter which gave the
formal basis of the Brueckner theory. 
The Goldstone linked-diagram theory applies to systems
with non-degenerate ground state, which is the case of nuclear matter
as well as of closed-shell finite nuclei.

The Brueckner  theory was seen to be the key to solve the paradox
on which the attention of many nuclear physicists was focused during the
early 1950s. 
In fact, it allowed to reconcile a description of the nucleus in terms
of an overall potential and the peculiar features of the two-body
nuclear force. 
This was very well evidenced in the paper by Bethe
\cite{Bethe56}, whose main purpose was indeed to establish that  the
Brueckner theory provided the theoretical foundation for the shell model.

After their first works, Brueckner and co-workers
published a series of papers on the same subject 
\cite{Brueckner55a,Brueckner55b,Brueckner58,Brueckner58a,Brueckner61}, 
where further analyses and developments of the method as well as
numerical calculations were given. 
Important advances as regards nuclear matter were provided by the
work of Bethe and Goldstone \cite{Bethe57} and 
Bethe {et al.} \cite{Bethe63}. 
For several years up to the 1960s, nuclear physicists
were indeed very active in this field, as may be seen from the review
papers by Day \cite{Day67}, Rajaraman and Bethe \cite{ Rajaraman67},
and Baranger \cite{Baranger69}, where comprehensive
lists of references can be found. 
For a recent review of the developments in this area  made Bethe and coworkers
we refer to  \cite{Brown06}.

As regards open-shell nuclei, we have shown in the previous section
that the model-space effective interaction may be obtained by way of a
linked-diagram expansion containing both folded
and non-folded Goldstone diagrams. 
An order-by-order perturbative calculation of such diagrams in
terms of $V_{NN}$ is not appropriate and 
one has to resort again to the reaction matrix $G$.

The renormalization of $V_{NN}$ through the Brueckner theory has
been the standard procedure to derive realistic effective
interactions since the pioneering works of the early 1960s. A survey of
shell-model calculations employing the $G$ matrix is given in Secs.~\ref{sec:SMCEP}
and \ref{sec:SMCMC}.

In Sec.~\ref{sec:GM2} we shall define the reaction matrix $G$ and discuss
the main problems related to its definition.
Then, in Sec.~\ref{sec:GM3}, we shall address the problem of how the reaction
matrix may be calculated, focusing attention on the
$G_T$ matrix for which plane waves are used as intermediate states. 
It is, in fact, this matrix which is most commonly used nowadays in realistic 
shell-model calculations. 
For simplicity, almost everywhere in this section we shall denote the $NN$
potential by $V$.

\subsubsection{Essentials of the theory}
\label{sec:GM2}

In the literature, the $G$ matrix is typically
introduced by way of the Goldstone expansion for the calculation
of the ground-state energy in 
nuclear matter and closed shell nuclei. 
As a first step, the ground-state energy is written as a linked-cluster 
perturbation series. 
Then,  all diagrams differing one from another only in the number
of $V$ interactions between two particles lines are summed. 
This corresponds to define a well-behaved two-body operator, the
reaction matrix $G$, that replaces the potential $V$  in
the series. 
A very clear and simple presentation of the $G$ matrix along this line
is provided in the paper by Day \cite{Day67}.

Here, we shall not discuss the details of the $G$-matrix theory,
but simply give the elements needed to make clear its definition 
in connection with the derivation of the model-space effective 
interaction for open-shell nuclei.
Therefore, in the following we refer, as in the previous section, to
an $A$-nucleon system  with the   Hamiltonian  given by
Eqs. (\ref{defh})-(\ref{defh1})  and  represented  as a 
doubly closed core plus valence nucleons moving in a limited number of
SP orbits  above the core. 
 
The two-body operator $G$ is defined by the integral equation
\begin{equation}
\label{eq:defg}
G(\omega) = V+V \frac{Q_{2p}}{\omega- H_{0}} G(\omega),
\end{equation}

\noindent
where $\omega$ is an energy variable known as ``starting energy'' and
$Q_{2p}$ an operator which projects onto particle-particle states, 
namely states composed of two SP levels above the doubly closed core. 
As we shall discuss later, this operator may be chosen in different ways
depending on the specific context  in which it is used.
Here it is worth noting  that its presence in  Eq. (\ref{eq:defg}) reminds us 
that  the $G$ matrix, differently from the Lippmann-Schwinger $T$ matrix, is 
defined in the nuclear medium.
It may be also noted that the starting energy $\omega$ is not a free
parameter.
Rather, its value is determined by the physical problem being studied.
For the $T$ matrix, $\omega$ denotes the scattering energy for two
particles in free space, but  depends on the nuclear medium for the 
$G$ matrix. 
In fact, we shall see that for a $G$ matrix to be used within the 
linked-diagram expansion  of the effective interaction  $\omega$ depends 
on the diagram where $G$ appears. Only in some cases it represents the energy 
of the two-particle incoming state, which was instead the meaning of the 
energy variable $\omega$ used in Sec.~\ref{sec:SMEI}. 

Let us start by writing  the $Q_{2p}$ operator as
\begin{equation}
\label{eq:defq}
Q_{2p}= \sum_{ij} C(ij)|\phi_{ij} \rangle \langle\phi_{ij}|, 
\end{equation}

\noindent
where the state $|\phi_{ij}\rangle$, which is the antisymmetrized product of
the two SP states $|i\rangle$ and $|j\rangle$, is an eigenstate of $H_{0}$
with energy $\epsilon_{i} + \epsilon_{j}$. 
The constant $C(ij)$ is 1 if the state
$|\phi_{ij}\rangle$ pertains to the space defined by  $Q_{2p}$, 0 otherwise. 
We now introduce the operator $P_{2p}$ which
projects on the complementary space
\begin{equation}
\label{eq:defp}
P_{2p}=1-Q_{2p}.
\end{equation}
 
Taking matrix elements of $G$ between states  of the
$P_{2p}$-space we have  
\begin{equation}
\label{eq:meg}
\langle \phi_{kl}|G(\omega)|\phi_{nm}\rangle =  \langle\phi_{kl}|V 
|\phi_{nm} \rangle + \sum_{ij} \langle\phi_{kl}|V|\phi_{ij}\rangle
  \frac{C(ij)}{\omega - \epsilon_{i}- \epsilon_{j}}
\langle\phi_{ij}|G(\omega)|\phi_{nm}\rangle,
\end{equation}

\noindent
which in a series expansion form becomes
\begin{eqnarray}
\langle\phi_{kl}| G(\omega)|\phi_{nm}\rangle  & = &
\langle\phi_{kl}|V|\phi_{nm} \rangle 
+ \sum_{ij} \langle\phi_{kl}|V|\phi_{ij}\rangle \frac{C(ij)}{\omega - 
 \epsilon_{i}- \epsilon_{j}} \langle\phi_{ij}|V|\phi_{nm} \rangle 
\nonumber \\
&& +  \sum_{iji'j'} \langle \phi_{kl}|V|\phi_{ij} \rangle
\frac{C(ij)}{\omega - \epsilon_{i}- \epsilon_{j}} \langle \phi_{ij}| V
|\phi_{i'j'}\rangle 
\frac{C(i'j')}{\omega -\epsilon_{i'}- \epsilon_{j'}} \langle 
\phi_{i'j'}|V|\phi_{nm} \rangle  + \cdots
\label{eq:expg} 
\end{eqnarray}

\begin{figure}[hbtp]
\begin{center}
\includegraphics[scale=0.8,angle=-90]{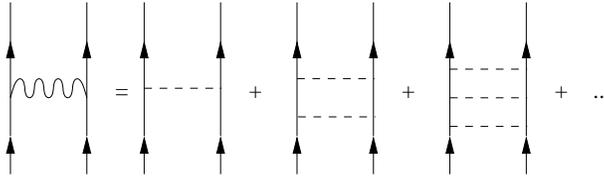}
\end{center}
\caption{Diagrammatic representation defining the matrix $G$.
\label{diagramGT}}
\end{figure}

The above expansion can be represented diagrammatically by
the series 
shown in Fig. \ref{diagramGT}.  
The diagrams on the
r.h.s. of this figure are known as ``ladder'' diagrams and 
each diagram corresponds  to a situation in which a pair of particles 
interacts a certain number of times with the restriction that
the intermediate states involved
in the scattering must be those defined through the $Q_{2p}$ operator.
In other words, two particles initially in a state of the
$P_{2p}$-space undergo a sequence of scatterings into states of the 
$Q_{2p}$-space and then after several such scatterings go back to
a state of the original space. 

For the sake of completeness, we introduce the correlated wave function
$|\psi_{nm}( \omega)\rangle$ \cite{Bethe57},
\begin{equation}
\label{eq:cwfv}
|\psi_{nm} (\omega) \rangle = |\phi_{nm}\rangle +
\frac{Q_{2p}}{\omega-H_{0}} V |\psi_{nm}(\omega)\rangle,
\end{equation}

\noindent
which once iterated becomes
\begin{equation}
\label{eq:cwfg}
|\psi_{nm} (\omega) \rangle= |\phi_{nm}\rangle +
\frac{Q_{2p}}{\omega-H_{0}} G(\omega)|\phi_{nm}\rangle,
\end{equation}
\noindent
where use has been made of the integral equation (\ref{eq:defg}). 
Eq. (\ref{eq:cwfg}) allows one to write the action of $G$ on an
unperturbed state as
\begin{equation}
\label{eq:gv}
G(\omega)|\phi_{nm}\rangle= V|\psi_{nm}(\omega)\rangle,
\end{equation}

\noindent
which makes evident that the $G$ operator may be considered an
effective potential. 
The correlated wave function (\ref{eq:cwfv}) and its properties are
extensively discussed in Refs.~\cite{Baranger69,Deshalit74,Bethe71}.

At this point, we go further in our discussion clarifying  
the meaning of the starting energy
$\omega$ and  illustrating some possible choices of the projection operator 
$Q_{2p}$.

\begin{figure}[hbtp]
\begin{center}
\includegraphics[scale=0.60,angle=-90]{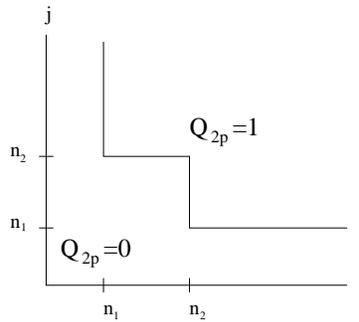}
\end{center}
\caption{A graph of the projection operator $Q_{2p}$ appropriate for 
open-shell nuclei.
\label{dfig:defq2}}

\end{figure}

Let us begin with $Q_{2p}$.  
We define the  $Q_{2p}$ operator by specifying its boundaries labelled
by the three numbers ($n_{1}$,$n_{2}$,$n_{3}$), each representing a 
SP level, the levels being numbered starting from the bottom of the 
potential well. 
Explicitly, using Eq. (\ref{eq:defq}) the operator $Q_{2p}$ is written as
$$
C(ij)= \left\{ \begin{array}{ll}
1 & \mbox{if $i,j > n_{2}$}\\
1 & \mbox{if $n_{1}< i \leq n_2$ and  $j>n_{2}$} \\
~& ~~~~~~~~ \mbox{or vice-versa}\\
0 & \mbox{otherwise.}
\end{array}
\right.
$$

The graph of Fig. (\ref{dfig:defq2}) makes more clear its definition. 
Note that the index $n_3$ denotes the number of levels in the full
space. In principle, it should be infinite; in practice, as we shall see in
Sec.~\ref{sec:GM3}, it is chosen to be a large but finite number.
As regards the indices  $n_1$ and  $n_2$, the only
mandatory requiriment is 
that none of them should be below the last occupied orbit of  the doubly 
closed core. It is customary to take  for $n_1$ the number the SP levels 
below the Fermi surface while  $n_2$ may be   
chosen  starting from the last SP valence level and going up. 
How this  choice is performed is better illustrated by an example.

\begin{figure}[hbtp]
\begin{center}
\includegraphics[scale=0.8,angle=-90]{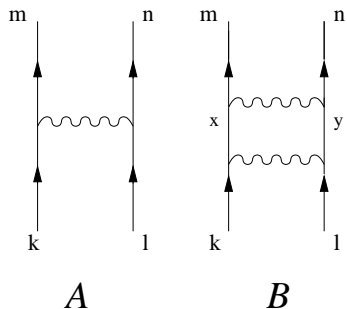}
\end{center}
\caption{Second-order ladder diagram contained in the $\hat Q$-box.
\label{diagramGT2}}
\end{figure}

Let us consider the nucleus $^{18}$O which  in the shell-model 
framework is described as
consisting of the doubly closed $^{16}$O and two valence neutrons which
are allowed  to occupy the three levels  of the $sd$ shell. 
The model-space effective interaction for $^{18}$O  may be derived by
using the linked-diagram expansion of Sec.~\ref{sec:SMEI}, with the $G$ matrix
replacing the $NN$ potential in all the irreducible, valence-linked
diagrams composing the $\hat Q$-box.
In so doing, one has to be careful to exclude from the   $\hat Q$-box 
those diagrams containing a 
ladder  sequence  already included in the  $G$ matrix.
We may take the  matrix $G$  with a  $Q_{2p}$ operator as specified, 
for instance, by $(3,6,\infty)$. In this  case,  
one of  two SP levels  composing the   intermediate state in the
calculation of the $G$ matrix  has  to be beyond the 
$sd$ shell while the other one may be also an $sd$ level.   
Then, when calculating  the $\hat{Q}$-box the included diagrams strictly
depend on the considered $G$ matrix. 
As an example, we have reported in Fig. \ref{diagramGT2} a
first- and second-order diagram of the $\hat Q$-box, the wavy lines denoting
the $G$-interactions. As discussed in Sec.~\ref{sec:SMEI22}, the incoming 
and outcoming 
lines of the A and B diagrams are levels of the $sd$ shell, while the 
the intermediate state of diagram B should have at least one passive line. 
This means that  none of the two SP states, $x$ or $y$, 
could be a $0s$ or a $0p$ level and
at least one of them must be beyond the $sd$ shell. Therefore, if we take the
$G$ matrix with the $(3,6,\infty)$  $Q_{2p}$ operator,  all possible B diagrams
of Fig. \ref{diagramGT2} are already contained in the A diagram. 
On the other hand,  if we increase the $n_2$ using a $G$ matrix with the  
$(3,10,\infty)$ $Q_{2p}$ operator,  the B diagrams  should  
explicitly appear in the calculation of the $\hat{Q}$-box providing that 
$x,y\leq 10$ and $x$ and/or $y\geq7$. 
A detailed discussion on the points illustrated here is 
in Ref.~\cite{Krenciglowa76}. 

\begin{figure}[hbtp]
\begin{center}
\includegraphics[scale=0.7,angle=-90]{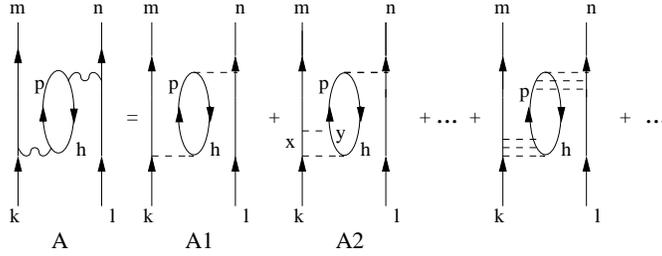}
\end{center}
\caption{Illustration of the summation procedure to replace $V$ vertices with
$G$ vertices.
\label{fig:vbyg}}
\end{figure}

We now come to the energy variable $\omega$ which may be seen as the
``starting energy'' at which $G$ is computed. We know that any  
$G$-vertex  in a  $\hat Q$-box diagram is employed as  the substitute 
for the  ladder series of Fig.~\ref{diagramGT} and therefore  
the corresponding starting energy depends on the diagram as a whole 
and in particular on  the location of the $G$-vertex in the diagram.
For instance, the $G$ matrices  in the three vertices of  diagrams A and B 
of Fig. \ref{diagramGT2} have to be calcutated at 
$\omega= \epsilon_{k}+\epsilon_{l}$, corresponding to the energy
of the two-particle incoming state.
As an other example, 
we have shown in Fig. \ref{fig:vbyg} the diagram A. 
To illustrate how $\omega$ is evaluated we have written it as
the sum of the ladder sequence A1, A2,...,A$n$,....
The lowest vertex of diagram A corresponds to
$\langle\phi_{mp}|G(\omega)| \phi_{kh}\rangle$ and its starting energy
can be determined by looking at the two lowest vertices in diagram A2
whose contribution may be written as
\begin{equation}
\frac{\langle \phi_{mp} | V | \phi_{xy} \rangle \langle \phi_{xy} | V |
\phi_{kh}\rangle}
{ \epsilon_{k} + \epsilon_{h} - \epsilon_{x} - \epsilon_{y}} ~~.
\label{starten}
\end{equation}

Note that the energy denominator of the $G$ matrix is given by
$\omega - H_0$, $H_0$ acting on the intermediate two-particle state
$|\phi_{xy}\rangle$. 
Therefore, to obtain the denominator of
Eq. (\ref{starten}) as $\omega -\epsilon_{x} - \epsilon_{y}$,
one has to choose $\omega= \epsilon_{k} + \epsilon_{h}$.
Similarly, for the  upper $G$-vertex of diagram A $\omega=
\epsilon_{k} +\epsilon_{l}+\epsilon_{h} - \epsilon_{m}$. 
In both the previous examples $\omega$ is different from the energy of the incoming state, which results to be $\epsilon_{k}+\epsilon_{l}$.

In concluding this subsection it is important to consider the dependence
of the $G$ matrix on the SP potential $U$, which explicitly appears in
the denominator through $H_{0}$.
The advantages of different choices have been
extensively investigated essentially in connection with the Goldstone
expansion and its convergence properties (see  the review paper by 
Baranger \cite{Baranger69}).

\begin{figure}[hbtp]
\begin{center}
\includegraphics[scale=0.65,angle=-90]{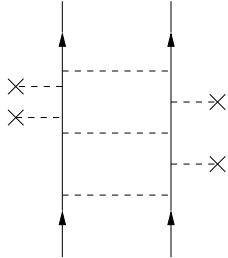}
\end{center}
\caption{Typical $U$ insertions to the $G$-matrix intermediate states.
\label{fig:defgt}}
\end{figure}

It is worth recalling  that, as a  general rule, in a
perturbation approach to nuclear structure calculations one starts by summing
and subtracting the SP potential $U$ to the Hamiltonian so as to have
as perturbation $H_{1}= V-U$.  
As a consequence, the diagrams of the $\hat Q$-box for the shell-model 
effective interaction contain $-U$ as well as $V$ vertices.
Diagrams where $-U$ vertices  are  attached to particle lines between 
successive $V$ interactions may be taken into account directly through the
reaction matrix. 
This means that each diagram of the expansion shown in
Fig. \ref{diagramGT} is replaced with a new diagram in which any
number of $-U$ insertions has been introduced in the particle lines. A typical diagram
is shown in Fig. \ref{fig:defgt}.  
In this way, a new reaction matrix may be introduced where the $-U$
interactions for the intermediate particle  states are summed up to
all orders. More precisely, by changing the 
propagator of the particle lines  from $H_{0}$ to
$H_{0}-Q_{2p}U Q_{2p}$ one obtains the reaction matrix with
plane waves as intermediate states, which is usually called  
$G_{T}$. 
Note that the term $Q_{2p}UQ_{2p}$ instead of $U$ is subtracted from $H_{0}$
since, as pointed out  in \cite{Baranger69}, particle and hole states
are defined with respect to the SP potential $U$.
In other words, the Hamiltonian $H_{0}-Q_{2p}UQ_{2p}$, which commutes
with $Q_{2p}$,  preserves the particle and hole states as defined by
$U$. 
The reaction matrix $G_{T}$ satisfies the  integral equation  
\begin{equation}
\label{eq:defgt}
G_{T}(\omega)=V+VQ_{2p}\frac{1}{\omega - Q_{2p}tQ_{2p}} Q_{2p} G_{T}(\omega),
\end{equation}

\noindent
where one has to recall that $t$ represents the kinetic energy. 

An extensive study of the properties and features of $G_T$ in
the framework of the shell model can be found in \cite{Krenciglowa76},
where arguments in favor of the use of $G_T$ are given. 
We find it interesting to briefly recall these arguments here.

First of all, the  $G_T$ matrix of Eq. (\ref{eq:defgt}) minimizes the
dependence on the SP potential. 
In fact, it depends only on $Q_{2p}$ while the reaction matrix $G$ of 
Eq. (\ref{eq:defg}) depends also on the spectrum of the SP states of 
the $Q_{2p}$ space.  

The second argument is essentially based on the paper by
Bethe {\em et al.} \cite{Bethe63}, where it was  shown that all $-U$ and
Brueckner-Hartree-Fock self-energy insertions cancel each other when the
self-energy is  calculated on energy shell.
This result may be plausibly extended \cite{Krenciglowa76} to low-energy
intermediate particle states not far off the energy shell, but such a 
cancellation  is very unlikely for the high-energy states.
On the other hand, 
in nuclear matter calculations \cite{Bethe71} it has been
shown that the sum of self-energy-insertion diagrams and other
three-body clusters is almost negligible, in other words they tend to
cancel each other.
If we assume that this nuclear-matter result still holds for finite
nuclei, 
we reasonably expect that only the $-U$ insertions survive for
high-energy intermediate states, so 
leading to the reaction matrix $G_T$. The use 
of the $G$ matrix of Eq. (\ref{eq:defg}) is instead suggested for low-lying 
states assuming  that  $-U$ and Brueckner-Hartree-Fock self-energy
insertions cancel approximately.
In other words the reaction matrix calculation may be performed 
following a two-step procedure, which is discussed in detail in
\cite{Krenciglowa76}. 
Here we only give  a sketch of such procedure.

Let us start by dividing  the intermediate states of the $Q_{2p}$ space 
into high-energy and low-energy states, the $Q_{2p}$ operator being
written as 
\begin{equation}
\label{eq:dpq}
Q_{2p}=Q_{2p}^{h} +Q_{2p}^{l},
\end{equation}
with orthogonalized plane waves used for $Q_{2p}^{h}$ and harmonic
oscillator wave functions for  $Q_{2p}^{l}$. 
The fist step consists in the calculation of the $G_{T}$ matrix of
Eq. (\ref{eq:defgt}) with the projection operator taken as
$Q_{2p}^{h}$. 
Then using this reaction matrix instead of the $NN$ potential the $G$
matrix of Eq. (\ref{eq:defg}) is calculated with a projection operator 
restricted to $Q_{2p}^{l}$. 
Actually, the second step is carried out by means of a pertubative
expansion which implies that the final reaction matrix, which we call  
$\tilde{G}_{T}$, is written as
\begin{equation}
\label{eq;newgtp}
\tilde{G}_{T}(\omega) = G_{T}(\omega) + G_{T} (\omega) \frac{Q_{2p}^{l}} 
{\omega- H_{0}} G_{T}(\omega)  
+G_{T}(\omega) \frac{Q_{2p}^{l}}{\omega- H_{0}} G_{T}(\omega) 
\frac{Q_{2p}^{l}}{\omega- H_{0}} G_{T}(\omega) + \cdots 
\end{equation}
Hence $\tilde{G}_{T}$ is obtained by summing to $G_{T}$  the second -, 
third-, $\cdots$ order  two-particle ladder diagrams  with $G_{T}(\omega)$ 
vertices and intermediate states restricted to the low-lying states
defined 
by the $Q_{2p}^{l}$ operator.
These ladder terms may be directly included in the evaluation of the 
$\hat Q$-box once the $G_{T}$ of Eq. (\ref{eq:defgt}) with $Q_{2p}^{h}$ as
projection operator has been calculated.
Note that in so doing 
no $-U$ or Brueckner-self-energy insertions are explicitly
included  in the $\hat Q$-box diagrams since, on the basis of the previous 
arguments, one may assume that they cancel each other.

It is worth to point out that no firm criteria exist on how to perform the 
partition of Eq. (\ref{eq:dpq}).
It is customary to take $n_{2}$ in the definition of the 
$Q_{2p}^{h}$ operator so that it encompasses the  shell above the model
space \cite{Krenciglowa76}.
For example, if we consider again $^{18}$O we specify 
$Q_{2p}$ by $(3,10,\infty)$ to define the $G_{T}$ matrix, then 
$Q_{2p}^l$ while corresponds to the $1p0f$-shell and the ladder terms
with two particle in this shell are taken into account
directly when the $\hat Q$-box is evaluated. 

\subsubsection{Calculation of the reaction matrix} \label{sec:GM3}

Starting from the introduction of the $G$ matrix, several approximate
techniques have been developed for the solution of Eq. (\ref{eq:defg}) 
A review of them
together with a critical discussion can be found in 
\cite{Day67,Baranger69,Hjorth95,Barrett99}.
Among the early methods we would just like to mention the separation
method of Moszkowski and Scott \cite{Moszkowski60} and the reference 
spectrum method of Bethe {\em et al.}
\cite{Bethe63}, which have both been extensively applied to both
nuclear matter and finite nuclei. For many years,
they were the primary methods for $G$-matrix calculation and were first
employed by Kuo and Brown
\cite{Kuo66} to calculate from the  Hamada-Johnston potential 
\cite{Hamada62} the $G$ matrix elements to be used in shell-model
calculations for $^{18}$O and $^{18}$F.
It is interesting to note that  during the 1970s significant advances
were made in the calculation of the reaction matrix as regards
the $G$ matrix of Eq. (\ref{eq:defg}) with harmonic oscillator
intermediate states as well as the $G_T$ matrix of
Eq. (\ref{eq:defgt}) having plane-wave intermediate states. 
In fact, accurate solutions of Eq. (\ref{eq:defg}) were proposed  by 
Barrett {\em et al. }\cite{Barrett71} and of  Eq. (\ref{eq:defgt}) by
Tsai and Kuo \cite{Tsai72}, both of them providing an exact treatment of the
projection operator
$Q_{2p}$.  

Here we shall only focus on the technique to compute the reaction
matrix of Eq. (\ref{eq:defgt}), which we have used in our shell-model calculations.
More precisely, we shall describe the method proposed by Tsai and Kuo
\cite{Tsai72},
which was further investigated and applied by Krengiglowa {\em et al.}
\cite{Krenciglowa76}.
This method, based on a matrix inversion in the momentum space,
allows, as 
mentioned above, an exact treatment of the projection $Q_{2p}$ operator 
except for the only approximation  of using a finite value instead of 
infinity for $n_{\rm 3}$ defining $Q_{2p}$. 

Let us start by noting that a solution of Eq. (\ref{eq:defgt}) is
\begin{equation}
G_T (\omega) = V + V Q_{2p} \frac{1}{Q_{2p}(e - V)Q_{2p}}Q_{2p} V,
\label{eq:defgtv}
\end{equation}

\noindent
as it can be easily shown using the operator identity
\begin{equation}
\frac{1}{A-B}=\frac{1}{A}+\frac{1}{A} B \frac{1}{A-B}~,
\label{eq:opid}
\end{equation}

\noindent
with $e=\omega-t$, $A=Q_{2p}~e~Q_{2p}$, and $B= Q_{2p}V Q_{2p}$.

The propagator of  Eq. (\ref{eq:defgtv}) was shown \cite{Tsai72} to
take the form
\begin{equation}
Q_{2p}\frac{1}{Q_{2p}(e-V)Q_{2p}}Q_{2p} = \frac{1}{e-V} - \frac{1}{e-V}
P_{2p} \frac{1}{P_{2p} \frac{1}{e-V} P_{2p}} P_{2p} \left. \frac{1}{e-V} 
\right. ,
\label{tkidentity}
\end{equation}

\noindent
where $P_{2p}$ is defined in Eq. (\ref{eq:defp}) . 
The identity (\ref{tkidentity})  allows us to rewrite
Eq. (\ref{eq:defgtv}) as 
\begin{equation}
G_T (\omega) = G_{TF} (\omega) + \Delta G (\omega) \label{GTeq},
\end{equation}

\noindent
with
\begin{equation}
\Delta G (\omega)  =  - V \frac{1}{e-V} P_{2p} \frac{1}{P_{2p} \frac{1}{e-V} 
 P_{2p}} P_{2p} \frac{1}{e-V} V
\end{equation}

\noindent
and
\begin{equation}
G_{TF} (\omega) =V+ V \frac{1}{e-V} V=V + V \frac{1}{e}G_{TF}(\omega),
\label{bruecknerGTF}
\end{equation}

\noindent
where we have used the identity (\ref{eq:opid}) with $Q_{2p}=1$ to
obtain relation (\ref{bruecknerGTF}).

An equivalent and very convenient expression of $\Delta G (\omega)$
which has also the advantage to be applicable when $V$ is a
hard-core potential is
\begin{equation}
\Delta G (\omega) = - G_{TF}(\omega) \frac{1}{e} P_{2p} 
\frac{1}{P_{2p} [ (1/e) + (1/e) G_{TF} (1/e) ] P_{2p}}
P_{2p} \frac{1}{e} G_{TF}(\omega).
\label{deltaG}
\end{equation}

The set of Eqs. (\ref{GTeq}, \ref{bruecknerGTF}, \ref{deltaG})
provides a very convenient way to calculate the $G_T$ matrix once a 
finite-$n_3$ approximation is made, namely a truncated $P_{2p}$-space
is taken. 
Let us define 
\begin{eqnarray}
G_R(\omega) & = & G_{TF} (1/e) ,\nonumber \\
G_L(\omega) & = & (1/e) G_{TF} ,\nonumber \\
G_{LR}(\omega) & = & (1/e) + (1/e) G_{TF} (1/e) ,
\end{eqnarray}

\noindent
then Eq. (\ref{GTeq}) in a matrix formalism can be written as
\begin{equation}
\langle \phi_{kl}| G_T (\omega) | \phi_{lm}\rangle = \langle
\phi_{kl}| G_{TF} (\omega) | \phi_{lm}\rangle - \sum_{ijkl \in P_{2p}}
\langle \phi_{kl} | G_L (\omega) | \phi_{ij} \rangle 
 \times \langle \phi_{ij}| G^{-1}_{LR} (\omega) | \phi_{kl}
\rangle \langle k~l| G_R (\omega) | \phi_{lm}\rangle~. \label{meGT}
\end{equation}

This equation makes clear the necessity of a finite-$n_3$
approximation. 
In fact, the $G^{-1}_{LR}$ matrix  appearing in this equation is
defined in the $P_{2p}$ space and represents the inverse of the
$P_{2p}G_{LR} P_{2p}$ matrix, which has to be a finite matrix.

The basic ingredient for the calculation of the $G_T$ matrix through
Eq. (\ref{GTeq}) is the matrix $G_{TF}$ which  is identical to the
$G_T$ defined by Eq. (\ref{eq:defgt}), except that is free from the
projection operator $Q_{2p}$.  
A very convenient way to calculate $G_{TF}(\omega)$ is to employ 
the momentum-space matrix inversion method \cite{Brown69}. 
Once the $G_{TF}$ matrix elements in the momentum space are known, the
matrices $G_{L}$, $G_{R}$, and $G_{LR}$ can be easily evaluated in the
same space.
As a final step, these matrices are all transformed in the harmonic
oscillator basis and the $G_T$ matrix is obtained by means of
Eq. (\ref{meGT}) though simple matrix operations.

Here we have only given a sketch of the procedure to solve
Eqs. (\ref{GTeq}), (\ref{bruecknerGTF}), (\ref{deltaG}).
A detailed description of the whole procedure  can be found   in
\cite{Krenciglowa76}, where the validity of an $n_3$-finite approximation
is discussed. 
It is worth mentioning that this point is also discussed in
Ref.~\cite{Hjorth95}. 

\subsection{The $V_{\rm low-k}$ approach} \label{sec:vlowk}

In this subsection we shall describe in detail the derivation and
main features  of the  low-momentum $NN$ interaction $V_{\rm low-k}$,
which has been introduced in
Refs.~\cite{Bogner01,Bogner02,Bogner03}. 
The idea underlying the construction of $V_{\rm low-k}$ from a
realistic model for $V_{NN}$ is based on the renormalization group
(RG) and the effective field theory (EFT) approaches
\cite{Lepage97,Bedaque02,vanKolck99,Beane01}. 
As mentioned in Sec.~\ref{sec:NN}, while the high-quality $NN$ potentials  all
reproduce the empirical deuteron properties and low-energy phase
shifts very accurately, they differ significantly from one another in
the high momentum behavior. 
A cutoff momentum $\Lambda$ that separates fast and slow modes is
then introduced and from the original $V_{NN}$ an effective potential 
$V_{\rm low-k}$, satisfying a decoupling condition between the low-
and high-momentum spaces, is derived by integrating out the high-momentum
components. 
The main result is that $V_{\rm low-k}$ is free from the
model-dependent high-momentum modes.
In consequence, it is a smooth potential which preserves exactly the
onshell properties of the original $V_{NN}$, and is suitable to be
used directly in nuclear structure calculations.
In Sec.~\ref{sec:res1}, we will show that the $V_{\rm low-k}$ approach provides 
a real alternative to the Brueckner $G$-matrix renormalization procedure.

\subsubsection{Derivation of the low-momentum $NN$ potential  1
$V_{\rm low-k}$} \label{sec:vlowk1}

In carrying out the above high-momentum integration, or decimation, an 
important requirement is that the low-energy physics of $V_{NN}$ is
exactly preserved by $V_{\rm low-k}$. 
For the two-nucleon problem, there is one bound state, namely the
deuteron. 
Thus one must require that the deuteron properties given by $V_{NN}$
are preserved by $V_{\rm low-k}$.
In the nuclear effective interaction theory, there are several well-developed
model-space reduction methods. 
One of them is the Kuo-Lee-Ratcliff (KLR) folded diagram method 
\cite{Kuo71,Kuo90}, which was originally formulated for discrete
(bound state) problems. 
A detailed discussion of this method has been given earlier in Sec.~\ref{sec:SMEI}. 

For the two-nucleon problem, we want the effective interaction 
$V_{\rm low-k}$ to preserve also the low-energy scattering phase
shifts, in addition  to the deuteron binding energy. 
Thus we need an effective interaction for scattering (unbound)
problems. 
A convenient framework for this purpose is the $T$-matrix equivalence 
approach, as described below.

We use a  continuum model space specified by 
\begin{equation}
P=\int d \mbox{\boldmath $p$} \mid \mbox{\boldmath $p$}  \rangle \langle 
\mbox{\boldmath $p$} \mid ~,~~~~p \leq \Lambda 
\end{equation} 

\noindent
where $\mbox{\boldmath $p$}$ is the two-nucleon relative momentum and $\Lambda$  the cutoff
momentum which is also known as the decimation momentum. 
Its typical value is about 2 fm$^{-1}$ as we shall discuss later. 
Our purpose is to look for an effective interaction $PV_{\rm eff}P$,
with $P$ defined above, which preserves certain properties of the
full-space interaction $V_{NN}$ for both bound and unbound states. 
This effective interaction is referred to as $V_{\rm low-k}$.

We start from the full-space half-on-shell $T$-matrix
(written in a single partial wave channel)
\begin{equation}
T(k',k,k^2) =  V_{NN}(k',k) + \mathcal{P} \int _0 ^{\infty} q^2 dq  
V_{NN}(k',q)
  \frac{1}{k^2-q^2  } T(q,k,k^2 ). \label{hostmat}
\end{equation}
This is the $T$-matrix for the two-nucleon problem with the 
Hamiltonian
\begin{equation}
 H=H_0+V_{NN}= t+V_{NN}~,
\label{H2N}  
\end{equation} 
\noindent
$t$ being the relative kinetic energy.
We then define a $P$-space low-momentum $T$-matrix by 
\begin{equation}
T_{\rm low-k }(p',p,p^2) =  V_{\rm low-k }(p',p) + \mathcal{P} \int _0
^{\Lambda} q^2 dq  V_{\rm low-k }(p',q) 
 \frac{1}{p^2-q^2} T_{\rm low-k} (q,p,p^2), 
\label{hostmateff} 
\end{equation}

\noindent
where $(p',p) \leq \Lambda$ and the integration interval is from 0 to 
$\Lambda$. 
In the above, the symbol $\mathcal{P}$ in front of the integration sign
denotes principal value integration.
We require the equivalence condition
\begin{equation}
 T(p',p,p^2 ) = T_{\rm low-k }(p',p,p^2) ;~( p',p) \leq \Lambda . 
\label{constraintt} 
\end{equation}
The above equations define  the effective low-momentum interaction; it is 
required to preserve the low-momentum ($\leq \Lambda$) half-on-shell 
$T$-matrix. 
Since phase shifts are given by the full-on-shell $T$-matrix 
$T(p,p,p^2)$, low-energy phase shifts given by the above $V_{\rm 
low-k}$ are clearly the same as those of $V_{NN}$.

In the following, let us show that a solution of the above equations
may be found by way of the KLR folded-diagram method 
\cite{Kuo71,Kuo90} described  in Sec.~\ref{sec:SMEI} as a means to construct 
the shell-model effective interaction.
Within  this approach the $V_{\rm low-k}$ may be written in the operator form as
\begin{equation}
V_{\rm low-k} = \hat{Q} - \hat{Q'} \int \hat{Q} + \hat{Q'} \int \hat{Q} \int 
\hat{Q} - \hat{Q'} \int \hat{Q} \int \hat{Q} \int \hat{Q} + ~...~, 
\label{vlowkeq}
\end{equation}
\noindent
which contains a $\hat{Q}$-box whose explicit definition is given below.  

In the time-dependent formulation, the $T$-matrix of Eq. (\ref{hostmat})
can be written as $\langle k' \mid VU(0,-\infty ) \mid k \rangle$, 
$U(0,-\infty)$ being the time-evolution operator. 
In this way a diagrammatic analysis of the $T$-matrix can be made.  
A general term of $T$ may be written as  
\begin{equation}
\langle k' \mid \left[ V +V\frac{1}{e(k^2)}V +
V\frac{1}{e(k^2)}V\frac{1}{e(k^2)} V +...\right] \mid k \rangle ,
\end{equation}

\noindent
where $e(k^2)\equiv (k^2- H_0)$.

Note that the intermediate states (represented by 1 in the numerator)
cover the entire space.
In other words, we have $1=P+Q$ where $P$ denotes the model-space
projection operator and $Q$ its complement. 
Expanding it out in terms of $P$ and $Q$, a typical term of $T$ 
is of the form 
\begin{equation}
V\frac{Q}{e} V\frac{Q}{e}V\frac{P}{e}V\frac{Q}{e}V \frac{P}{e}V . 
\end{equation}

Let us now define the $\hat Q$-box as 
\begin{equation}
\langle k' \mid \hat Q(k^2)\mid k \rangle =  \langle k' \mid \left[
V + V\frac{Q}{e(k^2 )}V \right. 
 +\left. V\frac{Q}{e(k^2)} V\frac{Q}{e(k^2 ) }V +... \right] \mid k \rangle,
\end{equation}

\noindent
where all intermediate states belong to $Q$.
 One readily sees that the $P$-space portion of the $T$-matrix can be 
regrouped in terms of a 
$\hat Q$-box series, more explicitly as 
\begin{equation}
\langle p' \mid T \mid p \rangle =  \langle p' \mid \left[
\hat Q+\hat Q \frac{P}{e(p^2)} \hat Q \right. 
 \left. + \hat Q \frac{P}{e(p^2)} \hat Q
\frac{P}{e(p^2)} \hat Q +... \right]\mid p \rangle . 
\end{equation}

\begin{figure}[hbtp]
\begin{center}
\includegraphics[scale=0.90,angle=0]{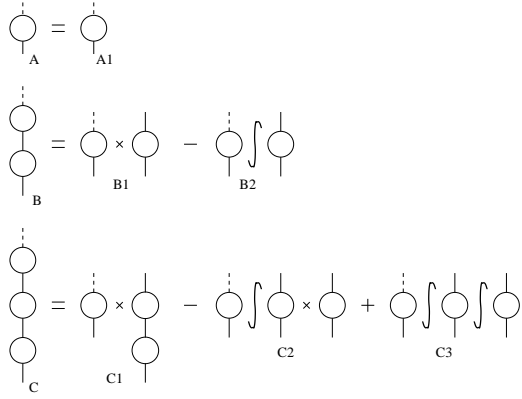}
\end{center}
\caption{Folded-diagram factorization of the half-on-shell $T$-matrix.
\label{vlkprima}}
\end{figure}

Note that all the $\hat Q$-boxes have the same energy variable, namely $p^2$. 
This regrouping is shown in Fig. \ref{vlkprima}, where each $\hat{Q}$-box
is denoted by a circle and the solid line represents the propagator 
$\frac{P}{e}$. 
The diagrams A, B and C are respectively the one- and two- and 
three-$\hat Q$-box terms of $T$, and clearly $T$=A+B+C+$\cdots$.
Note that the dashed vertical line is not a propagator; it is just a 
``ghost'' line to indicate the external indices.

Now a folded-diagram factorization for the $T$-matrix can be performed,
following the same procedure as in Sec.~\ref{sec:SMEI} \cite{Kuo71,Kuo90}. 
As is discussed in Sec.~\ref{sec:SMEI21}, diagram B of Fig. \ref{vlkprima} 
may be
 factorized into the product of two
parts (see B1) where the time integrations of the two parts are
independent from each other, each integrating from $-\infty$ to 0. 
In this way we have introduced a time-incorrect contribution which
must be corrected. 
In other words, B is not equal to B1, rather it is equal to B1 minus the 
folded-diagram correction B2.
Schematically the above factorization is written as
\begin{equation}
\hat Q \frac{P}{e} \hat Q=
\hat Q\times   \hat Q- \hat Q' \int \hat Q,
\end{equation}

\noindent
where the last term is the once-folded two $\hat Q$-box term.
The meaning of the correction term B2 can also be visualized in terms of the
differences in the $\hat{Q}$-box energy variables. 
With such energies explicitly written out, the above equation represents
\begin{equation}
\langle p' \mid \hat Q(p^2 ) \frac{P}{e(p^2 )}
 \hat Q(p^2) \mid p \rangle
 = \sum _{p''} \langle p' \mid \hat Q(p''^2 )
\mid p'' \rangle ~~
\langle p'' \mid \frac{P}{e(p^2 )} \hat Q(p^2)\mid p \rangle - 
\langle p' \mid \hat Q' \int \hat Q \mid p \rangle.
\end{equation} 

Note that the energy variable for the first $\hat Q$-box on the right
hand side of this equation is $p''^2$ instead of $p^2$.
Before factorization, the energy variable for this $\hat Q$-box
is $p^2$.
We recall that the integral sign represents a generalized folding
operation and  the leading $\hat
Q$-box of any folded term must be at least of second order in $V_{NN}$.

In the same way, we factorize the three-$\hat Q$-box term C as shown in the
third line of Fig. \ref{vlkprima}:
\begin{equation}
\hat Q \frac{P}{e} \hat Q \frac{P}{e} \hat Q  
=\hat Q\times   \hat Q\frac{P}{e}\hat Q - [\hat Q' \int \hat Q] \times \hat Q
+ \hat Q' \int \hat Q \int  \hat Q,
\end{equation}

Higher-order $\hat Q$-box terms are also factorized following the same
folded-diagram procedure.
Let us now collect the terms in the figure in a ``slanted'' way. 
The sum of terms A1, B2, C3... is just the low-momentum effective
interaction of Eq.~(\ref{vlowkeq}). 
The sum B1, C2, D3.... is $V_{\rm low-k}\frac{P}{e}\hat Q$. 
Similarly the sum C1+D2+E3+$\cdots$ is just 
$V_{\rm low-k}\frac{P}{e}\hat Q\frac{P}{e}\hat Q$. 
It is worth to mention that diagrams D1, D2, $\cdots$, E1, E2,
$\cdots$ are not shown in Fig. \ref{vlkprima}.
Continuing this procedure, it can be seen that 
the $P$-space portion of $T$-matrix is 
$V_{\rm low-k} + V_{\rm low-k} \frac{P}{e} T$. Namely, the
$V_{\rm low-k}$ defined by Eq. (\ref{vlowkeq}) is a solution of 
Eqs. (\ref{hostmat})-(\ref{constraintt}). 

An advantageous method to calculate the series in Eq. 
(\ref{vlowkeq}) is the Lee-Suzuki one, as described in Sec.~\ref{sec:SMEI3}.
Following this algebraic approach the  $V_{\rm low-k}$ is expressed as  
\begin{equation}
V_{\rm low-k} = P V_{NN} P + P V_{NN} Q \omega ,
\label{vlowkLS}
\end{equation}

\noindent
where the operator $\omega$ is given by the solution of the decoupling equation
(\ref{deceq2}). Note that in this latter equation the letter $\Omega$ was used 
instead of $\omega$ to avoid confusion with the energy variable $\omega$ 
defined in the same section. Here, since there is no problem of ambiguity, 
the same letter of the original paper by Lee and Suzuki \cite{Lee80} has been used and, 
for the sake of clarity, Eq.~(\ref{deceq2}) is 
re-written for the Hamiltonian (\ref{H2N})

\begin{equation}
Q V_{NN} P + Q H Q \omega - \omega P H P - \omega P V_{NN} Q \omega = 0
~~. \label{deceq3} 
\end{equation}

The iterative technique presented in Sec.~\ref{sec:SMEI3} to  solve the decoupling 
equation is formulated   for degenerate model spaces. Therefore, it is  not 
useful  in this case since $PH_0P=PtP$ is obviously
non-degenerate. 
We can then resort to the iterative technique for non-degenerate model
spaces proposed by Andreozzi \cite{Andreozzi96} to solve the decoupling equation
[see Eq.~(\ref{deceq3})] of the Lee-Suzuki method. This procedure, which  
will be referred to the Andreozzi-Lee-Suzuki (ALS)
procedure, is quite convenient for obtaining the low-momentum $NN$ interaction.

We give now a sketch of this procedure. 
Let us define the operators:

\begin{eqnarray}
p(\omega) & = & PHP + PHQ \omega ,\\
q(\omega) & = & QHQ - \omega PHQ ,
\end{eqnarray}
\noindent
in terms of which
the basic equations of the ALS  iterative  procedure read

\begin{eqnarray}
x_0  & = & - (QHQ)^{-1} QHP , \nonumber \\
x_1 & = & q (x_0)^{-1} x_0 p (x_0) , \nonumber \\
~& ~.~.~.&~ \nonumber \\
x_n & = & q ( x_0+x_1+...+x_{n-1})^{-1} x_{n-1} p ( x_0+x_1+...+x_{n-1}).
\end{eqnarray}

Once the iterative procedure has converged, $x_{n} \rightarrow 0$, 
\cite{Andreozzi96},
the operator $\omega$ is given by

\begin{equation}
\omega_n= \sum^n_{i=0} x_i ,
\end{equation}

\noindent
and the 
$V_{\rm low-k}$ is obtained by Eq.~(\ref{vlowkLS}).

In applying the ALS method, we have employed a momentum-space
discretization procedure by introducing an adequate number of Gaussian  mesh
points \cite{Krenciglowa76}. 

\subsubsection{Phase-shift equivalent Hermitian $V_{\rm low-k}$'s}
\label{sec:vlowk2}

It should be pointed out that the low-momentum $NN$ interaction given
by the above folded-diagram expansion is not
Hermitian. 
This is not convenient for applications, and one would like to have an 
interaction which is Hermitian.
We shall describe how a family of Hermitian $V_{\rm low-k}$'s can be
derived if we relax the half-on-shell $T$-matrix preservation.
This subject has been studied in \cite{Holt04a}, where a
general framework for constructing hermitian effective interactions
is introduced, and it is shown how this generates a family of
hermitian phase-shift equivalent low-momentum $NN$ interactions.

In the following, we outline the essential steps for the derivation 
of a Hermitian $V_{\rm low-k}$.
We denote the folded-diagram  low-momentum $NN$ interaction obtained
from the ALS procedure as $V_{LS}$.
The model-space secular equation for $V_{LS}$ is 
\begin{equation}
P(H_0+V_{LS})P | \chi _m \rangle =E_m | \chi _m \rangle, 
\label{lsseculeq}
\end{equation}

\noindent
where $\{E_m\}$ is a subset of the eigenvalues $\{E_n\}$ of the
full-space Schr{\"o}dinger equation, $(H_0+V) |\Psi _n \rangle =E_n |
\Psi _n \rangle$, and $| \chi _m \rangle$ is the $P$-space projection
of the full-space wave function $| \Psi _m \rangle$, namely $ |\chi _m
\rangle =P |\Psi _m \rangle $. 
The above effective interaction may be rewritten in terms
of the wave operator $\omega$ [see Eqs.(\ref{omegaop}),(\ref{omegapro1}),
(\ref{omegapro2})]
\begin{equation}
PV_{LS}P =Pe^{-\omega}(H_0 +V)e^{\omega}P-PH_0P.
\end{equation}

While the full-space eigenvectors $|\Psi _n \rangle$ are orthogonal to
each other, the model-space eigenvectors $|\chi _m \rangle$'s  are
clearly not so and consequently the effective interaction $V_{LS}$ is
not Hermitian. 
We now make a $Z$ transformation such that 
\begin{eqnarray}
Z | \chi_m \rangle &=& |v _m \rangle; \nonumber \\
\langle v_m \mid v_{m'} \rangle &=& \delta _{mm'};~~m,m'=1,~d, 
\label{ortho}
\end{eqnarray}

\noindent
where $d$ is the dimension of the model space. 
This transformation reorients the vectors $|\chi_m \rangle$'s such that
they become orthonormal.
We assume that the $|\chi _m \rangle$'s  are linearly independent
so that $Z^{-1}$ exists, otherwise the above transformation is not
possible. 
Since $|v_m \rangle$ and $Z$ exist entirely within the model space, we
can write $|v_m \rangle=P |v_m \rangle$ and $Z=PZP$. 

Using Eq. (\ref{ortho}), we transform Eq. (\ref{lsseculeq}) into
\begin{equation}
Z(H_0+V_{LS})Z^{-1} |v_m \rangle =E_m | v_m \rangle,
\end{equation}

\noindent
which implies
\begin{equation}
Z(H_0+V_{LS})Z^{-1}=\sum _{m\epsilon P} E_m \mid v_m \rangle 
\langle v_m \mid .
\end{equation}
Since $E_m$ is real (it is an eigenvalue of $(H_0+V)$ which is Hermitian)
and the vectors $|v_m \rangle$ are orthonormal, $Z(H_0+V_{LS})Z^{-1}$
is clearly Hermitian. 
The non-Hermitian secular equation (\ref{lsseculeq}) is now transformed into a 
Hermitian model-space eigenvalue problem
\begin{equation}
P(H_0 +V_{herm})P |v_m \rangle =E_m |v_m \rangle, 
\label{hermseculeq}
\end{equation}

\noindent
with the Hermitian  effective interaction given by 
\begin{equation}
V_{herm}=Z(H_0+V_{LS})Z^{-1}-PH_0P ,
\label{hermseculeq1}
\end{equation}

\noindent
or equivalently
\begin{equation}
V_{herm}=Ze^{-\omega}(H_0+V)e^{\omega}Z^{-1}-PH_0P .
\end{equation}

To calculate $V_{herm}$, we must first have the $Z$ transformation. 
Since there are certainly many ways to construct $Z$, this generates
a family of Hermitian effective interactions, all originating from $V_{LS}$.
For example, we can construct $Z$ using the familiar Schmidt 
orthogonalization procedure.
We denote the hermitian effective interaction obtained using this
  $Z$ transformation as $V_{schm}$.

We can also construct $Z$ analytically in terms of the wave operator.
From the properties of the wave operator $\omega$,  we have
\begin{equation}
\langle \chi _m \mid (1+\omega ^+ \omega )\mid \chi _{m'} \rangle
=\delta _{mm'} .
\end{equation}
It follows that an analytic choice for the $Z$ transformation is
\begin{equation}
Z=P (1+\omega ^+ \omega)^{1/2}P.
\end{equation}
This leads to the Hermitian effective interaction
\begin{equation}
V_{\rm okb}  =  P (1+\omega ^+ \omega)^{1/2}P(H_0+V_{LS})P
 (1+\omega ^+ \omega)^{-1/2}P - PH_0P,
\end{equation}

\noindent
or equivalently
\begin{equation}
V_{\rm okb} = P (1+\omega ^+ \omega)^{-1/2}(1+\omega^+)(H_0+V)
(1+\omega) 
  (1+\omega ^+ \omega)^{-1/2}P-PH_0P .
\end{equation}
This is just the Okubo Hermitian effective interaction \cite{Okubo54}.
The Hermitian effective interaction  of Suzuki and Okamoto
\cite{Suzuki83,Suzuki94} is equivalent to the above.

There is another interesting choice for the transformation $Z$. 
As pointed out in  \cite{Andreozzi96}, the positive definite 
operator $P(1+\omega ^+ \omega)P$ admits the Cholesky decomposition, 
namely
\begin{equation}
P (1+\omega ^+ \omega)P= PL L^TP,
\end{equation}

\noindent
where $L$ is a lower triangular matrix, $L^T$ being its
transpose. 
Since $L$ is real and within the $P$-space, we have 
\begin{equation} 
Z=L^T ,
\end{equation}

\noindent
and the corresponding Hermitian effective interaction from Eq. 
(\ref{hermseculeq1}) is
\begin{equation}
V_{\rm chol}=PL^TP(H_0+V_{LS})P(L^{-1})^TP -PH_0P.
\end{equation}
We have found that this Hermitization procedure is numerically 
more convenient than the other methods. 
In the practical applications presented in this review we
have employed this procedure.
From now on, we shall denote the Hermitian low-momentum $NN$ interaction
as $V_{\rm low-k}$.

Using a solvable matrix model, it has been shown that the above
Hermitian effective interactions $V_{\rm schm},V_{\rm okb}$, and 
$V_{\rm chol}$ can be in general quite different \cite{Holt04a} from
each other and from $V_{LS}$, especially when $V_{LS}$ is largely 
non-Hermitian.
For the $V_{NN}$ case, it is fortunate that the $V_{\rm low-k}$
corresponding to $V_{LS}$ is only slightly non-Hermitian.
As a result, the Hermitian low-momentum $NN$ interactions $V_{\rm
schm}, V_{\rm okb}$, and $V_{\rm chol}$ are all quite similar to each other
and to the one corresponding to $V_{LS}$, as discussed in \cite{Holt04a}.

We have just shown that a family of Hermitian effective interactions 
$V_{\rm low-k}$'s can be generated starting from the Lee-Suzuki 
transformation. 
In the following, we shall prove that all such interactions preserve
the full-on-shell $T$-matrix $T(p,p,p^2)$, $p\leq \Lambda$, and
consequently they are all phase-shift equivalent

Let us consider two $T$-matrices
\begin{equation}
 T_1(p^2)=V_1+V_1g(p^2)T_1(p),
\end{equation}

\noindent
and
\begin{equation}
T_2(p'^2)=V_2+V_2g(p'^2)T_2(p'),
\end{equation}

\noindent
which are defined starting from two different potentials, 
$V_1$ and $V_2$, and/or two different propagators, 
$g(p^2)={\mathcal P}(p^{2}_1-H_0)^{-1}$ and 
$g(p'^2)= {\mathcal P}(p'^{2}_2-H_0)^{-1}$. 
These $T$-matrices are related by the well known two-potential formula
\cite{Bethe63} 
\begin{equation}
T_2^{\dagger}(p'^2)
=  T_1(p^2) + T_2^{\dagger}(p'^2)
[g(p'^2) - g(p^2)] T_1(p^2)  
 +  X_2(p'^2)[V_2-V_1] X_1(p^2),
\end{equation}
\noindent
where the operator $X=e^{\omega}$ [see Eq.~(\ref{omegaop})]
is defined so as to have $T = VX$.

Applying the above relation to the half-on-shell $T$-matrix
elements in  momentum space, we have
\begin{equation}
\langle p' | T_2^{\dagger}(p'^2) |p \rangle
= \langle p' | T_1(p^2) |p\rangle + \langle p' | T_2^{\dagger}
(p'^2) [g(p'^2) -  g(p^2) ] T_1(p^2) |p\rangle + 
\langle \psi_2(p') | (V_2-V_1)
|\psi_1 (p) \rangle . \label{phaseshifteq1}
\end{equation}
Here the true and unperturbed wave functions are related by 
$|\psi_1 (p)\rangle =X_1(p^2) |p \rangle $ and similarly for $\psi _2$. 
Using the above relation, we shall now show that the phase shifts of
the full-space interaction $V_{NN}$ are preserved by the Hermitian interaction
$V_{\rm low-k}$  for momentum $\leq \Lambda$.
Let us denote the last term of Eq. (\ref{phaseshifteq1}) as $D(p',p)$.
We use $V_{\rm low-k}$ for $V_2$ and  $V_{NN}$ for $V_1$.
Recall that the eigenfunction of $(H_0+V_{\rm low-k})$ is $|
v_m \rangle$ 
[see Eq. (\ref{lsseculeq})] and that for $H\equiv (H_0+V_{NN})$ is $|
\Psi_m \rangle$. 
We define a wave operator
\begin{equation}
U_P=\sum _{m \in P} |v_m \rangle \langle \Psi_m |.
\end{equation}

\noindent
Then  $|v_m \rangle =U_P |\Psi _m \rangle$  and 
$PV_{\rm low-k}P=U_P(H_0+V_{NN})U_P^{\dagger}-PH_0P$. 
For our present case, $\langle \psi _2(p')|$ is $\langle v_{p'}|$ and 
$|\psi _1(p)\rangle$ is $|\Psi _p\rangle$. 
Since $\langle v_{p'}|U_P =\langle \Psi _{p'} |$, we have
\begin{equation}
D(p',p)=
\langle \Psi_{p'} | (HU_P^{\dagger}-U_P^{\dagger}H) |\Psi _p \rangle
= (p'^2-p^2)\langle v_{p'}|\Psi _p \rangle.
\end{equation}
Clearly $D(p,p)=0$. 
The second term on the right hand side of Eq. (\ref{phaseshifteq1})
vanishes when $p'=p$. 
Hence 
\begin{equation}
\langle p | T_{\rm low-k}(p ^2) |p \rangle
=\langle p | T(p ^2) |p \rangle,~p\leq \Lambda,
\end{equation}

\noindent
where $T_{\rm low-k}$ is the $T$-matrix for
$(PH_0P+V_{low-k})$ and $T$ for $(H_0+V_{NN})$. 
Thus we have shown that the phase shifts of $V_{NN}$ are preserved by
$V_{\rm low-k}$.
Recall that our $T$-matrices are real, because of the principal-value
boundary conditions employed.

\subsubsection{Properties of $V_{\rm low-k}$} \label{sec:vlowk3}

We have shown above that low-energy observables given by the
full-momentum potential $V_{NN}$ are preserved by the low-momentum
potential $V_{\rm low-k}$. 
This preservation is an important point, and it is worth checking
numerically that the deuteron binding energy and low-energy phase
shifts given by $V_{NN}$ are indeed reproduced by $V_{\rm low-k}$.  
In this subsection we would like to report some results in this regard, 
together with some discussion about the differences between the various modern
potentials.

Before presenting the results, let us first address an important question
concerning the choice of the momentum cutoff $\Lambda$ in the
derivation of $V_{\rm low-k}$. 
Phase shifts are given by the full-on-shell $T$-matrix 
$T(p,p,p^2)$, hence, for a chosen $\Lambda$, $V_{\rm low-k}$ can
reproduce phase shifts up to $E_{lab}= 2 \hbar ^2 \Lambda^ 2/M$, $M$
being the nucleon mass. 
Realistic $NN$ potentials are constructed to fit empirical phase
shifts up to $E_{lab}\approx 350$ MeV \cite{Stoks93}, which is the
pion production threshold. 
It is reasonable then to require $V_{\rm low-k}$ to reproduce phase
shifts also up to this energy. 
Thus one should use $\Lambda$ in the vicinity of 2 fm$^{-1}$.

\begin{table} 
\caption{Calculated binding energies $BE_d$ of the deuteron (MeV) for 
different values of the cutoff $\Lambda$ (fm$^{-1}$). 
The $V_{\rm low-k}$ has been calculated from the CD-Bonn potential for 
which $BE_d=2.224575$ MeV \cite{Machleidt01b}.\label{binddeut}}
\begin{center}
\begin{tabular}{lcccccc} \hline
$\Lambda$ & 0.5 & 1.0 & 1.5 & 2.0 & 2.5 & 3.0  \\ \hline
  & 2.2246 & 2.2246 & 2.2246 & 2.2246 & 2.2246 & 2.2246 \\ \hline
\end{tabular}
\end{center}
\end{table}

We first check the deuteron binding energy $BE_d$ given by 
$V_{\rm low-k}$.
For a range of $\Lambda$, such as 0.5 fm$^{-1}\leq \Lambda \leq 3$ fm$^{-1}$,
$BE_d$ given by $V_{\rm low-k}$ agrees very accurately (to 4
places after the decimal point) with that given by $V_{NN}$. 
In Table \ref{binddeut}, we present the results for the 
$V_{\rm low-k}$ derived from the CD-Bonn $NN$ potential \cite{Machleidt01b}.

We have also checked the phase shifts and the half-on-shell 
$T(p',p,p^2)$ matrix elements with $(p',p)\le \Lambda$.

In Table~\ref{phaseshifts} we report the neutron-proton $^1S_0$ phase
shifts calculated both with the CD-Bonn potential and its $V_{\rm
low-k}$ (with a cutoff momentum $\Lambda= 2~{\rm fm}^{-1}$).
It is worth noting that the degree of accuracy shown in Table
\ref{phaseshifts} allows $V_{\rm low-k}$ to reproduce the
$\chi^2/datum$ of the original potential up to 
$E_{\rm lab}= 2 \hbar^2 \Lambda^2/M$.
\begin{table}[hbtp]
\caption{$np$ $^1S_0$ phase shifts (deg) as predicted by the CD-Bonn
potential and its $V_{\rm low-k}$ ($\Lambda=2.0~{\rm fm}^{-1}$).
\label{phaseshifts}}
\begin{center}
\begin{tabular}{lccc} \hline
   $E_{\rm lab}$ & CD-Bonn & $V_{\rm low-k}$ & Expt. \\ \hline 
         1   & 62.1 & 62.1 & 62.1 \\ 
         10  & 60.0 & 60.0 & 60.0 \\ 
         25  & 50.9 & 50.9 & 50.9 \\ 
         50  & 40.5 & 40.5 & 40.5 \\ 
         100 & 26.4 & 26.4 & 26.8 \\ 
         150 & 16.3 & 16.3 & 16.9 \\ 
         200 & 8.3  & 8.3  & 8.9  \\ 
         250 & 1.6  & 1.6  & 2.0  \\ 
         300 & -4.3 & -4.3 & -4.5 \\ \hline
\end{tabular}
\end{center}
\end{table}

In Table \ref{Qvalues}, for sake of completeness, we present also the
deuteron quadrupole moments for different values of the cutoff
momentum $\Lambda$, calculated with the corresponding $V_{\rm
  low-k}$'s, derived from the Argonne $V_{18}$ potential.
These values are compared with the one predicted by the original
potential. 
\begin{table}[hbtp]
\caption{Deuteron quadrupole moments (fm$^2$) as predicted by the Argonne
  $V_{18}$ potential and its $V_{\rm low-k}$ for different values of
  $\Lambda$ (fm$^{-1}$)
\label{Qvalues}}
\begin{center}
\begin{tabular}{cll} \hline
Argonne $V_{18}$ & $\Lambda$ &  $V_{\rm low-k}$ \\ \hline
 0.270 &  1.8 & 0.268 \\
 & 2.0 & 0.269 \\
 & 2.2 & 0.270 \\
 & 2.4 & 0.270 \\
 & 2.6 & 0.270 \\
 & 2.8 & 0.270 \\ \hline
\end{tabular}
\end{center}
\end{table}

\begin{figure}[hbtp]
\begin{center}
\includegraphics[scale=0.6,angle=0]{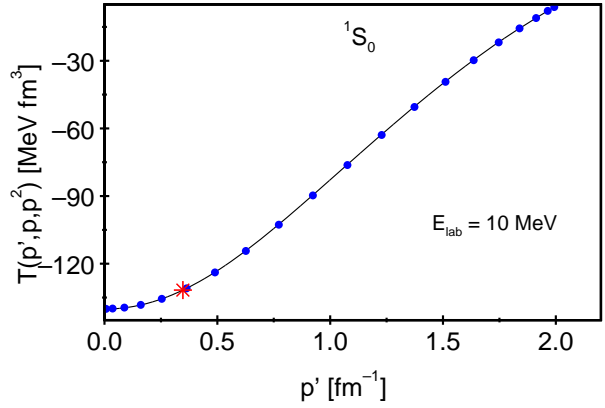}
\end{center}
\caption{(Color online) Comparison of the half-on-shell $T(p',p,p^2)$ matrix elements 
(continuous line) with those of $T_{\rm low-k}(p',p,p^2)$ for 
$p=0.35$ fm$^{-1}$ ($E_{\rm lab}=10$ MeV). 
The asterisk indicates the on-shell matrix element.
The cutoff momentum is $\Lambda=2$ fm$^{-1}$. 
\label{vlkseconda}}
\end{figure}

\begin{figure}[hbtp]
\begin{center}
\includegraphics[scale=0.6,angle=0]{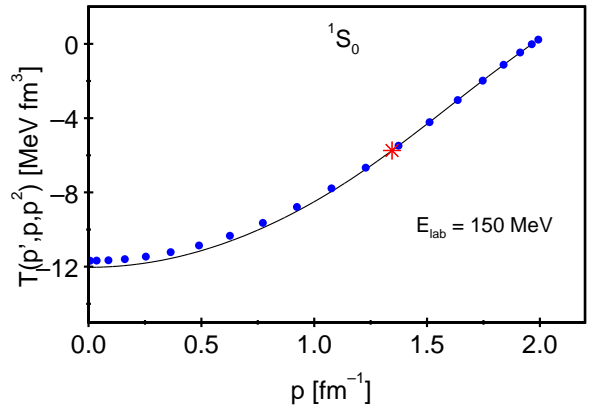}
\end{center}
\caption{(Color online) As in Fig. \ref{vlkseconda}, but for $p=1.34$ fm$^{-1}$ 
($E_{\rm lab}=150$ MeV).
\label{vlkterza}}
\end{figure}

In Figs. \ref{vlkseconda} and \ref{vlkterza} we compare $np$ $^1S_0$  
$T(p',p,p^2)$ matrix elements given by the CD-Bonn potential and its 
$V_{\rm low-k}$ with a cutoff momentum $\Lambda=2.0$ fm$^{-1}$ and
$p=0.35,~1.34$ fm$^{-1}$.
From these figures it can be seen that, even if the half-on-shell
$T$-matrix is not exactly preserved by the ALS transformation, it is 
reproduced to a good degree of accuracy.

\begin{figure}[hbtp]
\begin{center}
\includegraphics[scale=0.6,angle=0]{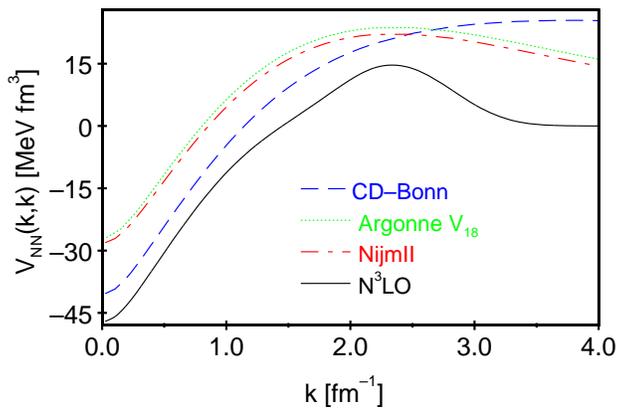}
\end{center}
\caption{(Color online) Matrix elements in the $^1S_0$ channel for different $NN$
potentials. \label{vlkquarta}}
\end{figure}

As mentioned earlier, modern $NN$ potentials all possess a strong
short-range repulsion. 
As a result, the $k$-space matrix elements $\langle k' \mid V_{NN}
\mid k \rangle$ are large for large momenta, as illustrated in
Fig. \ref{vlkquarta}, where we compare the diagonal matrix elements of
the phase-shift equivalent potentials NijmII, Argonne $V_{18}$, CD-Bonn, and
N$^3$LO. 

\begin{figure}[hbtp]
\begin{center}
\includegraphics[scale=0.6,angle=0]{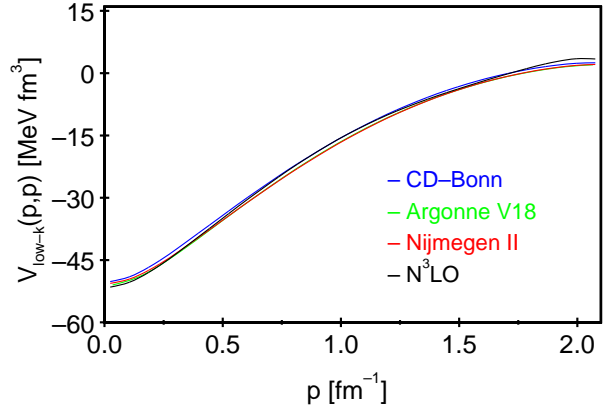}
\end{center}
\caption{(Color online) As in Fig. \ref{vlkquarta}, but for the respective
 $V_{\rm low-k}$'s with a cutoff momentum $\Lambda=2.1$ fm$^{-1}$.
\label{vlkquinta}}
\end{figure}

It is of interest to examine the behavior of $V_{\rm low-k}$.
In Fig. \ref{vlkquinta}, we display some momentum-space matrix
elements of $V_{\rm low-k}$ derived from the above $V_{NN}$'s. 
Here, we see that $V_{\rm low-k}$ matrix elements vary smoothly and
become rather small near the cutoff momentum $\Lambda =2.1$ fm$^{-1}$. 
Clearly, $V_{\rm low-k}$ is a smooth potential, no longer having the
strong short-range repulsion contained in $V_{NN}$.

From Fig. \ref{vlkquarta} it is evident that the matrix elements of
$V_{NN}$ for the various potentials considered are all very
different, even if the latter are onshell equivalent. 
It is interesting to note that the $V_{\rm low-k}$'s  derived
from them are very close to each other, as shown in Fig. \ref{vlkquinta}. 
This suggests that the various $V_{\rm low-k}$'s are almost identical.

\begin{table}
\caption{Calculated $P_D$'s with different $NN$ potentials and 
with the corresponding $V_{\rm low-k}$'s.
\label{pd}}
\begin{center}
\begin{tabular}{lccccc} \hline
  & NijmII & Argonne $V_{18}$  & CD-Bonn & N$^3$LO \\ \hline
 Full potential       & 5.63 & 5.76 & 4.85 & 4.51 \\
 $V_{\rm low-k}$      & 4.32 & 4.37 & 4.04 & 4.32 \\ \hline
\end{tabular}
\end{center}
\end{table}

In Ref.~\cite{Coraggio05} this $V_{\rm low-k}$ property has been
evidenced through the study of the ground-state properties of some 
doubly-magic nuclei starting from the above mentioned $NN$ potentials.
As a matter of fact, it has been found that the calculated binding
energies per nucleon and the rms charge radii are scarcely sensitive
to the choice of the potential from which the $V_{\rm low-k}$ is derived.
This insensitivity may be traced to the fact that when renormalizing 
the short-range repulsion of the various potentials, the differences 
existing between their offshell properties are attenuated.
In this regard, let us consider the offshell tensor force strength.
This is related to the $D$-state probability of the deuteron $P_D$,
which implies that, when comparing $NN$ potentials, offshell
differences are seen in $P_D$ differences (see for instance 
Ref.~\cite{Machleidt94}).
In Table \ref{pd} the predicted $P_D$'s for each of the potentials
under consideration are reported and compared with those calculated
with the corresponding $V_{\rm low-k}$'s ($\Lambda=2.1~{\rm
fm}^{-1}$). 
It can be seen that while the $P_D$'s given by the full potentials are
substantially different, ranging from 4.5 to 5.8 \%, they become very
similar after renormalization. 
This is an indication that the starting ``onshell equivalent''
potentials have been made almost ``offshell equivalent'' by the 
renormalization procedure.

\begin{figure}[hbtp]
\begin{center}
\includegraphics[scale=0.5,angle=90]{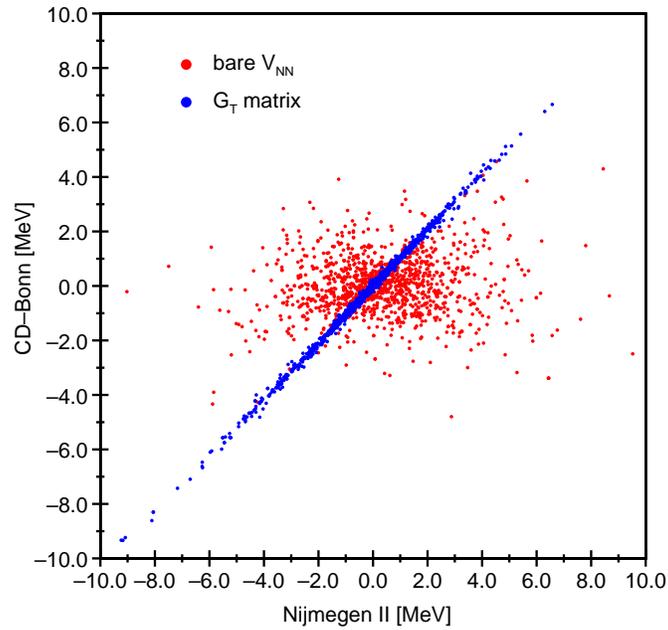}
\end{center}
\caption{(Color online) Correlation plots between Nijmegen II and CD-Bonn 
matrix elements and between the corresponding $G_{T}$ matrix elements 
(see text for details). The red and blue dots correspond to bare $V_{NN}$
and $G_{T}$-matrix elements, respectively.
\label{gmatconf}}
\end{figure}

It is worth to point out that  when the short-range repulsion of different
$V_{NN}$'s is renormalized by means of the $G$-matrix approach the
differences between these potentials are strongly quenched too.
This is  illustrated in Fig. \ref{gmatconf}. There we report the correlation
plot between Nijmegen II and CD-Bonn matrix elements in the HO basis
with the oscillator parameter $\hbar \omega=14$ MeV as well 
the correlation plot between the
corresponding $G_{T}$ matrix elements with a Pauli operator tailored
for the $sd$ shell (see Sec.~\ref{sec:GM2}) and a starting energy of
-5 MeV.
We see that while the bare matrix elements are quite
different this is not the case for the $G_{T}$ matrix ones.

Before closing this section, a brief discussion of $V_{\rm low-k}$ in the 
framework of the  RG-EFT approach may be in order. 
A main point of the RG-EFT approach  is that low-energy physics
is not sensitive to fields beyond a cutoff scale $\Lambda$.
Thus for treating low-energy physics, one just integrates out
the fields beyond $\Lambda$, thereby obtaining a low-energy
effective field theory. 
In RG-EFT, this integration out, or decimation, generates an infinite
series of counter terms \cite{Lepage97} which is a simple power series
in momentum

The question as to wether
the decimation of high-momentum modes in the  derivation of $V_{\rm low-k}$
generates a series of counter terms has been recently investigated in  
Refs.~\cite{Holt04b} and \cite{Kuo05}. In these studies it is assumed that 
 $V_{\rm low-k}$ can be expressed as 

\begin{equation}
V_{\rm low-k}(q,q')\simeq V_{NN}(q,q')+V_{\rm counter}(q,q'); 
~(q,q')\leq \Lambda ~~, \label{eqct1}
\end{equation}

\noindent
where the counter potential is given as a power series 
\begin{equation}
V_{\rm counter}(q,q') = C_0+C_2q^2+C_2'q'^2+C_4(q^4+q'^4) 
                    + C_4'q^2q'^2+C_6(q^6+q'^6)+C_6'q^4q'^2 
                   + C''_6q^2q'^4+~... \label{eqct2} 
\end{equation}

The counter term coefficients are determined using standard fitting 
techniques so that the right hand side of Eq.~(\ref{eqct1}) provides a
best fit to the left-hand side of the same equation. In Ref.~\cite{Holt04b}
this fitting was performed over all partial wave channels and a
consistently good agreement was obtained. 

An examination of the counter terms was made in \cite{Holt04b,Kuo05},
where the coefficients of some of them are also listed.
A main result is that the counter terms are all rather small except
for $C_0$ and $C_2$ of the $S$ waves. 
This is consistent with the RG-EFT approach where the counter term
potential is given as a delta function plus its derivatives \cite{Lepage97}. 

\section{Realistic shell-model calculations}\label{sec:SMC}

\subsection{The early period} \label{sec:SMCEP}

As discussed in the Introduction, while the first attempts to derive the shell-model effective interaction from the free $NN$ potential go back to some forty years ago, the practical value of this approach has emerged only during the last decade or so. On this basis, we may call the ``early period" that extending over the previous three decades. It is the aim of this section to give a brief survey of the main efforts and developments made in the field during this long period of time.

The first attempt to perform a realistic shell-model calculation may be considered the one by Dawson, Talmi and Walecka \cite{Dawson62}. They chose $^{18}$O as testing ground and approximately solved a Bethe-Goldstone equation for the wave function of the two valence neutrons using the free $NN$ potential of Brueckner-Gammel-Thaler 
\cite{Brueckner58} which contains a hard core.
This initial approach, in which the effective interaction was simply taken to be the Brueckner reaction matrix, was followed by several authors \cite{Kahana65,Kahana69,Kallio64,Engeland64}. In this context, the merit of the various methods used to obtain the nuclear reaction matrix elements was a matter of discussion 
\cite{Badhuri65,Kallio65,Kuo65,Wong67}.

A quantitative estimate of the core polarization effects was first given by 
Bertsch \cite{Bertsch65}, who found, for the Kallio-Kolltveit force 
\cite{Kallio64}, that they can produce a correction to the interaction of two valence nucleons in 
$^{18}$O and $^{42}$Sc as large as 30\% of the $G$-matrix interaction. Soon afterwards a turning point in the development of the field was marked by the 
work of Kuo and Brown \cite{Kuo66}, who derived an $sd$-shell effective interaction from the Hamada-Johnston potential \cite{Hamada62}. After having carefully evaluated the  reaction matrix elements for this potential, they took into account the corrections arising from the one-particle-one-hole ($1p-1h$) excitation of the $^{16}$O core (the so-called ``bubble diagram" $G_{\rm 1p1h}$). Comparison of the calculated and experimental spectra of $^{18}$O and $^{19}$F provided clear evidence of the crucial role played by the core polarization. 

This remarkable achievement prompted the derivation of effective interactions for nuclei in other mass regions, such as the calcium \cite{Kuo68a}, nickel \cite{Lawson66,Kuo67} and lead region \cite{Kuo68b,Herling72,McGrory75}.  These effective interactions were all deduced from the Hamada-Johnston potential by use of the 
$G$-matrix theory and including core-polarization effects. In this context, the individual effects of the various second-order diagrams were studied \cite{Brown67,Kuo68a} and the conclusion was reached that the inclusion of the core polarization diagram, $G_{\rm 1p1h}$, is generally sufficient to obtain a satisfactory agreement between calculated and experimental spectra \cite{Kuo74}.

The substantial soundness of the effective interactions derived along the lines of the Kuo-Brown approach 
is evidenced by the fact that, with empirical modifications, they have been used in a number of successful 
shell-model calculations to date. This has brought about various ``semi-realistic" versions of the original
 Kuo-Brown interaction. In this connection, it is worth pointing out that these empirical
modifications concern essentially the monopole parts of the effective interaction, as described in detail
in Ref.~\cite{Poves81}, where the  KB1, KB2, and KB3 interactions were introduced and applied to the
study of $fp$-shell nuclei. 
This kind of approach, which is still favored by several
 authors, especially in the context of large scale shell model calculations, is outside the scope of
 the present article. We only mention here that recently three-body monopole corrections to 
realistic forces have been considered as a simple way to avoid a full treatment of three-body 
forces~\cite{Zuker03}.
 For a detailed discussion of the various effective interactions and the results
 of several applications, the reader is referred to the review papers by 
Brown \cite{Brown01}, Otsuka {\em et al.}
\cite{Otsuka01}, Talmi \cite{Talmi03}, and Caurier {\em et al.} 
\cite{Caurier05}, where a full list of references 
can also be found.  

It is interesting to note that in spite of the very encouraging  results produced by the first 
generation of realistic shell-model calculations, the general attitude of the workers in the field 
was rather pessimistic. This is well reflected in the final statements made by Kirson in his Summary 
talk \cite{Kirson75} at the  1975 Tucson International  Conference:
\begin{quote}
``To summarize my summary, I would say that there are definitely major obstacles in the 
way of doing really convincing calculations. We have a theory, we are aware of the weakness 
of our computations, and we should certainly invest the effort needed to plug the more 
obvious holes. But we should recognize the lack of precision inherent in our inability to go 
to high orders in perturbation theory and adopt more qualitative methods of extracting 
information and gaining understanding. New ideas, new viewpoints are much needed. We should 
also try to withstand the temptations of seductive $^{18}$O, and pay some attention 
to other systems."
\end{quote}

The aim of the Tucson Conference \cite{Barrett75} was to define 
the current status of effective interaction and operator theory in the mid 1975s.
The main issues of this Conference were expounded in the subsequent excellent  review 
by Ellis and Osnes \cite{Ellis77}. As it emerges from the first three talks 
\cite{Brandow75,Johnson75,Ratcliff75} presented at the Tucson Conference and 
as mentioned  in Sec.~\ref{sec:SMEI}, in the years around 1970 the theoretical bases 
for an energy-independent
 derivation of the effective interaction were laid down within the framework of the
time-independent \cite{Brandow67} as well as time-dependent 
\cite{Johnson71,Kuo71} 
perturbation theory. 
At  the same time, the problems inherent in the practical application of 
the theory became 
a matter of discussion. In this context, 
considerable effort was made to study the convergence properties
of the perturbative approach to the derivation of the effective interaction. 
As representative
examples for these studies, we may mention here the works of Refs. 
\cite{Barrett70,Vary73,Schucan72,Schucan73}. In the first of these works all 
significant terms through 
third order
and a few selected fourth-order terms in the $G$ matrix were calculated and the convergence 
order-by-order of the perturbation series was investigated, while in the second one the focus
was on the convergence rate of the sum over intermediate-particle states in the second-order 
core polarization contribution to the effective shell-model interaction. In the third work
the authors raised the problem of the divergence of the perturbation series due to the occurrence 
of the so-called ``intruder states". 

A most important turn in the field of realistic shell-model calculations resulted from the
explicit inclusion \cite{Shurpin77,Shurpin83} of the folded diagrams in the derivation 
of the effective interaction within the framework of the Kuo, Lee, and Ratcliff theory 
\cite{Kuo71}. In these works, starting from a 
realistic $G$ matrix, 
effective interactions for $sd$-shell nuclei were derived by means of the 
 folded-diagram expansion, and  employed in calculations for $A=18$ to 24. 
Note that in Ref.  \cite{Shurpin83} the very convenient method proposed in 
\cite{Suzuki80}
(see Sec.~\ref{sec:SMEI3}) was used to sum up the folded diagram series. 
It turned out that the contribution of the folded diagrams to the effective 
interaction is quite large, 
typically about 30\%. In the earlier study \cite{Shurpin77} the $G$ matrix was obtained starting from 
the Reid soft core potential while in the later study \cite{Shurpin83} both this potential and the 
Paris potential were used. It may be worth noting that this latter paper was the first one where the 
Paris potential was employed in a microscopic calculation of finite open-shell nuclei.

As pointed out in Sec.~\ref{sec:NN1}, the advent of advanced meson-theoretic $NN$ potentials, such as the Paris and Bonn ones, was instrumental in reviving the interest in the realistic effective interactions during the 1980s. In this context, a main issue was the sensitivity of nuclear structure calculations to the strength of the tensor-force component contained in the $NN$ potential, for which a practical measure is the predicted $D$-state probability of the deuteron $P_D$ (see Sec.~\ref{sec:NN1}). 

Earlier work on this problem was done by Sommermann {\em et al} 
\cite{Sommermann81} who reexamined the intermediate state convergence problem for the core polarization diagram in the expansion of an effective shell-model interaction using the Bonn-J{\"u}lich (MDFP$\Delta$2) potential \cite{Holinde77,Holinde78} in comparison to the Reid soft-core interaction \cite{Reid68}. It is a remarkable achievement of this work to have shown that the so-called Vary-Sauer-Wong effect, namely the slow rate of convergence found for the Reid potential \cite{Vary73,Kung79}, was of negligible importance for the Bonn-J{\"u}lich
potential, as a consequence of the much weaker tensor force contained in the latter.
  
Some ten years later, using a $G$-matrix folded-diagram method, the $sd$-shell effective interaction was derived from various potentials and applied to calculate the spectra of some light $sd$-shell nuclei \cite{Jiang92}. In particular, the three variants of the OBE parametrization of the Bonn full model, Bonn A, Bonn B, and Bonn C, differing in the strength of the tensor force (see Sec.~\ref{sec:NN1})  
as well as the Paris potential, were considered. It turned out that the best agreement with experiment was achieved with the weakest tensor force potential (Bonn A). 

The development of more realistic $NN$ potentials, such as the Bonn and Paris potentials, led also in the early 1990s to new studies of the role of the third- and higher-order perturbative contributions to the effective interaction, which had been investigated some twenty years before in the work of 
Ref.~\cite{Barrett70}. Within the framework of the folded diagram theory, the inclusion of the third-order diagrams in $G$ in the calculation of the $\hat Q$-box was studied and the order-by-order convergence of the effective interaction in terms of the mass number $A$ was examined \cite{Hjorth92a,Hjorth92b,Hjorth92c,Hjorth96}. The main findings of the last study of this series \cite{Hjorth96} were that the convergence of the perturbation expansion seems to be rather insensitive to the mass number and that the effects of third-order contributions in the $T=1$ channel are almost negligible. We refer to Ref.~\cite{Hjorth95} for a 
review of the status of the theory of realistic effective interactions in the mid 1990s.  

We have given above a brief review of the great deal of work done in the field of realistic shell-model calculations during what we have called, somewhat arbitrarily, the ``early period". Before finishing this subsection, a word is in order concerning this choice of presentation, which, in our opinion, reflects how hard it has been to make this kind of calculations a reliable tool for quantitative nuclear structure studies.

We draw the reader's attention to the fact that, after the calculations performed by Kuo and collaborators from the mid 1960's up to the mid 1970's 
\cite{Kuo68a,Lawson66,Kuo67,Kuo68b,Herling72,McGrory75}, practically no other calculations
were done for about two decades having as only aim a detailed description of nuclear structure properties. 
This lack of confidence in the predictive power of realistic effective interactions is well evidenced by the fact that in almost all the papers on this subject published from the mid 1970's up to about the early 1990's, the focus has been on nuclei of the $sd$-shell, considered as a necessary testing ground for the effective interaction before embarking on a more ambitious program of applications. In this connection, the warning by Kirson (see above citation) against the temptations of seductive $^{18}$O acquires a particular significance.

\subsection{Modern calculations} \label{sec:SMCMC}

As discussed in the previous subsection, during the 1980s a substantial progress was made in the practical application of shell model with realistic effective interactions. This has resulted in a new generation of realistic shell-model calculations oriented to the study of the spectroscopic properties of nuclei in various mass regions. 

The first attempts in this direction focused on the Sn isotopes
\cite{Engeland95,Hjorth95,Andreozzi96a}, in particular on the neutron deficient ones.
In the work of Ref.~\cite{Engeland95} the Bonn A potential was used while in 
that of Ref.~ \cite{Hjorth95} results obtained with the stronger tensor-force 
versions of this potential (Bonn B and Bonn C) were also reported. In both these works all two-body diagrams through third order in the $G$ matrix were included in the 
$\hat Q$-box. In the study of Andreozzi {\em et al.} \cite{Andreozzi96a} both 
the Paris and 
Bonn A potentials were employed  and the $\hat Q$-box was taken to be 
composed of diagrams through second order in $G$ (a comparative discussion of 
the above studies can be  found in \cite{Andreozzi96a}).  The results of these initial ``modern" calculations, 
in particular those obtained with 
the Bonn A potential, turned out to be in remarkably good agreement with the experimental 
data.   

These achievements encouraged further work along these lines. In the few following years realistic shell-model calculations were carried out for a number of nuclei around doubly magic $^{100}$Sn \cite{Coraggio00}, $^{132}$Sn \cite{Andreozzi97,Covello97,Holt97,Holt98}, and $^{208}$Pb \cite{Coraggio98,Andreozzi99,Coraggio99}. Note that in all these calculations, except the one of Ref.~\cite{Holt98} where the CD-Bonn 1996 potential 
\cite{Machleidt96} was employed, the effective interaction was derived from the Bonn A potential. We refer to Covello {\em et al.} \cite{Covello01} for a 
survey of the status of realistic shell-model calculations at the end of the 1990s. Suffice it to say  here that the rms deviation for the spectra of 15 medium- and heavy-mass nuclei reported in this paper is in 7 cases less than 100 keV, only in one case reaching 165 keV.

The success achieved by these calculations gave a clear-cut answer to the crucial, long-standing problem of how accurate a description of nuclear structure properties can be provided by realistic effective interactions, opening finally the way to a more fundamental approach to the nuclear shell model than the traditional, empirical one.

As is well known, much attention is currently being focused on nuclei in the regions of shell closures off stability, as they allow to explore for possible shell-structure modifications when approaching the proton and neutron drip lines.
The experimental study of these nuclei is very difficult, but in recent years new data have become available for some of them, which provide a challenging testing ground for realistic effective interactions.
Starting in the early 2000s, several ``exotic" nuclei around doubly magic 
$^{100}$Sn and $^{132}$Sn have been studied employing realistic effective interactions derived from the high-precision CD-Bonn $NN$ potential \cite{Machleidt01b}. In the first paper of this series \cite{Coraggio02a} shell-model calculations were performed for Sn isotopes beyond $N=82$, in particular the heaviest Sn isotope known to date, $^{134}$Sn. In this work, similarly to what had been done in their  previous papers, the authors employed a $G$-matrix folded-diagram formalism. A very good agreement was found between the calculated  results and the available experimental data.

A most important turn in the derivation of the effective interaction from the free $NN$ potential has resulted from the development of the $V_{\rm low-k}$ approach (see Sec.~\ref{sec:vlowk}). This approach 
to renormalize the short-range repulsion of $V_{NN}$ has been recently used in several calculations. Attention has been focused on exotic nuclei neighboring $^{100}$Sn 
\cite{Coraggio04}, and $^{132}$Sn \cite{Coraggio02b,Genevey03,Scherillo04,Coraggio05,Coraggio06a,Covello06}. Actually, most of these works are concerned with nuclei in the latter region, for which there are new interesting experimental data. The agreement between theory and experiment has turned out to be very good in all cases. This supports confidence in the predictions of these realistic calculations, which may therefore stimulate and be helpful to, future experiments. To illustrate this important point, 
we shall review in Sec.~\ref{sec:res3} 
some selected results for $^{132}$Sn neighbors with neutrons beyond the 82 shell.       

While most of the realistic shell-model calculations carried out in last few years have employed effective interactions derived from the CD-Bonn $NN$ potential, some attempts to put to the test chiral potentials have also been made. Representative examples for these studies are the works of 
Refs.~\cite{Coraggio02c,Coraggio03,Coraggio05b,Coraggio06b}, all of them making use of the $V_{\rm low-k}$ renormalization method. 
Actually, only in \cite{Coraggio02c}, where the NNLO Idaho potential \cite{Entem02c} was used, shell model calculations for various two valence-particle nuclei were performed. In the other three works \cite{Coraggio03,Coraggio05b,Coraggio06b} ground-state properties of doubly magic nuclei $^{4}$He, $^{16}$O, and $^{40}$Ca were calculated within the framework of the Goldstone expansion \cite{Goldstone57} with the N$^3$LO potential \cite{Entem03}. 
The results obtained in \cite{Coraggio02c} turned out to be comparable to those produced by effective interactions derived from the CD-Bonn potential. Clearly, this is related to the important question of how much nuclear structure results depend on the $NN$ potential one starts with. We shall discuss this point in Sec.~\ref{sec:res2}.
 
\subsection{Results}\label{sec:res}

\subsubsection{Comparison between $G$-matrix and $V_{\rm low-k}$ approaches} \label{sec:res1}

As evidenced by the discussion given in Secs.~\ref{sec:SMCEP} and 
\ref{sec:SMCMC}, for about forty years
the Brueckner $G$ matrix represented practically the
only tool to renormalize the repulsive core contained in the $NN$ potential.
After its introduction, the $V_{\rm low-k}$ potential has been readily seen as an 
alternative to the $G$ matrix.  In fact, $V_{\rm low-k}$  is a smooth free-space potential 
and its matrix elements between
unperturbed two-particle states can be directly used as input to the $\hat Q$-box folded-diagram
procedure of Sec.~\ref{sec:SMEI}. Therefore, a main issue has been to assess 
the merit  of  the 
$V_{\rm low-k}$ approach in practical applications.

First work along this line was done by Kuo {\em et al.}~\cite{Kuo02}, who carried 
out shell-model calculations
for $^{18}$O using two-body matrix elements of effective interactions derived from the Bonn A and Paris
potentials.  Other studies~\cite{Covello02,Bogner02,Covello03} were then performed which, 
starting from the CD-Bonn potential, extended  the comparison between the $G$ matrix
and $V_{\rm low-k}$ approaches to nuclei heavier than $^{18}$O. In particular, 
the calculations of Ref.~\cite{Covello02} concern $^{132}$Sn neighbors, while
results for   $^{18}$O, $^{134}$Te, and  $^{135}$I are presented in 
\cite{Bogner02}.
A comparison between the  $G$ matrix 
and $V_{\rm low-k}$ spectra for the heavy-mass nucleus $^{210}$Po can be found 
in~\cite{Covello03}.
In all these calculations the cutoff parameter $\Lambda$ is chosen in the vicinity of 2 fm$^{-1}$.

It has been a remarkable finding of all these studies  that the    
$V_{\rm low-k}$ results are as good or even slightly better than the $G$-matrix ones. 
The rather close similarity between the results of the two approaches finds an exception for  
the Paris potential \cite{Kuo02}. We shall comment on this point in 
Sec.~\ref{sec:res2}. It is also important
to point out that the  $V_{\rm low-k}$ approach appears to work equally well in  different mass regions. 
A detailed description of the above mentioned calculations can be found in 
the cited references.
Here, for the sake of completeness, we report
two examples concerning $^{18}$O  and $^{134}$Te, which are both taken from 
\cite{Bogner02}, where  more details about the calculations can be found.

\begin{figure}[hbtp]
\begin{center}
\includegraphics[scale=0.75]{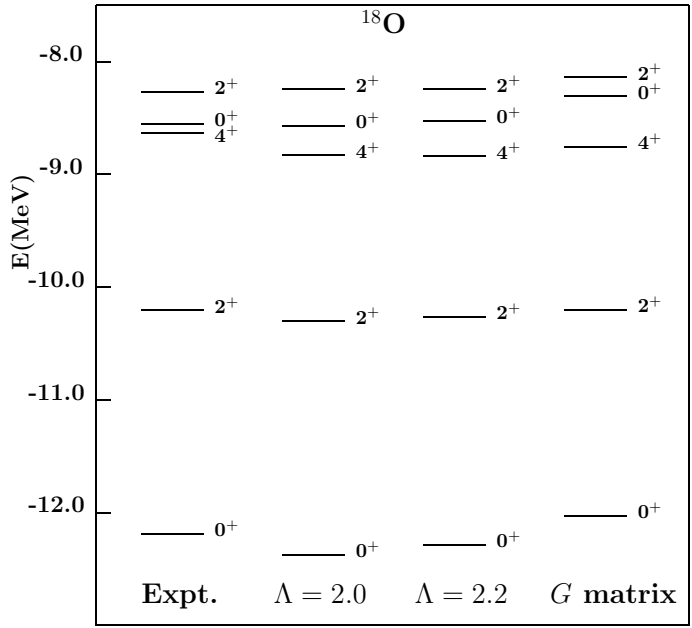}
\end{center}
\caption{Low-lying states in $^{18}$O.
\label{O18}}
\end{figure}

\begin{figure}[hbtp]
\begin{center}
\includegraphics[scale=0.85]{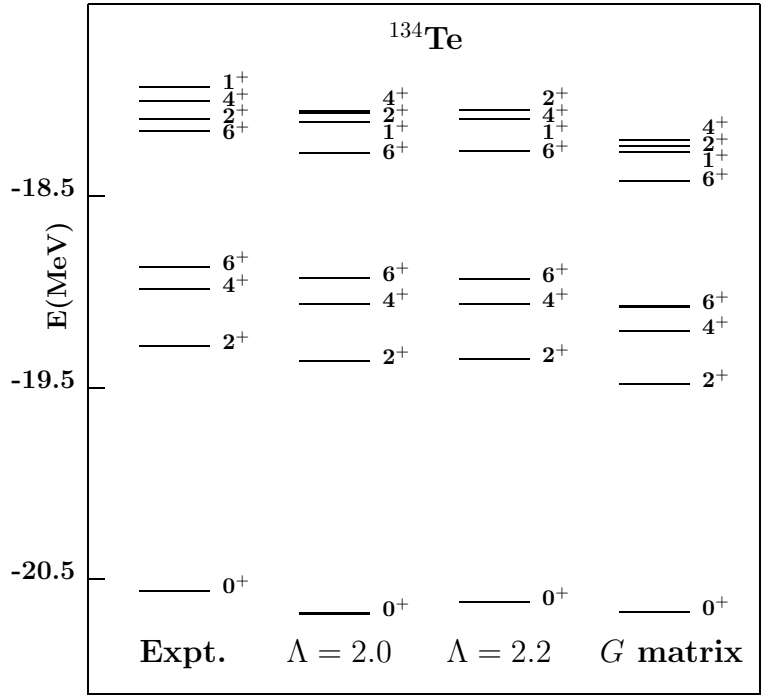}
\end{center}
\caption{Low-lying states in $^{134}$Te.
\label{Te134}}
\end{figure}

The experimental and
calculated spectra of  $^{18}$O including the five lowest-lying states, 
and   those  of   $^{134}$Te with the eight lowest-lying states are 
reported in 
Figs.~\ref{O18} and \ref{Te134}, respectively.
The calculated spectra    have been obtained from
the CD-Bonn potential through both the $V_{\rm low-k}$ and $G$-matrix 
folded-diagram approaches.  As for the $V_{\rm low-k}$ calculations, two 
values of the cutoff momentum have been used, $\Lambda=2.0$ and 2.2 fm$^{-1}$. 
For $^{18}$O we have adopted a model space with the two valence neutrons in the 
$(0d_{5/2},0d_{3/2},1s_{1/2})$  shell, while for $^{134}$Te
the  two valence protons are allowed  to occupy the levels of  the 
$(0g_{7/2},1d_{5/2},1d_{3/2},2s_{1/2},0h_{11/2})$ shell. In both cases      
the SP energies  are taken from experiment, in particular from the spectra of
$^{17}$O  and $^{133}$Sb for the $sd$ and $sdgh$ shells,  respectively. 
The energy of the $2s_{1/2}$ proton
level, which is still missing, is from the study by 
Andreozzi~{\em et al.}~\cite{Andreozzi97}.
Note that in  Figs.~\ref{O18} and \ref{Te134} the energies are relative to 
the ground state  of the neighboring doubly closed nucleus  $^{16}$O 
and $^{132}$Sn, respectively, with 
the mass-excess values for $^{17}$O and $^{133}$Sb needed for the absolute
scaling of the SP energies coming from Refs.~\cite{Audi93} and 
\cite{Fogelberg99}.  
As for $^{134}$Te,  the Coulomb contribution  
was taken into account by adding to the CD-Bonn effective interaction the 
matrix element of the Coulomb force between  two 
$g_{\frac{7}{2}}$ protons with $J=0$.

As pointed out above, we see from  Figs.~\ref{O18} and \ref{Te134} that 
the $V_{\rm low-k}$ and $G$-matrix results 
are rather close to each other and in quite good agreement with experiment. 
It is worth noting that the $V_{\rm low-k}$ results 
for the two values of $\Lambda$ do not differ significantly. We shall come back 
to this point later.

In this context, we may also mention the work of Refs.~\cite{Schwenk05} and 
\cite{Schwenk06}, where harmonic-oscillator matrix elements of
the $V_{\rm low-k}$ potential and $G$ matrix are compared in 4 major shells. It was found that 
both the $T=0$ and $T=1$ matrix elements are very similar.

On the above  grounds, the conclusion is reached that the $V_{\rm low-k}$ approach is a 
reliable way to renormalize the bare $NN$
potential before employing  it in the derivation of the shell-model effective interaction.  
    
Another relevant point to be stressed is that the $V_{\rm low-k}$ 
is far easier to be used than the $G$ matrix. In fact, $V_{\rm low-k}$ is a soft $NN$ potential and as such
does not depend either  on the starting energy or on the model space,
as is instead the case of the
$G$ matrix, which is defined in the nuclear medium. 
This has the desirable  consequence that the $V_{\rm low-k}$ matrix element to be used in an
interaction vertex of a  given
$\hat Q$ box diagram is  completely determined by the incoming and outcoming lines attached to the 
vertex under consideration. Hence, it does not depend on any other part of the diagram.  Also, it is worth
noting that the same $V_{\rm low-k}$ can be used in different mass regions.

Finally, when considering the features of the $V_{\rm low-k}$ approach, one has also to keep in mind
that,  as discussed in Sec.~\ref{sec:vlowk3}, different 
$NN$ potentials lead to low-momentum potentials which are remarkably similar to each other. In the same section,
we have showed that  the differences between the different
potentials are also quenched within the $G$ matrix approach.
This will be further discussed in Sec.~\ref{sec:res2}.

We conclude by mentioning   that the $V_{\rm low-k}$ approach has also proved to be an advantageous alternative to
the $G$ matrix in the calculation of the ground-state properties of doubly closed nuclei.
As is well known, 
the traditional approach to this problem is the  Brueckner-Hartree-Fock (BHF) theory. In the study of Ref.~\cite{Coraggio03} 
it has been showed that the $V_{\rm low-k}$ is suitable for being used directly in the Goldstone expansion. Good results for
$^{16}$O and $^{40}$Ca were obtained by solving  the Hartree-Fock (HF) 
equations for $V_{\rm low-k}$ and then using the resulting self-consistent basis to 
compute higher-order
Goldstone diagrams.
In this connection, it should be mentioned also the work of Ref.~\cite{Kuckei03},
where  use was made of  the $ V_{\rm low-k}$ potential in a HF and BHF approach to 
evaluate the 
binding energy and the saturation density of nuclear matter and finite nuclei. Their conclusion was that
for nuclear matter at  small and medium densities as well as for finite nuclei HF and BHF results with $V_{\rm low-k}$
are nearly identical. 

\subsubsection{Calculations with different $NN$ potentials} \label{sec:res2}

As discussed in Sec.~\ref{sec:SMC}, a  main problem relevant to realistic shell-model calculations has been  
to assess the accuracy of the effective interaction derived from the $NN$ potential.
However, it is also an important issue what information comes out on the $NN$ potential from this kind of 
calculations. As we have seen in
Sec.~\ref{sec:NN}, from the early 1950s on several different potentials have been constructed. The potentials of 
the last generation all fit equally well the $NN$ phase shifts, but,
except for the OPE tail, they  are quite different from each other.
Actually, these potentials are well-constrained by the deuteron properties and the $NN$ scattering
data for $E_{\rm lab} < 350$ MeV and therefore their offshell behavior is subject to ambiguities.
This is reflected  in the different values predicted for the $D$-state probability of the  deuteron 
$P_{D}$  (see Sec.~\ref{sec:NN}) as well as in the different binding energies calculated for $A=3,4$ systems \cite{Nogga00}. 
In fact, for $A > 2$ the interaction may be off the 
energy shell since two nucleons may have different energies before and after they interact.

In this context, it is of great interest to perform nuclear structure 
calculations
to see to which extent the differences between $NN$ potentials persist 
once the short-range repulsion
is renormalized through the $G$-matrix or the $V_{\rm low-k}$  approach.  
This problem has long been investigated  by studying properties of both 
nuclear matter and doubly closed nuclei. A list of references on this subject 
can be found in the work of Ref.~\cite{Coraggio05b}, which
we have briefly discussed in Secs.~\ref{sec:vlowk3} and \ref{sec:SMCMC}.
As for open-shell nuclei, some attempts have been made within the framework 
of the shell model.

In the early work by Sommermann {\em et al.}~\cite{Sommermann81}, the 
comparison between 
Bonn-J\"{u}lich and Reid  results aimed at  
investigating the intermediate-state convergence problem, as discussed in
Sec.~\ref{sec:SMCEP}. The paper by Shurpin {\em et al.}~\cite{Shurpin83} 
had instead as a  
main purpose to compare the spectra of several $sd$-shell nuclei 
calculated with
two different potentials,  Paris and Reid.
The results showed that the Paris potential was as good as, if not better than,
the Reid potential. The idea of 
comparing shell-model calculations with different $NN$ potentials was revived at the end of the 1980s.
In the works of Refs.~\cite{Jiang89,Jiang92} and \cite{Maglione91} $sd$-shell nuclei were studied using 
as input several different $NN$ potentials, all constructed before the 1990s. These works were 
based on the use of the reaction matrix $G$ to derive the effective interaction through  
the $\hat Q$-box folded-diagram method, including up to second-order diagrams. 
The outcome of these studies was that the effective interaction matrix elements 
become more attractive  when  a $NN$ potential 
with a weaker tensor-force strength is used. Jiang {\em et al.}~\cite{Jiang92} 
noted that the nonlocality of the
potential gives a further contribution in this direction. All these calculations indicated in general  
that the best agreement with experiment was produced by the Bonn potential. 

A confirmation of the important role played by
weaker tensor force potentials came out from the paper of
Ref.~\cite{Andreozzi96a}, where  
Paris and Bonn A results for neutron deficient Sn isotopes are compared. 
A better agreement with experiment was obtained with the
Bonn A potential.
The merit  of potentials with weak tensor force was also  evidenced in
\cite{Hjorth95}, who performed  realistic shell-model
calculations for nuclei in the mass regions of $^{16}$O, $^{40}$Ca, and $^{100}$Sn using 
the three variants (A, B, and C)  of the Bonn potential.

A comparative study of shell-model results obtained from different $NN$ 
potentials, including the high-precision  CD-Bonn potential, was made  for the 
first time in \cite{Covello99}. In this paper, the nucleus $^{134}$Te was used as a 
testing ground to investigate the effects of the CD-Bonn potential with respect to 
those of the three potentials Bonn A, Paris, and Nijm93.
As in all previous calculations, it was found that a weak tensor force leads to a better description of 
nuclear structure properties. Furthermore, it turned out that 
the calculated spectra, except that obtained from  the Paris potential, are  only slightly different from each other. 

\begin{figure}[hptp]
\begin{center}
\includegraphics[scale=0.8]{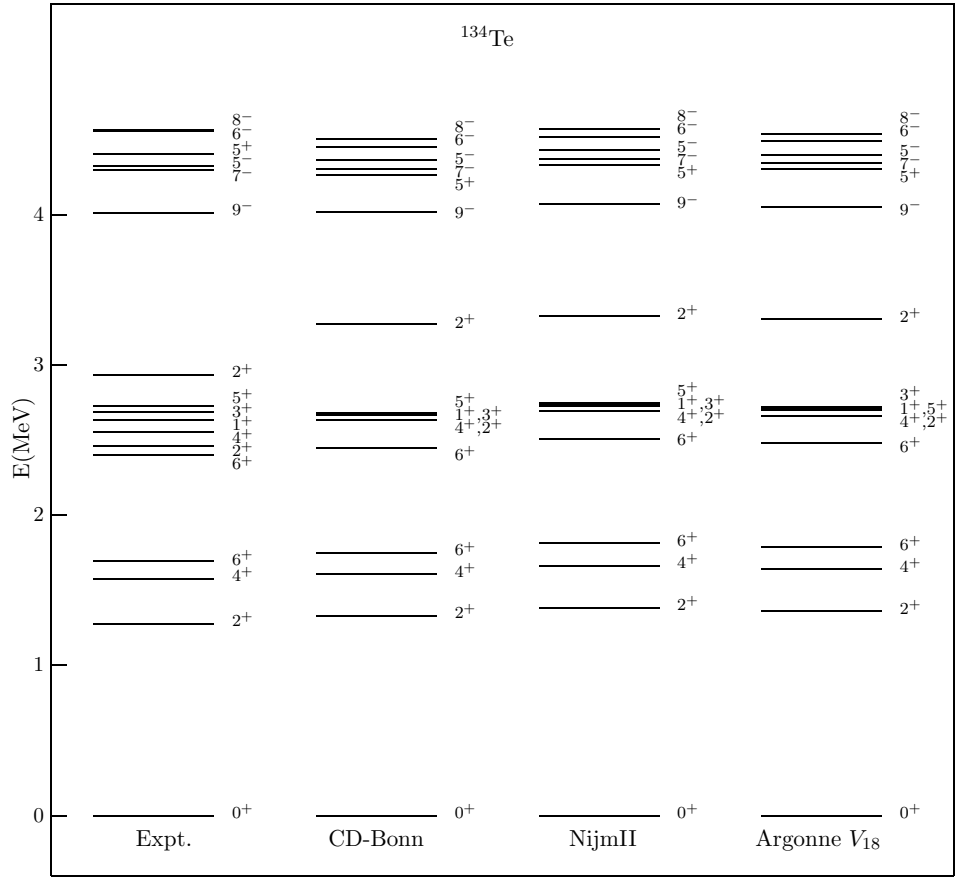}
\end{center}
\caption{Spectrum of $^{134}$Te. Predictions by various $NN$ potentials 
are compared with experiment.
\label{psequiv}}
\end{figure}

All the above calculations made use of the reaction matrix $G$
and their results provide evidence  that the differences existing between 
various potentials are quenched by
the  $G$-matrix renormalization procedure
(see Fig. \ref{gmatconf}). 
One may wonder what happens when the $V_{\rm low-k}$ potential is used. The 
first answer to this question  may be found in \cite{Kuo02}, 
where the $V_{\rm low-k}$ approach
is used to calculate the Paris and Bonn A spectra for $^{18}$O. It is interesting to note  that 
these two spectra are much closer to each other than those obtained with the $G$ matrix 
approach \cite{Jiang92}. This reflects the fact that the $V_{\rm low-k}$ procedure is 
more effective in reducing the differences between the
two potentials than the $G$-matrix approach.
The matrix elements in the $^{3}S_{1}$ channel for both  Paris and Bonn A potentials
are reported  in Fig.~1 of \cite{Kuo02}, where it is shown that they
become similar after the $V_{\rm low-k}$ renormalization, the latter  being much larger for
the strongly  repulsive Paris potential.

In Sec.~\ref{sec:vlowk3} (see Figs.~\ref{vlkquarta} and \ref{vlkquinta}), 
we have reported a similar  comparison in the
$^{1}S_{0}$ channel for
the chiral potential N$^{3}$LO and the three high-precision potentials 
CD-Bonn, NijmII, and Argonne $V_{18}$. The differences between the matrix elements of these potentials 
 practically disappear when 
the $V_{\rm low-k}$ matrix elements are considered. 
On these grounds one expects that shell-model effective interactions derived from 
phase-shift equivalent 
$NN$ potentials through the  
$V_{\rm low-k}$ approach should lead to very similar results.

This we have verified by calculating the spectra of $^{134}$Te starting
from the CD-Bonn, NijmII, and Argonne $V_{18}$ potentials. 
For all the three $V_{\rm low-k}$'s derived from these $NN$
potentials, use has been made 
of the same cutoff momentum $\Lambda=2.2$~fm$^{-1}$. Other details on how 
these calculations  have been performed may be found in Sec.~\ref{sec:res3}, where 
we present  results for nuclei
beyond $^{132}$Sn, which have been obtained using the CD-Bonn potential.

In Fig.~\ref{psequiv}, the three calculated spectra are shown 
and compared with the
experimental one. 
Note that, with respect to Fig.~\ref{Te134}, the energies are 
relative to the ground state and higher lying levels are included. We see that the calculated spectra are very similar,
the differences between  the predicted  energies not exceeding 80 keV. It is also seen that the agreement
with experiment is very good for all the three potentials. In fact, the rms
deviation is 115, 128, and 143 keV for CD-Bonn, Argonne $V_{18}$, and NijmII, respectively. 

Finally, it is of interest to make a few comments regarding the value of the 
cutoff momentum $\Lambda$.
In Sec.~\ref{sec:vlowk3}  the plausibility of using a value in the vicinity of 
2 fm$^{-1}$ was discussed. It was also shown in 
Sec.~\ref{sec:res1} that shell-model results do not change significantly 
when varying  $\Lambda$ from 2 to 2.2 fm$^{-1}$.
A preliminary study of the dependence  of shell-model results on $\Lambda$ 
can be found in the work of Ref.~\cite{Covello05}, 
where results for $^{134}$Te are reported using the  
CD-Bonn, Argonne $V_{18}$, and NijmII potentials with $\Lambda$ varying over
a rather large range (1.5-2.5 fm$^{-1}$). It has been  found that the best agreement 
with experiment for each of these three potentials  
corresponds to a somewhat different value of $\Lambda$. However, changes 
around this value do not 
significantly modify the quality of agreement for all the  three potentials.  
As an example, the rms deviation for NijmII remains below 150 keV for any value
of $\Lambda$ between 1.8 and 2.2 fm$^{-1}$  
This subject deserves certainly further studies. 

\subsubsection{Neutron rich nuclei beyond $^{132}$Sn: comparing theory and experiment} \label{sec:res3}

The study of exotic nuclei is a subject of a current experimental
and theoretical interest. Experimental information for nuclei in the 
vicinity of $^{78}$Ni, $^{100}$Sn, $^{132}$Sn, which have been long
inaccessible to spectroscopic studies, is now available thanks to new
advanced facilities and techniques. 
The advent  of  radioactive  ion beams is opening new perspectives 
in this kind of studies. Nuclei in the regions of shell closures 
are a source of direct
information on shell structure. New data  offer therefore
the opportunity to test the shell model 
and look for a possible 
evolution of shell structure when going toward proton or neutron drip lines. 

This is stimulating shell-model studies in these regions. 
Several realistic shell-model calculations have
been recently  performed  for  nuclei around $^{100}$Sn and 
$^{132}$Sn, as mentioned in Sec.~\ref{sec:SMCMC}.
We report here some selected results on  $^{132}$Sn neighbors with $N>82$. 
More precisely, we consider the three nuclei $^{134}$Sn,  $^{134}$Sb, and  
$^{135}$Sb, which
have been very recently studied in \cite{Covello06}, 
\cite{Coraggio06a}, and  \cite{Coraggio05}, to which we refer for a more 
detailed discusssion.

\begin{figure}
\begin{center}
\includegraphics[scale=0.85]{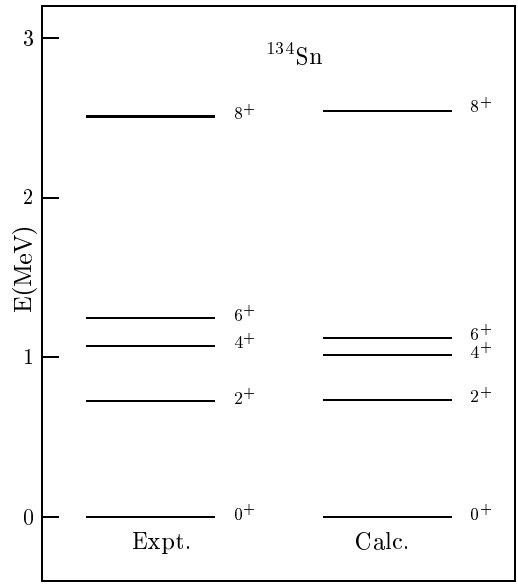}
\end{center}
\caption{Experimental and calculated spectra of $^{134}$Sn.
\label{134Sn}}
\end{figure}

The experimental data which have been now become available for 
these neutron-rich
nuclei may suggest a modification in the shell structure. They are, in fact, 
somewhat different from what 
one might expect by extrapolating the existing results for $N<82$,
and as a possible explanation a change in the  single-proton
level scheme has been suggested. The latter,
caused by a more diffuse nuclear surface, may be seen as a precursor
of major effects which should show up
at larger neutron excess.

We shall see now that
the properties of these  three nuclei are all well accounted for by a  
shell-model calculation with a  two-body effective interaction 
derived from a modern $NN$ 
potential. In so doing, we evidence the merit of realistic shell-model
calculations, showing, at the same
time, that there is no need to invoke new effects  to describe 
few-valence-particle
nuclei just above $^{132}$Sn. In this connection,
it is worth mentioning  that different shell-model calculations 
have been performed for the above nuclei 
\cite{Shergur02,Shergur05,Shergur05a,Chou92,Sarkar01,Korgul02}.
None of these studies, however, has succeeded in accounting simultaneously for the peculiar
features of all the three nuclei.

\begin{figure}
\begin{center}
\includegraphics[scale=0.95]{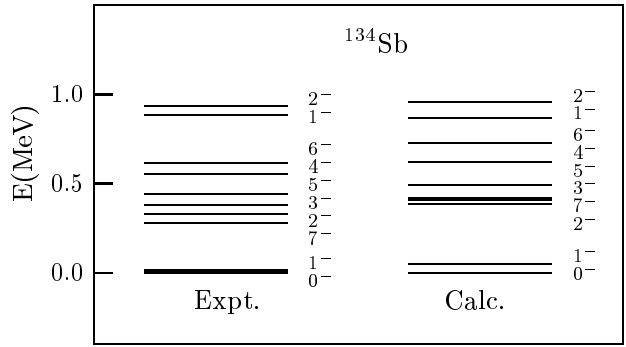}
\end{center}
\caption{Experimental and calculated spectra of $^{134}$Sb.
\label{134Sb}}
\end{figure}

In our calculations, we assume that $^{132}$Sn is a closed core and let the valence
neutrons occupy the six levels $0h_{9/2}$, $1f_{7/2}$, $1f_{5/2}$, $2p_{3/2}$,
$2p_{1/2}$, and  $0i_{13/2}$ of the 82-126 shell, while for the proton
the model space includes the five  levels  $0g_{7/2}$, $1d_{5/2}$, $1d_{3/2}$, $2s_{1/2}$,
and $0h_{11/2}$ of the 50-82 shell.
The proton and neutron single-particle energies have been taken from the experimental spectra of  
$^{133}$Sb and $^{133}$Sn, respectively, 
the only exceptions being those of the $s_{1/2}$ proton 
and   $i_{13/2}$ neutron  levels which are still
missing. The energy of the former, as mentioned in Sec.~\ref{sec:res1},
is  from \cite{Andreozzi97}, while that of the latter from \cite{Coraggio02a}.
The two-body matrix elements of the effective interaction 
have been derived from the CD-Bonn $NN$ potential. This potential is 
renormalized  by constructing the $V_{\rm low-k}$ with a cutoff momentum  
$\Lambda$ =2.2 fm$^{-1}$. 
The effective
interaction has been then derived within the framework of  the 
$\hat {Q}$-box plus folded diagram method
including  diagrams up
to second order in $V_{\rm low-k}$. The computation  of these diagrams is 
performed within the harmonic-oscillator basis using  intermediate states composed of all possible hole 
states and particle states restricted  to the five  shells above the Fermi surface. This
guarantees the stability of the results when increasing the number of intermediate particle states.
The oscillator parameter used is $\hbar \omega = 7.88$ MeV and for protons the Coulomb force has been
explicitly added to the $V_{\rm low-k}$ potential.

\begin{figure}[hbtp]
\begin{center}
\includegraphics[scale=0.85]{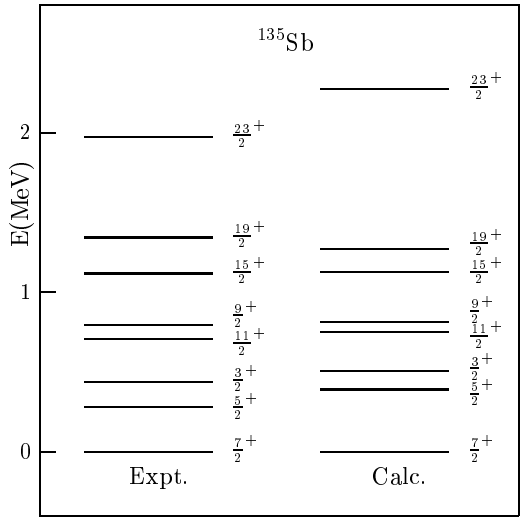}
\end{center}
\caption{Experimental and calculated spectra of $^{135}$Sb.
\label{135Sb}}
\end{figure}

The experimental and calculated spectra of  $^{134}$Sn,  $^{134}$Sb, and  $^{135}$Sb
are compared in Figs.
\ref{134Sn}, \ref{134Sb}, and \ref{135Sb}, respectively.
Note that both the calculated and experimental  spectra  of $^{134}$Sb
include all levels  lying below 1 MeV excitation energy, while  for $^{135}$Sb
all the yrast states have been reported. 
As for $^{134}$Sn, only the levels reported in Fig.~\ref{134Sn} have been 
observed, while
in the calculated spectrum  five states, predicted in  between the $6^+$ and $8^+$
states, 
are omitted.

 The two isotones with $A=134$ 
are the most appropriate 
systems  to study the effects of the
effective interaction. In fact,  $^{134}$Sn is  a direct source of information on
the neutron-neutron channel while $^{134}$Sb on the proton-neutron one. The nucleus 
$^{135}$Sb represents a further interesting test, since it allows to investigate at once
the role of the 
effective interaction in both channels. 

From the above figures, we see that the experimental levels in all the three nuclei 
are very well reproduced by the theory, 
the discrepancies being  of the order of few tens of keV for most of the states.
Note that the very low-energy  positions 
of both the first-excited $2^+$ and $1^-$ states in  $^{134}$Sn and  $^{134}$Sb,
respectively, are well accounted for, as is also the case for 
the low excitation energy of the $5/2^{+}$ state in $^{135}$Sb. We refer to the above mentioned papers 
for more details on  each of the three calculations and for a discussion of the structure of the states 
in $^{134}$Sb and $^{135}$Sb. As for the latter, it is shown in \cite{Coraggio05}
that it is the admixed nature
of the $5/2^{+}$ state that explains its anomalously low position and the 
strongly hindered $M1$ transition to the ground state.
The role  of the effective interaction as well as of the $5/2^+$ single-proton
energy in determining this situation has been examined,
with particular attention to the effects induced by the proton-neutron interaction.
The relevant role of the latter was also evidenced  in $^{134}$Sb
\cite{Coraggio06a}. For this nucleus, a direct relation can be
established between the diagonal matrix elements of the proton-neutron
effective interaction and the calculated energies reported in 
Fig.~\ref{134Sb}, since the corresponding wave functions are 
characterized by very little configuration mixing. In particular, the eight 
lowest-lying states can be identified with the eight members of the 
$\pi g_{7/2} \nu f_{7/2}$ multiplet while the two higher states,
with $J^{\pi}=1^-$ and $2^-$,  are the only members of 
$\pi d_{5/2} \nu f_{7/2}$ multiplet having an experimental counterpart.

\begin{figure}[hbtp]
\begin{center}
\includegraphics[scale=0.80]{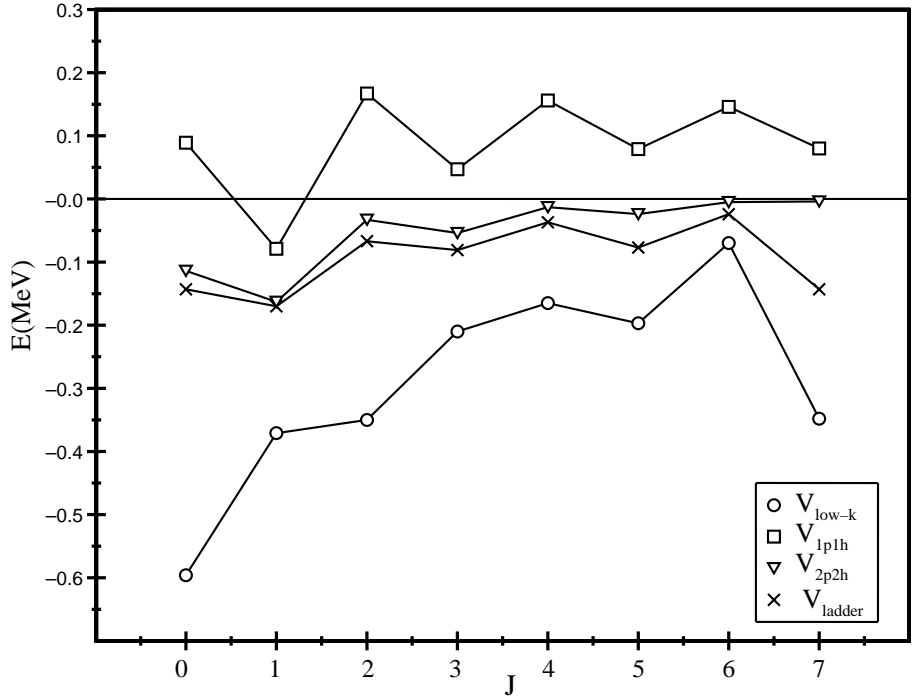}
\end{center}
\caption{Diagonal matrix elements of the $V_{\rm low-k}$ and contributions 
from the two-body second-order diagrams for the $\pi g_{7/2} \nu f_{7/2}$
configuration.
\label{vcontr}}
\end{figure}

In this connection, it may be interesting to summarize  here the
analysis of the proton-neutron matrix elements reported 
in the work of Ref.~\cite{Coraggio06a}. Focusing on the $\pi g_{7/2} \nu f_{7/2}$ 
configuration, we examine the various terms 
which contribute to the effective interaction, namely the $V_{\rm low-k}$ 
and all the second-order two-body diagrams composing
the $\hat Q$ box. It is worth pointing out that the effective interaction $V_{\rm eff}$ 
is obtained by summing to the 
$\hat Q$-box the folded-diagram series [see Eq. (\ref{heffop})], but the 
contribution of the latter provides a common 
attenuation for all matrix elements, thus not affecting the following discussion.

\begin{figure}[hbtp]
\begin{center}
\includegraphics[scale=0.85]{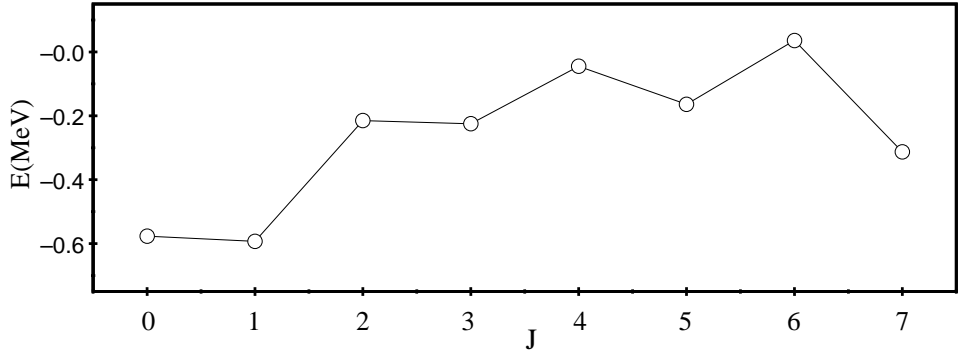}
\end{center}
\caption{Diagonal matrix elements of the two-body effective interaction for 
the $\pi g_{7/2} \nu f_{7/2}$
configuration.
\label{veff}}
\end{figure}

The diagonal matrix elements of $V_{\rm low-k}$ for the $\pi g_{7/2} \nu f_{7/2}$
configuration
are  reported in Fig.~\ref{vcontr} as a function of $J$ together with the 
contributions arising from the 
$1p-1h$, $2p-2h$, and ladder excitations. 
In Fig.~\ref{veff} we present
for the same configuration the  diagonal matrix elements  of the effective 
interaction  which we have obtained  starting from this  $V_{\rm low-k}$. Note that
the behavior of   $V_{\rm eff}$  shown in Fig.~\ref{veff} is  quite similar to that 
of  the $\pi g_{7/2} \nu f_{7/2}$ multiplet   in $^{134}$Sb, whose members, as mentioned above, 
are almost pure. This behavior, however,  is quite  different from that 
of $V_{\rm low-k}$, especially as regards
the $0^{-}-1^{-}$ spacing. This means that  the second-order two-body contributions
accounting for configurations  left out of the chosen model space are of key 
importance
to renormalize  the $V_{\rm low-k}$ and get the correct behavior. In particular,
a crucial role is played by the core-polarization
effects  arising from the $1p-1h$ excitations. 

At this point one may wonder  why our effective interaction is able to
explain the properties of these neutron-rich nuclei above $^{132}$Sn
while  the  realistic shell-model calculations of Refs.~\cite{Shergur02, Shergur05, Shergur05a}
do not. In the latter calculations the effective interaction has been derived  within the $G$-matrix
folded-diagram approach including  in the $\hat Q$-box diagrams up to third order in $G$  
and intermediate states with at most $2 \hbar \omega$ excitation energy \cite{Hjorth96}. 
With this interaction, a satisfactory
agreement with experiment was obtained only for $^{134}$Sn when using experimental SP energies. 
A downshift of the $\pi d_{5/2}$ level with respect to the  
$\pi g_{7/2}$ one  by 300 keV turned out to be  necessary to obtain  a good description of 
$^{135}$Sb \cite{Shergur02,Shergur05a}, but did not help for $^{134}$Sb \cite{Shergur05}.
We have verified that the differences between the results presented in this 
section and those of Refs.~\cite{Shergur02, Shergur05, Shergur05a} can be 
traced mainly to the different dimension  of the
intermediate state space. In fact, we have found that including intermediate states only  
up to $2 \hbar \omega$ excitation energy in our derivation of the effective interaction, the results of  
the two calculations become very similar.

In summary, we have shown that modern realistic shell-model calculations 
are able to describe with 
quantitative accuracy the spectroscopic properties of $^{134}$Sn,  
$^{134}$Sb, and  $^{135}$Sb, in particular, their
peculiar features. This leads to the conclusion that a consistent 
description of   $^{132}$Sn neighbors with neutrons beyond 
the $N=82$ shell is possible. In this connection, it is worth mentioning  two
more recent works \cite{Covello07a,Covello07b}, where shell-model results for other nuclei around doubly magic
$^{132}$Sn are reported, including predictions for $^{136}$Sn. 

\subsubsection{The role of many-body forces} \label{sec:res4}

In this subsection, we shall discuss the role that the $3N$ forces, both 
real and effective,  play in shell-model calculations 
based on the results we obtain  
for the even-mass 
$N=82$ isotonic chain.
This study has been performed with the same input for all nuclei of the 
isotonic chain, namely  the SP energies  and the matrix elements 
of the two-body effective interaction employed in the calculations discussed in
Sec.~\ref{sec:res3}.
Here, the shell-model calculations have been carried out by
using the Oslo code \cite{Engeland91}.

Let us first consider the two valence-proton  system  $^{134}$Te.
As mentioned in Sec.~\ref{sec:NN3}, a genuine $3N$ potential, which may 
affect the one-
and two-body components of the shell-model effective interaction  
$H_{\rm eff}^{1}$
of Eq.~(\ref{heffop}), 
has never been considered explicitly.
The contributions of the $3N$ 
potential to the one-body term of  
$H_{\rm eff}^{1}$ may  be assumed  as  
taken into account when,  as usual,the  
SP energies are taken from the experimental data.
As regards the contribution of the $3N$ potential to 
the two-body component of $H_{\rm eff}^{1}$, it may  
be reduced with an appropriate choice of the cutoff momentum $\Lambda$
of the $V_{\rm low-k}$.
In 
\cite{Schwenk05}, it was suggested to choose  $\Lambda$
around 2 fm$^{-1}$, since this value seems to minimize the role of the 
$3N$ potential when calculating the $^3$H and $^4$He binding energies. In this 
context, it is worth recalling that here  we have used experimental
SP energies  and an effective interaction derived from a $V_{\rm low-k}$ 
with $\Lambda=2.2$ fm$^{-1}$. 

For  the ground-state energy of  $^{134}$Te, relative 
to the doubly  magic $^{132}$Sn, we obtain a value of -20.796 MeV to be compared 
with the experimental one -20.584 MeV. As regards the energy spectrum, 
the rms deviation is 0.115 MeV (see Fig. \ref{psequiv}).  
These results seem to validate for many-nucleons systems  the 
hypothesis made in \cite{Schwenk05} about the choice of $\Lambda$.

\begin{figure}[hbtp]
\begin{center}
\includegraphics[scale=0.70,angle=90]{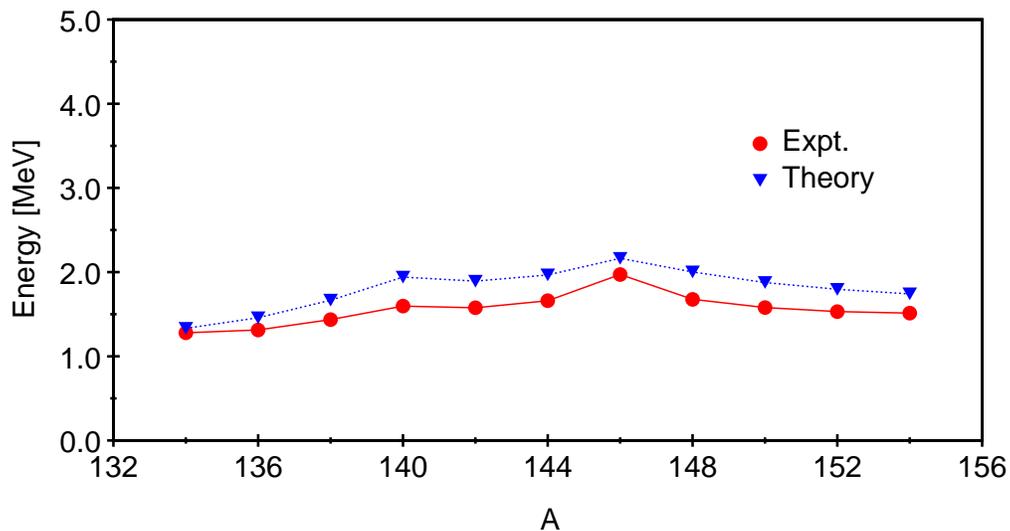}
\end{center}
\caption{Excitation energies of the $J=2^+$ yrast states for $N=82$ 
isotones as a function of mass number $A$.
\label{j02n82}}
\end{figure}
\vspace{1.3truecm}

Let us now come to nuclei with more than two valence nucleons. We know that 
in this case
many-body terms appear in the shell-model 
effective interaction, even if the original potential is only a two-body one. 
The effects of these terms
have not  yet been quantitatively evaluated 
\cite{Dean04,Kuo90,Muther85,Polls83}. However, it is generally  
believed that effective many-body forces have little influence on the 
excitation energies, while the binding energies
may be significantly affected \cite{Ellis05}. In this connection, it may be 
mentioned that  Talmi
\cite{Talmi93} has shown that the experimental binding energies of the tin isotopes can be reproduced 
quite accurately employing an empirical two- and three-body interaction.

\begin{figure}[hbtp]
\begin{center}
\includegraphics[scale=0.70,angle=90]{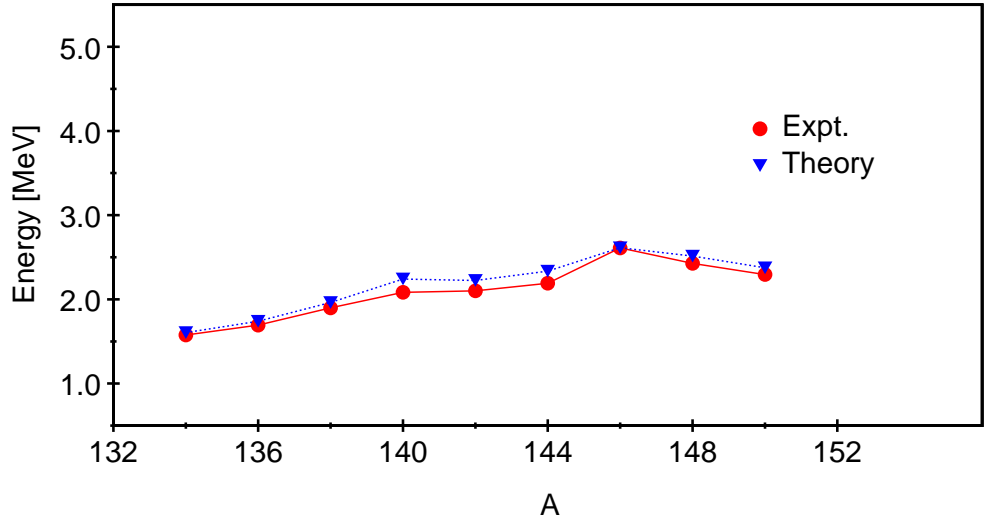}
\end{center}
\caption{Same as in Fig.\ref{j02n82} but for  $J=4^+$.
\label{j04n82}}
\end{figure}

\begin{figure}[hbtp]
\begin{center}
\includegraphics[scale=0.70,angle=90]{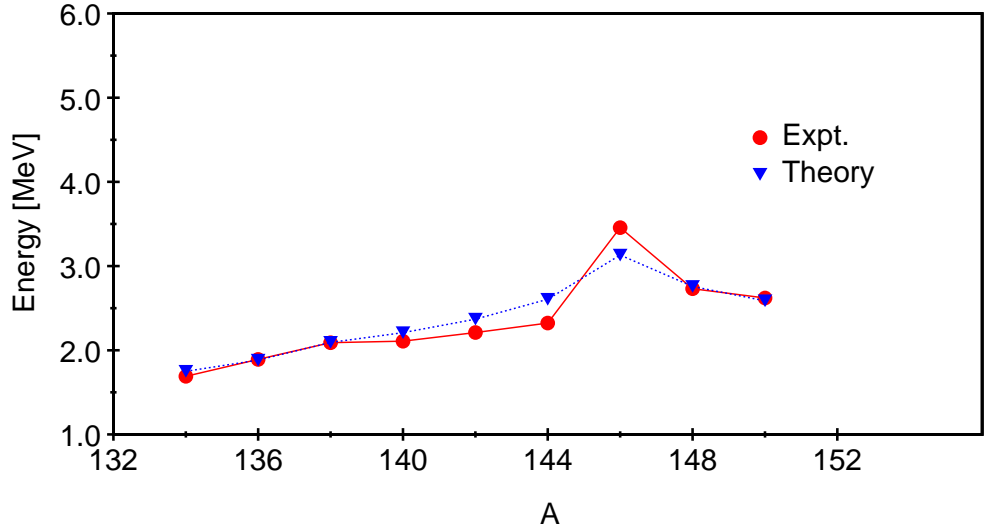}
\end{center}
\caption{Same as in Fig.\ref{j02n82} but for  $J=6^+$.
\label{j06n82}}
\end{figure}

In order to study the effective many-body forces, we have calculated the 
excitation energies of the lowest-lying yrast  states in the even $N=82$ 
isotones.  
In Figs. \ref{j02n82}, \ref{j04n82}, and \ref{j06n82}, the calculated energy 
of the  $J^{\pi}=2^{+}$, $4^{+}$, and $6^{+}$
states, respectively, are reported  as a function of the mass number $A$ and 
compared with the experimental data. 
It is seen that the  agreement between theory and experiment is of the same
 quality all over the isotonic chain, 
thus confirming that effective many-body forces play a negligible role
 in the description of the relative energy spectra.

\begin{figure}[hbtp]
\begin{center}
\includegraphics[scale=0.70]{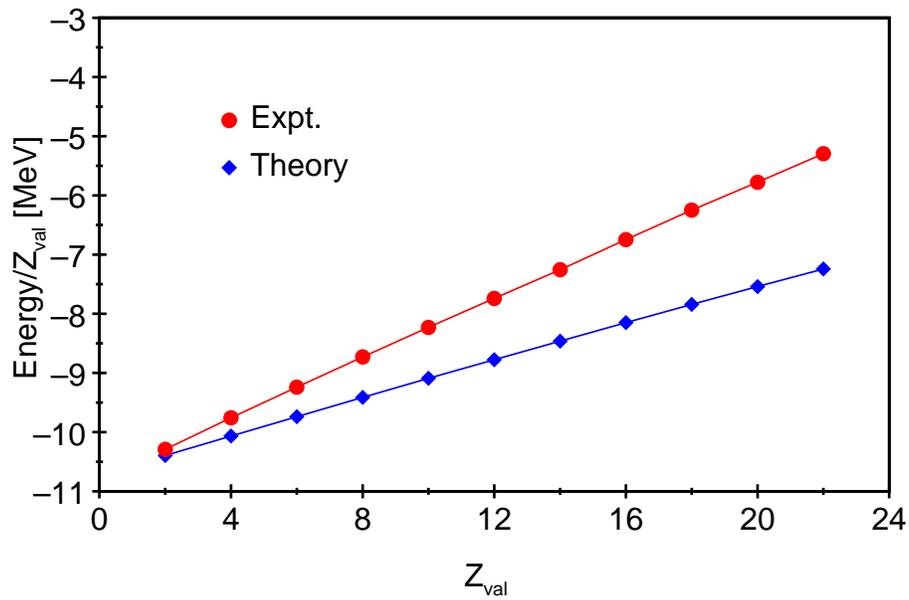}
\end{center}
\caption{Ground-state energies per valence proton of $N=82$ isotones
  as a function of the valence-proton number  $Z$.
\label{j00n82}}
\end{figure}

In Fig. \ref{j00n82}, we report the ground-state energy per valence nucleon.
It appears that the experimental and theoretical behaviors diverge 
when increasing the number of protons. As in the case of tin isotopes \cite{Engeland99}, the binding 
energy is increasingly overestimated. This deficiency may be healed by adding a small repulsive monopole 
contribution, whose origin may be traced to effective many-body forces \cite{Talmi93}.

\section {Summary and conclusions} \label{sec:sum}

In this paper, we have tried to give a self-contained review of the present 
status of shell-model calculations employing two-body effective interactions 
derived from the free nucleon-nucleon potential. A main feature of this kind of
 calculations, which are commonly referred to as realistic shell-model 
calculations, is that no adjustable parameter appears 
in the determination of the effective interaction. This removes the uncertainty
 inherent in the traditional use of empirical interactions.

As discussed in the Introduction, only in the last few years has this approach, which started in the 1960s, acquired a reputation for being able to provide an accurate description of nuclear structure properties. This means that to place in its proper perspective the current status of the field it is helpful to look back and survey the difficulties faced and the progress achieved over about four decades. The present review, while aimed at giving an insight into modern realistic shell-model calculations, is also concerned with the above aspects in regard to both the $NN$ potential and the many-body methods for deriving the effective interaction.

As regards the first issue, in Sec.~\ref{sec:NN} we have discussed in some detail the main advances in the 
understanding of the $NN$ interaction from the early stages of development until the current
 efforts based on chiral effective field theory. This we have done to show how the advances in this 
field have been intimately related to the development of realistic nuclear structure calculations. 
The present stage of development is  certainly enough to provide a sound starting point for a 
quantitative description of nuclear structure properties. It remains a main task for the near 
future to 
assess the role of the three-body forces in many-nucleon systems. A brief discussion on this matter 
is given in Sec.~\ref{sec:res4}  based on the results for the even $N=82$ 
isotones.  

In Sec.~\ref{sec:SMEI} we have described how the shell-model effective interaction can be derived 
within the framework of degenerate time-dependent perturbation theory.

The problem of the short-range repulsion contained in the $NN$ potential has been discussed in Sec.~\ref{sec:G&V}, where, after an outline of the traditional Brueckner $G$-matrix method, we have described  
a new approach to the renormalization of $V_{NN}$, which consists in constructing a low-momentum $NN$ potential, $V_{\rm low-k}$, that preserves the physics of  $V_{NN}$ up to a cutoff momentum $\Lambda$. Since $V_{\rm low-k}$ is already a smooth potential, it can be used directly as input for the derivation of the shell-model effective interaction instead of the usual $G$ matrix. While its merit in this respect has been assessed by several calculations 
(Sec.~\ref{sec:res1}) the $NN$ potential $V_{\rm low-k}$ is currently 
attracting much attention. On the one hand, it is being applied in various contexts, for instance in the study of few-body 
problems \cite{Fujii04,Nogga04}. On the other hand, its properties are being studied in detail, in particular as regards the dependence on the original potential model \cite{Holt04,Coraggio05b,Covello05}. Concerning shell-model calculations, results obtained with different $NN$ potentials using the $V_{\rm low-k}$
approach have been presented in Sec.~\ref{sec:res2}.

A main motivation for this review has been the practical success achieved by realistic shell-model calculations in the last few years, after a long period of hope and dismay. In Sec.~\ref{sec:SMC}. we have first surveyed the main efforts in this field going from the early to  present days. We have then presented, by way of illustration, some selected results of recent calculations for nuclei around doubly magic $^{132}$Sn. These neutron-rich nuclei, which lie well away from the valley of stability, offer the opportunity for a stringent test of the matrix elements of the effective interaction. The very good agreement with the available experimental data shown in Sec.~\ref{sec:res3} supports confidence in the predictive power of realistic shell-model calculations in the regions of shell closures off stability, which are of great current interest.

As already pointed out above, in this review our focus has been on shell-model calculations employing 
genuine free $NN$ potentials.
As a consequence of this choice, even a brief summary of the many successful shell-model calculations employing modified versions of free $NN$ potentials would have been beyond the scope of the present article. Here, we have also not touched upon another important aspect of shell-model work, namely the development of new methods, such as the Monte Carlo shell model (MCSM), as well as of high-quality codes, such as ANTOINE \cite{Caurier99}, which has greatly extended the feasibility of large-scale calculations. However, both these issues are discussed in
detail in the recent article by Caurier {\em et al.} \cite{Caurier05} 
while for a specialized review of the MCSM we refer to Ref.~\cite{Otsuka01}.

To conclude, we may say that the stage for realistic shell-model calculations
 is by now well set, which is a major step towards the understanding
 of the properties of complex nuclei in terms of the forces among nucleons. 
 In  this context, the $V_{\rm low-k}$ approach provides a new effective tool 
to handle the short-range repulsion of the free $NN$ potential.

\section*{Acknowledgments}

Helpful discussions with Gerry Brown, Scott Bogner, Jason Holt and
Achim Schwenk are gratefully acknowledged. This work was supported in
part by the Italian Ministero dell'Istruzione, dell'Universit\`a e
della Ricerca (MIUR), and by the U.S. DOE Grant No.
DE-FG02-88ER40388.

\bibliographystyle{h-elsevier} 

\bibliography{BiblioCoraggioppp} 

\end{document}